\documentclass[aps,prd,showpacs,superscriptaddress,groupedaddress,nofootinbib]{revtex4-2}  
\usepackage[T1]{fontenc}
\usepackage{lmodern} 
\usepackage{graphicx}  
\usepackage{dcolumn}   
\usepackage{bm}        
\usepackage{amssymb}   
\usepackage{amsmath}
\usepackage{bbold}
\usepackage{array}
\usepackage{makecell}
\usepackage{braket}
\usepackage{longtable}
\usepackage{supertabular,booktabs}

\usepackage{titlesec} 
\usepackage[colorlinks,citecolor=blue,urlcolor=blue,hypertexnames=true]{hyperref}
\usepackage{subfigure}

\usepackage[usenames,dvipsnames,svgnames,table]{xcolor} 

\renewcommand{\arraystretch}{2}

\newcommand{\be}{\begin{equation}}
\newcommand{\ee}{\end{equation}}
\newcommand{\bea}{\begin{eqnarray}}
\newcommand{\eea}{\end{eqnarray}}
\newcommand{\bml}{\begin{subequations}}
\newcommand{\eml}{\end{subequations}}
\newcommand{\bfig}{\begin{figure}}
\newcommand{\efig}{\end{figure}}

\newcommand{\slashed}{\hspace{-1.1ex}/}

\newcommand{\bmat}{\begin{pmatrix}}
\newcommand{\emat}{\end{pmatrix}}
\usepackage{graphicx,slashed,booktabs,xcolor,multirow,float,
amsfonts,bbold,mathtools,sidecap,tikz,bm,enumitem}
\usepackage{multirow}
\usepackage{titlesec}
\usepackage{hyperref}
\usepackage{amssymb}
\usepackage{pifont}

\titleclass{\subsubsubsection}{straight}[\subsection]

\newcounter{subsubsubsection}[subsubsection]
\renewcommand\thesubsubsubsection{\thesubsubsection.\arabic{subsubsubsection}}

\titleformat{\subsubsubsection}
  {\normalfont\normalsize\bfseries}{\thesubsubsubsection}{1em}{}
\titlespacing*{\subsubsubsection}
{0pt}{3.25ex plus 1ex minus .2ex}{1.5ex plus .2ex}

\makeatletter
\renewcommand\paragraph{\@startsection{paragraph}{5}{\z@}%
  {3.25ex \@plus1ex \@minus.2ex}%
  {-1em}%
  {\normalfont\normalsize\bfseries}}
\renewcommand\subparagraph{\@startsection{subparagraph}{6}{\parindent}%
  {3.25ex \@plus1ex \@minus .2ex}%
  {-1em}%
  {\normalfont\normalsize\bfseries}}
\def\toclevel@subsubsubsection{4}
\def\toclevel@paragraph{5}
\def\toclevel@paragraph{6}
\def\l@subsubsubsection{\@dottedtocline{4}{7em}{4em}}
\def\l@paragraph{\@dottedtocline{5}{10em}{5em}}
\def\l@subparagraph{\@dottedtocline{6}{14em}{6em}}
\makeatother

\begin{document}


\definecolor{lime}{HTML}{A6CE39}
\DeclareRobustCommand{\orcidicon}{\hspace{-2.1mm}
\begin{tikzpicture}
\draw[lime,fill=lime] (0,0.0) circle [radius=0.13] node[white] {{\fontfamily{qag}\selectfont \tiny \,ID}}; \draw[white, fill=white] (-0.0525,0.095) circle [radius=0.007]; 
\end{tikzpicture} \hspace{-3.7mm} }
\foreach \x in {A, ..., Z}{\expandafter\xdef\csname orcid\x\endcsname{\noexpand\href{https://orcid.org/\csname orcidauthor\x\endcsname} {\noexpand\orcidicon}}}
\newcommand{\orcidauthorA}{0000-0002-0459-3873}
\newcommand{\orcidauthorB}{0000-0001-9434-0505}
\newcommand{\orcidauthorC}{0000-0003-1081-0632}


\title{\textcolor{Sepia}{\textbf \huge\Large Quantum loop effects on the power spectrum and constraints on primordial black holes }}


\author{{\large  Sayantan Choudhury\orcidA{}${}^{1}$}}
\email{sayantan\_ccsp@sgtuniversity.org,  \\ sayanphysicsisi@gmail.com (Corresponding author)}
\author{\large Sudhakar~Panda\orcidB{}${}^{2,3}$}
\email{panda@niser.ac.in }
\author{ \large M.~Sami\orcidC{}${}^{1,4,5}$}
\email{ sami\_ccsp@sgtuniversity.org,  samijamia@gmail.com}

\affiliation{ ${}^{1}$Centre For Cosmology and Science Popularization (CCSP),\\
        SGT University, Gurugram, Delhi- NCR, Haryana- 122505, India,}
\affiliation{${}^{2}$School of Physical Sciences,  National Institute of Science Education and Research, Bhubaneswar, Odisha - 752050, India,}
\affiliation{${}^{3}$ Homi Bhabha National Institute, Training School Complex, Anushakti Nagar, Mumbai - 400085, India,}
\affiliation{${}^{4}$Center for Theoretical Physics, Eurasian National University, Astana 010008, Kazakhstan.}
	\affiliation{${}^{5}$Chinese Academy of Sciences,52 Sanlihe Rd, Xicheng District, Beijing.}

\begin{abstract}
We present a detailed exposition on the prospects of the formation of Primordial Black Holes (PBHs) during Slow Roll (SR) to Ultra Slow Roll (USR) sharp transitions in the framework of single-field inflation. We use an effective field theory (EFT) approach in order to keep the analysis model-independent and applicable to both the canonical and non-canonical cases. We show in detail how renormalizing the power spectrum to one loop order in $P(X,\phi)$ theories severely limits the prospects for PBH formation in a single-field inflationary framework. We demonstrate that for the allowed range of effective sound speed, $1<c_s<1.17$, the consistency of one-loop corrected power spectrum leaves a small window for black hole masses, $M_{\rm PBH}\sim \mathcal{O}(10^2-10^3)$gm to have sufficient e-foldings, $\Delta {\cal N}_{\rm Total}\sim {\cal O}(54-59)$ for inflation. We confirm that adding an SR regime after USR before the end of inflation does not significantly alter our conclusions. Our findings for sharp transition strictly rule out the possibility of generating large masses of PBHs from all possible models of single field inflation (canonical and non-canonical). \textcolor{black}{
Our results are at least valid for the situation where constraints from the loop effects are computed using either Late-Time (LT) or Adiabatic-Wave function
(AF) scheme followed by Power Spectrum (PS) renormalization schemes. }

\end{abstract}

\maketitle
\tableofcontents
\newpage

\section{Introduction}
Primordial black holes (PBHs) have been under an active consideration for past several years \cite{Zeldovich:1967lct,Hawking:1974rv,Carr:1974nx,Carr:1975qj,Chapline:1975ojl,Carr:1993aq,Kawasaki:1997ju,Yokoyama:1998pt,Kawasaki:1998vx,Rubin:2001yw,Khlopov:2002yi,Khlopov:2004sc,Saito:2008em,Khlopov:2008qy,Carr:2009jm,Choudhury:2011jt,Lyth:2011kj,Drees:2011yz,Drees:2011hb,Ezquiaga:2017fvi,Kannike:2017bxn,Hertzberg:2017dkh,Pi:2017gih,Gao:2018pvq,Dalianis:2018frf,Cicoli:2018asa,Ozsoy:2018flq,Byrnes:2018txb,Ballesteros:2018wlw,Belotsky:2018wph,Martin:2019nuw,Ezquiaga:2019ftu,Motohashi:2019rhu,Fu:2019ttf,Ashoorioon:2019xqc,Auclair:2020csm,Vennin:2020kng,Nanopoulos:2020nnh,Gangopadhyay:2021kmf,Inomata:2021uqj,Stamou:2021qdk,Ng:2021hll,Wang:2021kbh,Kawai:2021edk,Solbi:2021rse,Ballesteros:2021fsp,Rigopoulos:2021nhv,Animali:2022otk,Correa:2022ngq,Frolovsky:2022ewg,Escriva:2022duf,Kristiano:2022maq,Karam:2022nym,Riotto:2023hoz,Kristiano:2023scm,Riotto:2023gpm,Ozsoy:2023ryl,Ivanov:1994pa,Afshordi:2003zb,Frampton:2010sw,Carr:2016drx,Kawasaki:2016pql,Inomata:2017okj,Espinosa:2017sgp,Ballesteros:2017fsr,Sasaki:2018dmp,Ballesteros:2019hus,Dalianis:2019asr,Cheong:2019vzl,Green:2020jor,Carr:2020xqk,Ballesteros:2020qam,Carr:2020gox,Ozsoy:2020kat,Baumann:2007zm,Saito:2008jc,Saito:2009jt,Choudhury:2013woa,Sasaki:2016jop,Raidal:2017mfl,Papanikolaou:2020qtd,Ali-Haimoud:2017rtz,Di:2017ndc,Raidal:2018bbj,Cheng:2018yyr,Vaskonen:2019jpv,Drees:2019xpp,Hall:2020daa,Ballesteros:2020qam,Ragavendra:2020sop,Carr:2020gox,Ozsoy:2020kat,Ashoorioon:2020hln,Ragavendra:2020vud,Papanikolaou:2020qtd,Ragavendra:2021qdu,Wu:2021zta,Kimura:2021sqz,Solbi:2021wbo,Teimoori:2021pte,Cicoli:2022sih,Ashoorioon:2022raz,Papanikolaou:2022chm,Wang:2022nml,Mishra:2019pzq,ZhengRuiFeng:2021zoz,Cohen:2022clv,Arya:2019wck,Bastero-Gil:2021fac,Correa:2022ngq,Gangopadhyay:2021kmf,Cicoli:2022sih,Brown:2017osf,Palma:2020ejf,Geller:2022nkr,Braglia:2022phb,Kawai:2022emp,Frolovsky:2023xid,Aldabergenov:2023yrk,Aoki:2022bvj,Frolovsky:2022qpg,Aldabergenov:2022rfc,Ishikawa:2021xya,Gundhi:2020kzm,Aldabergenov:2020bpt,Cai:2018dig,Fumagalli:2020adf,Cheng:2021lif,Balaji:2022rsy,Qin:2023lgo,Riotto:2023hoz}. The original motivation to investigate their formation in the early universe was to understand the origin of supermassive black holes in the universe, including the one at the centre of our own galaxy, which could be seeded by PBHs.
Their formation due to the collapse of large fluctuations in the early universe seems to be a realistic possibility. Inflation, soon after its invention, became a natural arena where such large fluctuations suitable to PBH formation could be generated quantum mechanically. These exotic objects might be hiding real secrets of nature. To this effect, let us note that, in spite of the remarkable successes of modern cosmology, there are distinguished unresolved issues related to each of the four stages of the evolution of the universe; for instance, a lack of understanding of baryon asymmetry (radiation era) and the nature and origin of dark matter in the matter dominated regime. Remarkably, these objects could address the mentioned  outstanding issues of modern cosmology. Indeed, PBHs after their formation, undergo Hawking evaporation, and the products of such a process could give rise to dark matter and/or observed baryon asymmetry in the universe.

It may be noted that the process of PBH formation during inflation is under intense scrutiny in the literature at present \cite{Kristiano:2022maq,Riotto:2023hoz,Kristiano:2023scm,Riotto:2023gpm}.  It has been  argued using canonical single field framework of inflation that one-loop corrections \footnote{To know more about one-loop effects on inflation and de Sitter space see the refs. \cite{Adshead:2008gk,Senatore:2009cf,Senatore:2012nq,Pimentel:2012tw,Sloth:2006az,Seery:2007we,Seery:2007wf,Bartolo:2007ti,Seery:2010kh,Bartolo:2010bu,Senatore:2012ya,Chen:2016nrs,Markkanen:2017rvi,Higuchi:2017sgj,Syu:2019uwx,Rendell:2019jnn,Cohen:2020php,Green:2022ovz,Premkumar:2022bkm}.} severely limit their formation during sharp transitions from Slow Roll (SR) to Ultra Slow Roll (USR) \cite{Choudhury:2023vuj}. It was further pointed out in ref.\cite{Choudhury:2023jlt} that 
bringing in the non-canonical features would further worsen the outcome. To this effect, it is more than desirable to use a generic framework of single-field inflation which is model-independent and applies to both the canonical and non-canonical cases. See refs. \cite{Choudhury:2023hvf,Choudhury:2023kdb,Choudhury:2023hfm,Bhattacharya:2023ysp,Choudhury:2023fwk,Motohashi:2023syh,Firouzjahi:2023ahg,Franciolini:2023lgy,Firouzjahi:2023aum,Cheng:2023ikq,Tasinato:2023ukp} for more details on this issue where the authors have studied various possibilities along the same subject line having smooth and sharp transitions. Effective field theory (EFT) of inflation \cite{Weinberg:2008hq,Cheung:2007st,Choudhury:2017glj,Delacretaz:2016nhw} is a suitable scenario
that relies upon an effective action that is valid below the UV cut-off scale.  Additionally, the structure of the EFT action is constrained by the underlying symmetry.  The principal objective is constructing a UV complete theory that is allowed by the underlying symmetries. However, to date, we really do not have a proper understanding of the full-proof version of the UV complete theory, which can be constructed by utilizing fundamental physical principles at very high energy scales where the UV cut-off scale is fixed.  In principle, this cut-off can be fixed at the scale, $\Lambda_{UV}=M_{pl}\sim 10^{19}{\rm Gev}$. The theory above the scale $\Lambda_{UV}$ is completely unknown and treated as a black box in this type of analysis. In a technical sense, this is often referred to as the hidden sector of the theory in the corresponding EFT literature. We really don't need to think about this sector in principle, as at the end of the day all such degrees of freedom should be path integrated out to formulate a viable description of the EFT made of only visible sector field degrees of freedom. This visible sector description of the EFT setup should be valid below the mentioned UV cut-off scale, or the maximum can run up to the mentioned scale, i.e. $\Lambda_{EFT}\leq\Lambda_{UV}$. Consequently, the coupling parameters that appear in front of each EFT operator and describe the visible sector of the theory can also run from the lower scale to the mentioned cut-off scale, because each of them is in principle a function of the running scale of the underlying theory. The behaviour, pole structure, and dependence of the mentioned coupling parameters with respect to the scale of the underlying theory can be computed by solving Renormalization Group (RG) flow equations if we have a full understanding of the construction of a UV complete theory from fundamental physical principles. For more details on these aspects see refs. \cite{Assassi:2013gxa,Baumann:2019ghk,Green:2022ovz,Cohen:2020php,Green:2020txs,Gorbenko:2019rza,Burgess:2022rdo,Burgess:2021luo,Burgess:2020tbq,Burgess:2020qsc,Burgess:2020fmg,Burgess:2017ytm,Burgess:2015ajz,Burgess:2014eoa,Collins:2012nq,Boyanovsky:2011xn,Burgess:2010dd}.

There are three possibilities using which one can construct a basic EFT set-up, which is appended below point-wise:
\begin{enumerate}
    \item \underline{\bf Top-down approach:}
    In this description, one needs to start with the UV complete theory, which is valid at $\Lambda_{EFT}\leq\Lambda_{UV}$. Since the proper understanding of the construction of the UV complete theory is not fully known yet, it is extremely difficult to start with this particular approach to construct an EFT theory. Apart from having difficulties in the technical construction, a lot of efforts have been made to construct an EFT from string theory because, out of other possible candidates, this is the most successful one that can be used to describe a UV complete version of quantum field theory. However, due to the enormous number of compactification schemes in the four-dimensional reduced version, one can have different types of EFT theories that can address various cosmological frameworks, starting from inflation to the late time acceleration of our universe. In this paper, we have not used this possibility to construct the EFT for primordial cosmology.
 \item \underline{\bf Intermediate approach:}
     In this description instead of starting with a UV complete theory, like String Theory, we use the concept of hidden sector and visible sector, which we have mentioned at the starting point of this section. In this description, we consider a minimalistic version where a heavy scalar field is interacting with a low-mass scalar field. Here the heavy field belongs to the hidden sector and the low mass field corresponds to the visible sector. To construct an EFT of the visible sector field in four dimensions for primordial cosmology one needs to integrate out heavy field from the UV complete two-field interacting theory. After performing this procedure one should have renormalizable (relevant and marginal) operators as well as non-renormalizable (irrelevant) higher mass dimensional operators, which all are suppressed by the underlying scale of EFT i.e. $\Lambda_{EFT}\leq\Lambda_{UV}$. All of the coefficients appearing in the front of these operators need to be fixed by the Wilsonian point of view of EFT to describe an underlying cosmological setup. However, this approach is strictly model dependent, extremely tedious and the starting point is a bit phenomenological. For this reason, in this paper, we have discarded the possibility to construct the EFT for primordial cosmology. 

      \item \underline{\bf Bottom-up approach:}
     Finally, in this description, instead of starting with a specific version of UV complete theory or utilizing a minimalistic phenomenological version of this type of theory (as mentioned in the intermediate approach), by making use of the underlying symmetry principle, we introduce a pure gravitational sector characterized by  Einstein theory written in a quasi de Sitter background. Additionally, two more time-dependent operators are inserted, which mimic the role of scalar field action in the standard general $P(X,\phi)$ theory \cite{Chen:2010xka,Chen:2006nt} description\footnote{We discuss the structure of such a $P(X,\phi)$ theory and its possible outcomes in the later part of this section as well as in the end part of this paper.}. On top of these mentioned contributions, we further introduce various combinations of gravitational fluctuations in terms of the perturbations on the temporal component of the metric tensor as well as on the extrinsic curvature. All such operators mimic the role of Wilsonian operators, which can cover both the renormalizable (relevant and marginal) operators as well as the non-renormalizable (irrelevant) higher mass dimensional operators. The good part of this approach is this setup doesn't rely on the appearance of any scalar field but on the specific structure of the kinetic contributions and the specific structure of the effective potential of the mentioned scalar field. This helps us make a completely model-independent prediction out of this prescribed theoretical setup. In this specific work, our job is to fix the Wilson coefficients, which in turn constrain the effective sound speed $c_s$ as well as the two-point primordial cosmological correlation function computed from this specific EFT setup used in this work. Now, one natural question comes to mind: since we have not used any specific structure of $P(X,\phi)$ model action, from where exactly the scalar curvature perturbation is generated, whose quantum loop corrected two-point correlation is the most desirable outcome of this work. The answer to this crucial question lies in St$\ddot{u}$ckelberg technique which creates Goldstone modes as an outcome of time diffeomorphism symmetry breaking in the unitary gauge. This is exactly similar to spontaneous symmetry breaking as appearing in the context of $SU(N)$ gauge theory. If we restrict our analysis to the linear regime, then one can make a direct connection between the mentioned Goldstone mode and the scalar curvature perturbation mode. Once this identification has been established, one can do the rest of the analysis in the language of scalar curvature perturbation within the framework of primordial cosmology. We have strictly followed this approach in this paper, which will be discussed in detail in the later half of this paper. In the end, to make a fruitful connection with the known models of $P(X,\phi)$ theories, one can further comment on how exactly the constraint on the effective sound speed $c_s$ will further fix the couplings of these models. We have discussed this possibility in detail at the end of this paper.
\end{enumerate}

         In terms of the magnitude of the sound speed parameter $c_s$, the framework  is categorized in the following class of theories:
 \begin{enumerate}
     \item \underline{\bf Ghost EFT:}  It is described by $c_s=0$,  which was proposed by  Arkani-Hameda, Creminelli, Mukohyama, and Zaldarriagaa studied particularly in the context of inflation in ref.\cite{Arkani-Hamed:2003juy}. This particular model is beyond the scope of our analysis performed in this paper. However, this might be a good direction to look for the outcomes in the present context.

     \item \underline{\bf Non-canonical and causal EFT:}  This are characterized by $0<c_s<1$ which is also a well-studied domain in cosmology in many contexts. It is important to note that, in the UV complete description, all classes of $P(X,\phi)$ theories minimally coupled to gravity having the general structure of $P(X,\phi)$\footnote{Here $P(X,\phi)$ in principle a general functional of the kinetic term $X=-1/2\left(\partial_{\mu}\phi\right)^2$ and the scalar field $\phi$. For an example, in the case of Tachyon and $K$ inflation this function is given by, $P(X,\phi)=-V(\phi)\sqrt{1-2\alpha^{'}X}$, and $P(X,\phi)=K(X)-V(\phi)$, where $V(\phi)$ is the effective potential in this description. } as well as various types of modified gravity theories can be described by this specific EFT setup where causality ($c_s<1$) and ghost-free criteria are strictly maintained. A few known examples are, Dirac-Born-Infeld (DBI), Tachyon, $K$ inflation, etc. See refs. \cite{Alishahiha:2004eh,Mazumdar:2001mm,Choudhury:2002xu,Panda:2005sg,Chingangbam:2004ng,Armendariz-Picon:1999hyi,Garriga:1999vw} for more details. 

     \item \underline{\bf Canonical and causal EFT:} It is described by $c_s=1$, which basically represents a scalar field theory minimally coupled to gravity having a structure, $P(X,\phi)=\left(X-V(\phi)\right)$, where $X$ is the kinetic term and $V(\phi)$ is the effective potential of a single scalar field $\phi$. This type of theory strictly satisfies the causality constraint, $c_s<1$ in the EFT description, and has been studied in the various contexts of cosmology. See refs. \cite{Choudhury:2017glj,Naskar:2017ekm,Choudhury:2015pqa,Choudhury:2014sua,Choudhury:2014kma,Choudhury:2013iaa,Baumann:2022mni,Baumann:2018muz,Baumann:2015nta,Baumann:2014nda,Baumann:2009ds} more details on these aspects.

     \item \underline{\bf Non-canonical and a-causal EFT:}
     This is characterized by $c_s>1$, which is satisfied by a very restricted class of EFTs available in the corresponding literature. This type of the theories has to be non-canonical, but most importantly breaks the causality constraint. For this specific reason, this is known as a-causal EFT where the superluminal effect is prominent. One of the most promising examples of this type of theory is GTachyon \footnote{Here GTachyon is treated to be a special class of $P(X,\phi)$ theory, which is actually a generalized version of Tachyon as appearing in String Theory. In such a model the effective sound speed varies in a large range, going from a causal to an acausal regime. The prospect of such a model in the present framework is discussed in detail in the Appendix \ref{app:P} in great detail. } in which one can understand this superluminal effect clearly in a part of the parameter space of theory. The other part of the parameter space supports causality and perfectly matches the predictions coming from the usual Tachyon. See ref.\cite{Choudhury:2015hvr} for more details.
 \end{enumerate}

In this work, we adhere to well known St$\ddot{u}$ckelberg technique in the unitary gauge where the Goldstone modes are created, which further generates the scalar comoving cosmological perturbation at the cost of breaking the time diffeomorphism symmetry. This  methodology is restricted to all classes of $P(X,\phi)$ theories which give rise to a single-field inflationary paradigm. Nonetheless, this setup can be generalized to multiple scalar field theories \cite{Senatore:2010wk,Khosravi:2012qg,Shiu:2011qw}. In what follows, we present a detailed analysis of one-loop corrections to the power spectrum within the framework of EFT of single-field inflation. The latter gives rise to important constraints on the process of formation of PBHs in the paradigm of single-field inflation, which is discussed in great detail in this paper.

The organization of this paper is as follows: In section \ref{s1} we have discussed the basics and the technical details of the Effective Field Theory (EFT) of Single Field Inflation. Further, in section \ref{s3a} we discuss the second-order perturbation from scalar modes and its connection with the Goldstone EFT-driven primordial cosmology. For this study, we consider two frameworks. Here Framework I is made up of one Slow Roll (SRI), then Ultra Slow Roll (USR), and then another Slow Roll (SRII) followed by the end of inflation. On the other hand, Framework II is made up of one Slow Roll (SR) and then Ultra Slow Roll (USR) at the end of which inflation ends.  Next, in section \ref{s4} we discuss the well-known in-in formalism using which we compute the one-loop corrected primordial power spectrum for scalar modes from both the mentioned frameworks. Further, in section \ref{s5} we compute the renormalization of the power spectrum which is derived in section \ref{s4c}. After that, in section \ref{s6} we discuss the important role Dynamical Renormalization Group (DRG) resummation technique \cite{Boyanovsky:1998aa,Boyanovsky:2001ty,Boyanovsky:2003ui,Burgess:2009bs,Dias:2012qy,Chen:2016nrs,Baumann:2019ghk,Burgess:2009bs,Chaykov:2022zro,Chaykov:2022pwd} using which we compute a softened version of the all-loop corrected power spectrum for the scalar modes. Then in section \ref{s7} we discuss the numerical results, outcomes, and further constraints from Framework I and Framework II in detail. Further in section \ref{s8} we discuss the stringent constraint on the PBH mass and its connection with the on-loop correction in the primordial power spectrum. Also in this section, we try to connect the underlying EFT setup with the known $P(X,\phi)$ theories of single field inflation, which helps us to give a realistic estimate and state the present status of these models in the light of PBH formation. Next, in section \ref{s9} we make a comparison between the outcomes obtained from Framework I and II in the light of PBH formation. Further, in section \ref{comp} we make a comparison between the outcomes obtained from Framework I and II in the light of PBH formation. a comparative study among the various outcomes from different studies which are obtained in the presence of smooth and sharp transition from different phases under consideration. Next, in section \ref{rs} we have provided a detailed discussion on the renormalization scheme dependence and the correctness of the derived result in this paper. Finally, in section \ref{s10} we conclude with some interesting future possible directions. Additionally, we provide four appendices \ref{s2}, \ref{app:P}, \ref{app:PBH}, \ref{app:A}, and \ref{app:B} where the details of the computations are given for more completeness. 

\section{The Effective Field Theory (EFT) of Single Field Inflation in a nutshell}
\label{s1}

If we like to describe the formulation in terms of a scalar field degrees of freedom, then it is important to note that, such contribution transforms into a scalar under the following full diffeomorphism symmetry:
\be x^{\nu}\longrightarrow x^{\nu}+\xi^{\nu}(t,{\bf x})~~~\forall~ \nu=0,1,2,3~.\ee   
In this context of discussion, the diffeomorphism parameter is represented by $\xi^{\nu}(t,{\bf x})$. Now, instead of utilizing the full symmetry one can consider the following two-fold parallel scenarios where the perturbation on the field $\delta \phi$ transforms:
\begin{enumerate}
    \item As a scalar only under the spatial part of the space-time diffeomorphism symmetry, and

    \item In a non-linear fashion only under the temporal part of the space-time diffeomorphism symmetry. 
\end{enumerate}
These specific changes can be expressed as follows:
\bea
{\textcolor{Sepia}{\bf Broken~ spatial~ diffeomorphism:}}~~	t&&\longrightarrow t,~x^{i}\longrightarrow   x^{i}+\xi^{i}(t,{\bf x})~~~\forall~ i=1,2,3
~~~\delta\phi\longrightarrow \delta\phi,\\	
{\textcolor{Sepia}{\bf Broken~ temporal~ diffeomorphism:}}~~	t&&\longrightarrow t+ \xi^{0}(t,{\bf x}),~x^{i}\longrightarrow x^{i}~~~\forall~ i=1,2,3
	~~~\delta\phi\longrightarrow \delta\phi +\dot{\phi}_{0}(t)\xi^{0}(t,{\bf x}).\quad\quad
\eea
In this context, the spatial and temporal diffeomorphism parameters are described by, $\xi^{0}(t,{\bf x})$ and $\xi^{i}(t,{\bf x})\forall i=1,2,3$, respectively. Here we adopt a gravitational gauge, \be \phi(t,{\bf x})=\phi_{0}(t),\ee where $\phi_{0}(t)$ represents the background time-dependent scalar field which is embedded in homogeneous isotropic spatially flat FLRW space-time.
Additionally, this necessitates that in this gauge choice:
\be \delta \phi(t,{\bf x})=0~,\ee   
The following key ingredients are required to create this EFT set up:
\begin{itemize}
	\item 
 A function of the gravitational space-time metric $g_{\mu\nu}$ must be required for EFT operators.  The Riemann tensor $R_{\mu\nu\alpha\beta}$,  the Ricci tensor $R_{\mu\nu},$ and the Ricci scalar $R$, which are also parts of the EFT action, may now be calculated using the derivatives of the space-time metric.

	\item 
 The polynomial powers of the metric's temporally perturbed component ($\delta g^{00}$), which are denoted by \be \delta g^{00}= \left(g^{00}+1\right),\ee is first needed to build the EFT action. Here, the time element of the background metric is represented by $g^{00}$.  This operator has to be invariant under the spatial diffeomorphism symmetry transformation.  
 \item 
 The background geometry is described by the spatially flat FLRW space-time metric with the quasi de Sitter solution in this context is denoted by:
\bea ds^2=a^2(\tau)\left(-d\tau^2+d{\bf x}^2\right),\eea
The scale factor for the quasi-de Sitter backdrop is solved in this case by the following expression:
 \be a(\tau)=-\frac{1}{H\tau} \quad \quad {\rm where} \quad -\infty<\tau<0,\ee 
 where $H$ is the Hubble parameter.  It is important to note that, here $\tau$ is the conformal time which is related to the physical time $t$ via the following equation in quasi de Sitter space \footnote{ Here we have used the relation $d\tau=dt/a$ to find out the connection between the physical and conformal time in quasi de Sitter space.}:
 \bea t=\frac{1}{H}\ln\left(-\frac{1}{H\tau}\right)\quad {\rm where}\quad 0<t<\infty.\eea
	
	\item 
 Additionally, in order to create the EFT action, the polynomial powers of the temporally perturbed component of the extrinsic curvature at constant time slice ($\delta K_{\mu\nu}$) are needed. This expression is given by \be \delta K_{\mu\nu}=\left(K_{\mu\nu}-a^2Hh_{\mu\nu}\right),\ee  where the extrinsic curvature ($K_{\mu\nu}$), unit normal ($n_{\mu}$),  and induced metric ($h _{\mu\nu}$) are defined in this context as:
\bea K_{\mu \nu}&=&h^{\sigma}_{\mu}\nabla_{\sigma} n_{\nu}\nonumber\\
&=&\left[\frac{\delta^{0}_{\mu}\partial_{\nu}g^{00}+\delta^{0}_{\nu}\partial_{\mu}g^{00}}{2(-g^{00})^{3/2}}
+\frac{\delta^{0}_{\mu}\delta^{0}_{\nu}g^{0\sigma}\partial_{\sigma}g^{00}}{2(-g^{00})^{5/2}}-\frac{g^{0\rho}\left(\partial_{\mu}g_{\rho\nu}+\partial_{\nu}g_{\rho\mu}-\partial_{\rho}g_{\mu\nu}\right)}{2(-g^{00})^{1/2}}\right],\\
h_{\mu \nu}&=&g_{\mu \nu}+n_{\mu} n_{\nu},\\
n_{\mu}&=&\frac{\partial_{\mu}t}{\sqrt{-g^{\mu \nu}\partial_{\mu}t \partial_{\nu}t}}
=\frac{\delta_{\mu}^0}{\sqrt{-g^{00}}},\\
\left[\delta K\right]^{m+2} &=&\delta K^{\mu_{1}}_{\mu_{2}}\delta K^{\mu_{2}}_{\mu_{3}}\delta K^{\mu_{3}}_{\mu_{4}}\cdots\delta K^{\mu_{m+1}}_{\mu_{m+2}}\delta K^{\mu_{m+2}}_{\mu_{1}}. \eea	
\end{itemize} 


Now, since computing the tree level and one-loop adjusted two-point correlation functions is of importance to us, we have limited it to the following abbreviated EFT action:
\bea
 S&=&\displaystyle\int d^{4}x \sqrt{-g}\left[\frac{M^2_{pl}}{2}R-c(t)g^{00}-\Lambda(t)+\frac{M^{4}_{2}(t)}{2!}\left(g^{00}+1\right)^2+\frac{M^{4}_{3}(t)}{3!}\left(g^{00}+1\right)^3~~~~~~~~\nonumber\right.\\&&
	\left.\displaystyle~~~~~~~~~~~~~~~~~~~~~~~~~~~~~~~~~~~~~~~~~~~~~~~~-\frac{\bar{M}^{3}_{1}(t)}{2}\left(g^{00}+1\right)\delta K^{\mu}_{\mu}-\frac{\bar{M}^{2}_{2}(t)}{2}(\delta K^{\mu}_{\mu})^2-\frac{\bar{M}^{2}_{3}(t)}{2}\delta K^{\mu}_{\nu}\delta K^{\nu}_{\mu}\right].
	\eea
 Here
the time-dependent coefficients $M_1(t)$, $M_3(t)$,  $\bar{M}_1(t)$,  $\bar{M}_2(t)$ and $\bar{M}_3(t)$ are mimicking the role of Wilson coefficients which we need to be fixed from our analysis. For more details on the construction of the EFT action and the related decoupling limit see the discussions appearing in the Appendix \ref{s2}.
	Further consider only the background contributions in the above-mentioned EFT action the corresponding Friedmann equations can be expressed as:
\bea
\left(\frac{\dot{a}}{{a}}\right)^2=H^2&=&\frac{1}{3M^2_{ pl}}\Bigg(c(t)+\Lambda(t)\Bigg)=\frac{{\cal H}^2}{a^2},\\
\frac{\ddot{a}}{{a}}=\dot{H}+H^2&=&-\frac{1}{3M^2_{ pl}}\Bigg(2c(t)-\Lambda(t)\Bigg)
=\frac{{\cal H}^{'}}{a^2},
\eea
where $'$ represents the derivative with respect to the conformal time $\tau$.  Additionally,  it is important to note that, $\displaystyle {\cal H}=\frac{a^{'}}{a}=aH.$  Solving these equations we found the following expressions for the time dependent parameters $c(t)$ and $\Lambda(t)$ at the background level:
\bea c(t)&=&-M^2_{pl} \dot{H}=-\frac{M^2_{pl}}{a^2}\Bigg({\cal H}^{'}-{\cal H}^2\Bigg),\\
\Lambda(t)&=&M^2_{pl} \left(3H^2+\dot{H}\right)=\frac{M^2_{pl}}{a^2}\Bigg(2{\cal H}^2+{\cal H}^{'}\Bigg).\eea
Finally,  substituting the expressions for the time dependent parameters $c(t)$ and $\Lambda(t)$ the EFT action can be written as:
\bea
 S&=&\displaystyle\int d^{4}x \sqrt{-g}\left[\frac{M^2_{pl}}{2}R+M^2_{pl} \dot{H} g^{00}-M^2_{pl} \left(3H^2+\dot{H}\right)+\frac{M^{4}_{2}(t)}{2!}\left(g^{00}+1\right)^2+\frac{M^{4}_{3}(t)}{3!}\left(g^{00}+1\right)^3~~~~~~~~\nonumber\right.\\&&
	\left.\displaystyle~~~~~~~~~~~~~~~~~~~~~~~~~~~~~~~~~~~~~~~~~~~~~~~~~~~~~~-\frac{\bar{M}^{3}_{1}(t)}{2}\left(g^{00}+1\right)\delta K^{\mu}_{\mu}-\frac{\bar{M}^{2}_{2}(t)}{2}(\delta K^{\mu}_{\mu})^2-\frac{\bar{M}^{2}_{3}(t)}{2}\delta K^{\mu}_{\nu}\delta K^{\nu}_{\mu}\right].
	\eea
Further, one needs to introduce the following deviation parameters, commonly known as the slow-roll parameters,
\bea
\epsilon &=&\bigg(1-\frac{{\cal H}^{'}}{{\cal H}^2}\bigg),\quad\quad
\eta =\frac{\epsilon^{'}}{\epsilon {\cal H}}.
\eea
The following conditions must be met in order to achieve inflation in the SR region, $
\epsilon\ll 1,|\eta|\ll 1.$

\section{Tree level scalar power spectrum from EFT}
\label{s3}

\subsection{Second order perturbation from scalar mode: Connecting with Goldstone EFT}
\label{s3a}

Let us now consider the decoupling limit $E>E_{mix}=\sqrt{\dot{H}}$, where all the mixing contributions can be easily neglected. In such a situation the second-order Goldstone action ($S^{(2)}_{\pi}$) is represented by the following expression:
   \bea 
  	S^{(2)}_{\pi}&\approx&\displaystyle \int d^{4}x ~a^3\left[-M^2_{pl}\dot{H}\left(\dot{\pi}^2-\frac{1}{a^2}(\partial_{i}\pi)^2\right)
   	+2M^4_2 \dot{\pi}^2\right]=\displaystyle \int d^{4}x ~a^3\left(\frac{-M^2_{pl}\dot{H}}{c^2_s}\right)\Bigg(\dot{\pi}^2-c^2_s\frac{\left(\partial_{i}\pi\right)^2}{a^2}\Bigg).~~~~~\quad\quad\eea 
In this context, the effective sound speed is defined in terms of the EFT Wilson coefficient, given by:
   \bea c_{s}\equiv \frac{1}{\displaystyle \sqrt{1-\frac{2M^4_2}{\dot{H}M^2_{pl}}}},\eea 
The effective sound speed can very easily be connect with a large class of $P(X,\phi)$ theories\footnote{Here it is important to note that, in the case of general $P(X,\phi)$ theories the effective sound speed can be computed as \cite{Chen:2010xka,Chen:2006nt}:
 \be c_s=\sqrt{\frac{P_{,\bar{X}}}{P_{,\bar{X}}+2\bar{X}P_{,\bar{X}\bar{X}}}}.\ee
 Here $\bar{X}$ is background value of $X$.
 This specific expression will be extremely useful when we try to make connection among realistic $P(X,\phi)$ models with the completely model independent version of EFT in terms of the dimensionless coupling parameter $\displaystyle \frac{M^4_2}{\dot{H}M^2_{pl}}$, which can be written as:
 \be \frac{M^4_2}{\dot{H}M^2_{pl}}=\frac{1}{2}\Bigg(1-\frac{P_{,\bar{X}}}{P_{,\bar{X}}+2\bar{X}P_{,\bar{X}\bar{X}}}\Bigg)=\Bigg(\frac{\bar{X}P_{,\bar{X}\bar{X}}}{P_{,\bar{X}}+2\bar{X}P_{,\bar{X}\bar{X}}}\Bigg).\ee
 For more details see the Appendix \ref{app:P} and Appendix \ref{app:PBH} we have have discussed the connection between Goldstone EFT and $P(X,\phi)$ in detail.}.

   It is important to note that, the spatial part of the metric fluctuation is described by:
   \bea g_{ij}\sim a^{2}(t)\left[\left(1+2\zeta(t,{\bf x})\right)\delta_{ij}\right]~~\forall~~~i=1,2,3,\eea
   where the scale factor in the FLRW quasi de Sitter background space-time is $a(t)=\exp(Ht)$. Additionally, the scalar comoving curvature perturbation is represented by the symbol $\zeta(t,{\bf x})$.
   
   The scale factor $a(t)$ in this case transforms in the following way under the broken time diffeomorphism:
    \bea a(t)\Longrightarrow  a(t-\pi(t,{\bf x}))&=& a(t)-H\pi(t,{\bf x})a(t)+\cdots\approx a(t)\left(1-H\pi(t,{\bf x})\right)~,\eea
    using which we get:
    \bea a^2(t)\left(1-H\pi(t,{\bf x})\right)^2&\approx & a^2(t)\left(1-2H\pi(t,{\bf x})\right)=a^{2}(t)\left(1+2\zeta(t,{\bf x})\right).\eea 
    This suggests that the following connecting relationship can be used to express the scalar curvature perturbation $\zeta(t,{\bf x})$ in terms of Goldstone modes $\pi(t,{\bf x})$
    :
     \be \zeta(t,{\bf x})\approx-H\pi(t,{\bf x})~.\ee
     Since this relationship is clearly identified 
    , now one can further write down the expression for the second-order perturbed action ($S^{(2)}_{\zeta}$) in terms of the co-moving curvature perturbation $\zeta(t,{\bf x})$ as:
     \bea 
  	S^{(2)}_{\zeta}&=&\displaystyle \int d^{4}x ~a^3\left(\frac{M^2_{pl}\epsilon}{c^2_s}\right)\Bigg(\dot{\zeta}^2-c^2_s\frac{\left(\partial_{i}\zeta\right)^2}{a^2}\Bigg).~~~~~\quad\quad\eea
   Further, for the simplification purpose instead of performing the rest of the computation in terms of the physical time coordinate, we are now going to use the conformal time coordinate, in terms of which the second-order perturbed action can be written as:
   \bea S^{(2)}_{\zeta}&=&M^2_{pl}\displaystyle \int d\tau\;  d^3x\;  a^2\;\left(\frac{\epsilon}{c^2_s}\right)\Bigg(\zeta^{'2}-c^2_s\left(\partial_i\zeta\right)^2\Bigg).~~~~\quad\eea  
   
   \subsection{Constructing Mukhanov Sasaki equation from Goldstone EFT}
   \label{s3b}
   
   A new variable is defined now for the comoving curvature perturbation field redefinition purpose, which is given by:
   \bea v=zM_{ pl}\zeta \quad{\rm with}\quad z=\displaystyle \frac{a\sqrt{2\epsilon}}{c_s},\eea
   which is commonly known as the Mukhanov Sasaki (MS) variable in the present context of discussion. The second order perturbed action discussed above has the following canonically normalized form when expressed in terms of the MS variable:
   \bea
S^{(2)}_{\zeta}=\frac{1}{2}\int d\tau\;  d^3x\;  \bigg(v^{'2}-c^2_s\left(\partial_iv\right)^2+\frac{z^{''}}{z}v^{2}\bigg).
   \eea
After that, using the following ansatz for the Fourier transformation, we will write the aforementioned action in the Fourier space:
   \bea
   v(\tau,{\bf x})=\int\frac{d^3{\bf k}}{(2\pi)^3}\; e^{i{\bf k}.{\bf x}}\;v_{\bf k}(\tau).
   \eea
One may further recast the aforementioned action in the form of the following Fourier transformed scalar modes,
     \bea
S^{(2)}_{\zeta}&=&\frac{1}{2}\int \frac{d^3{\bf k}}{(2\pi)^3}\;d\tau\;\bigg(|v^{'}_{\bf k}|^{2}-\omega^2(k,c_s,\tau)|v_{\bf k}|^{2}\bigg).
   \eea
   Then after varying the above mentioned action the MS equation for the scalar perturbed modes can be expressed by the following expression,
   \bea
 v^{''}_{\bf k}+\omega^2(k,c_s,\tau)v_{\bf k}=0\,,
   \eea
   where the expression for the effective time dependent frequency is given by the following expression, 
   \bea \omega^2(k,c_s,\tau):=\left(c^2_sk^2-\frac{z^{''}}{z}\right)\quad\displaystyle{\rm where}\quad\frac{z^{''}}{z}\approx\frac{2}{\tau^2}.\eea
 The mathematical structure of the general solution is now fixed using the following mentioned normalisation condition provided in terms of the Klein Gordon product for the scalar perturbed modes, 
   \be v^{'*}_{\bf k}v_{\bf k}-v^{'}_{\bf k}v^{*}_{\bf k}=i.\ee
  In the next two subsections we are going to solve the Mukhanov Sasaki equation for two frameworks, which is going to be extremely useful for the rest of the discussion of this paper.

 \subsection{Classical solution of Mukhanov Sasaki equation in Framework I}
 \label{s3c}
 
 In this paper, we consider a physical framework (Framework I) which is made up of the following phases mentioned in the chronological order point-wise:
   \begin{enumerate}
       \item \underline{\bf Region I:} First of all we consider a Slow Roll (SRI) region which persists for the conformal time scale $\tau<\tau_s$. At $\tau=\tau_s$ the SRI transit to an Ultra Slow Roll (USR) region. This implies that SRI ends at $\tau=\tau_s$ in this construction. 

       \item \underline{\bf Region II:} Next we consider a Ultra Slow Roll (USR) region which starts at the conformal time scale $\tau=\tau_s$ and ends at the scale $\tau=\tau_e$. In this construction $\tau=\tau_e$ is treated to be the second sharp transition scale at which USR to the second Slow Roll (SRII) sharp transition takes place.

      \item \underline{\bf Region III:} Finally we consider the second Slow Roll (SRII) region which starts at $\tau=\tau_e$ and the inflation ends after a short time span.
   \end{enumerate}
   Our job is to explicitly study the classical solution and its quantum effects from these mentioned three regions separately in the following subsections of this paper. 
    \subsubsection{Region I: First Slow Roll (SRI) region}
The following expression provides the general solution of the MS equation for the scalar perturbed mode, which is explicitly valid during the first SR (SRI) period ($\tau\leq\tau_s$):
   \bea
   v_{\bf k}(\tau)&=&\frac{\alpha^{(1)}_{\bf k}}{\sqrt{2c_sk}}\left(1-\frac{i}{kc_s\tau}\right)\; e^{-ikc_s\tau}+\frac{\beta^{(1)}_{\bf k}}{\sqrt{2c_sk}}\left(1+\frac{i}{kc_s\tau}\right)\; e^{ikc_s\tau},
   \eea
 where the proper selection of the initial condition fixes the solution in terms of the coefficients $\alpha^{(1)}_{\bf k}$ and $\beta^{(1)}_{\bf k}$, which are appearing in the above equation. In principle one can choose any arbitrary initial quantum vacuum states, provided the observational constraints from CMB for inflation has to be satisfied within the SR period. However, we decide to choose the most popular one, which is the well-known Bunch Davies quantum vacuum state. This is actually a Euclidean vacuum which in our context will fix the initial condition in the first SR (SRI) period and is characterized by the following equation, \bea \label{b1a}&&\alpha^{(1)}_{\bf k}=1, \\
 \label{b1b}&&\beta^{(1)}_{\bf k}=0.\eea
 Such a choice will be going to be extremely useful for the further computation performed in the rest of the paper. After imposing this mentioned initial condition, the scalar mode function can be expressed in terms of the following simplest abbreviation in the first SR (SRI) region:
 \bea
 v_{\bf k}(\tau)=\frac{1}{\sqrt{2c_sk}}\left(1-\frac{i}{kc_s\tau}\right)\; e^{-ikc_s\tau}.
 \eea
With the use of the previously mentioned solution of MS equation written for the scalar mode and fixed in terms of the Bunch Davies initial condition, particularly in the first SR regime, $\tau\leq \tau_s$, one can further write down the formula for the comoving curvature perturbation in terms of the conformal time, effective sound speed parameter $c_s$ and momentum modes as follows\footnote{It is important to note that, the loop corrections become significant in the sub-horizon region which is technically described by the limit $-kc_s\tau\gg 1$. On the contrary, in the super-horizon scale, which is given by, $-kc_s\tau\ll 1$ the corresponding scalar modes become completely frozen and behave classically. This further implies that, at the horizon crossing scale, which is governed by $-kc_s\tau= 1$ the mixture of quantum and classical effects, commonly referred to as the semi-classical approximations we need to use.}:
   \bea
 \zeta_{\bf k}(\tau)&=&\frac{v_{\bf k}(\tau)}{zM_{ pl}}=\left(\frac{ic_sH}{2M_{\rm pl}\sqrt{\epsilon}}\right)\frac{1}{(c_sk)^{3/2}}\left(1+ikc_s\tau\right)\; e^{-ikc_s\tau}.\quad
 \eea
In the SRI region, $\epsilon$ is approximately a constant quantity, however, it varies very slowly with the time scale.
 \subsubsection{Region II: Ultra Slow Roll (USR) region}
 
Now we will focus on the Ultra Slow Roll (USR) region which can be visualized in the conformal time scale window, $\tau_s\leq \tau\leq \tau_e$, within which the corresponding scalar modes are valid. In this description, the sharp transition scale from the first SR (SRI) to USR is identified by, $\tau_s$. On the other hand, the end of the USR period, as well as the inflationary paradigm, is described by the time scale, $\tau_e$. In the USR regime, one can explicitly write down the time dependence of the first slow-roll parameter, which can be expressed in terms of the first SR contribution as:
\be \epsilon(\tau)=\epsilon \;\left(\frac{a(\tau_s)}{a(\tau)}\right)^{6}=\epsilon  \;\left(\frac{\tau}{\tau_s}\right)^{6}\quad\quad\quad{\rm where}\quad\quad\tau_s\leq\tau\leq \tau_e.\ee  Here $\epsilon$ is the first slow-roll parameter in the SR region which we have defined explicitly in the first half of the discussion. The aforementioned mathematical form shows that this parameter is about a constant quantity at the point where from the first SR (SRI) to USR sharp transition takes place i.e. at $\tau=\tau_s$ we have $\epsilon(\tau_s)=\epsilon$, which is actually at the first SR (SRI) regime. The deviation from the constant behaviour appears after the first SR (SRI) to USR sharp transition, which happens at the scale $\tau>\tau_s$. Here it is important to note that, for the present computational purpose we have specifically considered a sharp transition from the first SR (SRI) to the USR region, which will be going to extremely useful information for the rest of the paper. Taking this specific time-dependent behaviour and non-constancy of the first slow-roll parameter the solution of the MS equation in the USR period for the comoving curvature perturbation can be expressed by the following simplified expression:
 \bea
 \zeta_{\bf k}(\tau)&=&\left(\frac{ic_sH}{2M_{ pl}\sqrt{\epsilon}}\right)\left(\frac{\tau_s}{\tau}\right)^{3}\frac{1}{(c_sk)^{3/2}}\times\bigg[\alpha^{(2)}_{\bf k}\left(1+ikc_s\tau\right)\; e^{-ikc_s\tau}-\beta^{(2)}_{\bf k}\left(1-ikc_s\tau\right)\; e^{ikc_s\tau}\bigg],
 \eea
 Here it is important to note that, in the above-mentioned solution obtained in the USR region the explicit structure of the momentum, sharp transition scale time $\tau_s$ and the effective sound speed dependent coefficients $\alpha^{(2)}_{\bf k}$ and $\beta^{(2)}_{\bf k}$ can be expressed in terms of the initial condition fixed in terms of Bunch Davies vacuum in the first SR (SRI) region via Bogoliubov transformations. This further implies the fact that in the USR region, the underlying structure of the vacuum state changes compared to the Bunch Davies initial state. Our objective is to determine these Bogoliubov coefficients $\alpha^{(2)}_{\bf k}$ and $\beta^{(2)}_{\bf k}$ in the USR region. This can be done by applying the following two boundary conditions, which in principle can be interpreted as Israel junction conditions that we need to apply at the first SR (SRI) to USR sharp transition scale, $\tau=\tau_s$, and given by the following expressions:
\begin{enumerate}
    \item $\left[\zeta_{\bf k}(\tau)\right]_{\rm SR, \tau=\tau_s}= \left[\zeta_{\bf k}(\tau)\right]_{\rm USR, \tau=\tau_s}$, which implies that the scalar modes obtained from the first SR (SRI) and USR region become continuous at the sharp transition point $\tau=\tau_s$,

    \item $\left[\zeta^{'}_{\bf k}(\tau)\right]_{\rm SR, \tau=\tau_s}= \left[\zeta^{'}_{\bf k}(\tau)\right]_{\rm USR, \tau=\tau_s}$, which implies that the momentum modes for the scalar perturbation obtained from the first SR (SRI) and USR region becomes continuous at the sharp transition point  $\tau=\tau_s$.
\end{enumerate}
After imposing the above-mentioned junction conditions we get two constraint equations and solving them we get the following closed form of the Bogoliubov coefficients $\alpha^{(2)}_{\bf k}$ and $\beta^{(2)}_{\bf k}$ in the USR region:
\bea \label{b2a}\alpha^{(2)}_{\bf k}&=&1-\frac{3}{2ik^{3}c^{3}_s\tau^{3}_s}\left(1+k^{2}c^{2}_s\tau^{2}_s\right),\\
\label{b2b}\beta^{(2)}_{\bf k}&=&-\frac{3}{2ik^{3}c^{3}_s\tau^{3}_s}\left(1+ikc_s\tau_s\right)^{2}\; e^{-2ikc_s\tau_s}.\eea
As the structure of the Bogoliubov coefficients is now fixed in terms of the momentum mode, effective sound speed, and the conformal time scale at the first SR (SRI) to USR sharp transition point $\tau=\tau_s$, the corresponding modified structure of the quantum vacuum state is also fixed. This information will be going to be extremely useful for the analysis performed in the rest of the paper.

\subsubsection{Region III: Second Slow Roll (SRII) region}

Now we will focus on the second Slow Roll (SRII) region which can be visualized in the conformal time scale window, $\tau_e\leq \tau\leq \tau_{\rm end}$. In this description, the sharp transition scale from the USR to SRII is identified by the time scale, $\tau_e$. On the other hand, the end of the USR period, as well as the inflationary paradigm, is described by the time scale, $\tau_e$ and $\tau_{\rm end}$, respectively. In the SRII regime, one can explicitly write down the time dependence of the first slow-roll parameter, which can be expressed in terms of the first SR (SRI) contribution as:
\be \epsilon(\tau)=\epsilon \;\left(\frac{a(\tau_s)}{a(\tau_e)}\right)^{6}=\epsilon  \;\left(\frac{\tau_e}{\tau_s}\right)^{6}\quad\quad\quad{\rm where}\quad\quad\tau_e\leq\tau\leq \tau_{\rm end}.\ee  Here $\epsilon$ is the first slow-roll parameter in the SRI region which we have defined explicitly in the first half of the discussion. The aforementioned mathematical form shows that this parameter is about a not constant quantity at the point where the USR to SRII sharp transition takes place i.e. at $\tau=\tau_e$, which is actually at the end of the USR regime. The deviation from the constant behaviour continues up to the time scale corresponding to the end of inflation, but this value will not going to change in between the time interval $\tau_e<\tau<\tau_{\rm end}$. Here we consider a sharp transition at the time scale, $\tau=\tau_e$. Taking this specific time-dependent behaviour and non-constancy of the first slow-roll parameter the solution of the MS equation changes accordingly in the present context.
As we have already pointed out, we need to consider another sharp transition from USR to the second SR  (SRII) region which is expected to happen at the sharp transition point $\tau=\tau_e$ and in the region $\tau>\tau_e$ the SR features persists with the non-constant value of the first slow-roll parameter. Hence, the solution of the MS equation in terms of the scalar modes in the region $\tau_e\leq\tau\leq \tau_{\rm end}$ \footnote{Here $\tau_{\rm end}$ corresponds to the conformal time scale at the end of inflation, which technically happens at the end of SRII phase. Also, the momentum scale associated with the conformal time scale $\tau_{\rm end}$ is given by $k_{\rm end}$, which will be extremely useful for the rest of the discussions.} can be expressed by the following equation:
\bea
 \zeta_{\bf k}(\tau)&=&\left(\frac{ic_sH}{2M_{ pl}\sqrt{\epsilon}}\right)\left(\frac{\tau_s}{\tau_e}\right)^{3}\frac{1}{(c_sk)^{3/2}}\times\bigg[\alpha^{(3)}_{\bf k}\left(1+ikc_s\tau\right)\; e^{-ikc_s\tau}-\beta^{(3)}_{\bf k}\left(1-ikc_s\tau\right)\; e^{ikc_s\tau}\bigg],
 \eea
 It is important to note that, in the above-mentioned solution obtained in the second SR region the explicit structure of the momentum, sharp transition scale time $\tau_e$, and the effective sound speed dependent coefficients $\alpha^{(3)}_{\bf k}$ and $\beta^{(3)}_{\bf k}$ can be expressed in terms of the boundary condition fixed in terms of new modified vacuum in the USR region via Bogoliubov transformations. This further implies the fact that in the second SR (SRII) region the underlying structure of the vacuum state changes compared to the previously computed vacuum state in the USR region. Our objective is to determine these Bogoliubov coefficients $\alpha^{(3)}_{\bf k}$ and $\beta^{(3)}_{\bf k}$ in the second SR (SRII) region explicitly. This can be done by applying the following two boundary conditions, which in principle can be interpreted as Israel junction conditions which we need to apply at the USR to the second SR (SRII) sharp transition scale, $\tau=\tau_e$, and given by the following expressions:
\begin{enumerate}
    \item $\left[\zeta_{\bf k}(\tau)\right]_{\rm USR, \tau=\tau_e}= \left[\zeta_{\bf k}(\tau)\right]_{\rm SR, \tau=\tau_e}$, which implies that the scalar modes obtained from USR and the second SR  (SRII) region become continuous at the sharp transition point $\tau=\tau_e$,

    \item $\left[\zeta^{'}_{\bf k}(\tau)\right]_{\rm USR, \tau=\tau_e}= \left[\zeta^{'}_{\bf k}(\tau)\right]_{\rm SR, \tau=\tau_e}$, which implies that the momentum modes for the scalar perturbation obtained from USR and the second SR region (SRII) becomes continuous at the sharp transition point  $\tau=\tau_e$.
\end{enumerate}
After imposing the above mentioned junction conditions we get two constraint equations and solving them we get the following closed form of the Bogoliubov coefficients $\alpha^{(3)}_{\bf k}$ and $\beta^{(3)}_{\bf k}$ in the second SR (SRII) region:
\bea \label{b3a}\alpha^{(3)}_{\bf k}&=&-\frac{1}{4k^6c^6_s\tau^3_s\tau^3_e}\Bigg[9\left(kc_s\tau_s-i\right)^2\left(kc_s\tau_e+i\right)^2 e^{2ikc_s(\tau_e-\tau_s)}\nonumber\\
&&\quad\quad\quad\quad\quad\quad\quad\quad\quad\quad\quad\quad\quad\quad\quad\quad-
\left\{k^2c^2_s\tau^2_e\left(2kc_s\tau_e-3i\right)-3i\right\}\left\{k^2c^2_s\tau^2_s\left(2kc_s\tau_s+3i\right)+3i\right\}\Bigg],\\
\label{b3b}\beta^{(3)}_{\bf k}&=&\frac{3}{4k^6c^6_s\tau^3_s\tau^3_e}\Bigg[\left(kc_s\tau_s-i\right)^2\left\{k^2c^2_s\tau^2_e\left(3-2ikc_s\tau_e\right)+3\right\}e^{-2ikc_s\tau_s}\nonumber\\
&&\quad\quad\quad\quad\quad\quad\quad\quad\quad\quad\quad\quad\quad\quad\quad\quad+i\left(kc_s\tau_e-i\right)^2\left\{3i+k^2c^2_s\tau^2_s\left(2kc_s\tau_s+3i\right)\right\}e^{-2ikc_s\tau_e}\Bigg].\eea

\subsection{Classical solution of Mukhanov Sasaki equation in Framework II}
\label{s3d}

   In this paper, we consider another physical framework (Framework II) which is made up of the following phases mentioned in the chronological order point-wise:
   \begin{enumerate}
       \item \underline{\bf Region I:} First of all we consider a Slow Roll (SR) region which appears at the conformal time scale $\tau<\tau_s$. At $\tau=\tau_s$ the SR sharp transition takes place to an Ultra Slow Roll (USR) phase. This implies that SR for this particular framework ends at $\tau=\tau_s$ in this construction. It is important to point out here that, this phase is exactly identical to the {\bf Region I} of the Framework I as discussed earlier in this paper. 

       \item \underline{\bf Region II:} Next we consider a Ultra Slow Roll (USR) region which starts at the conformal time scale $\tau=\tau_s$ and the USR phase as well as the inflation ends at the scale $\tau=\tau_e$. It is important to note that the ending of this phase is different from the {\bf Region II} of Framework I as discussed earlier in this paper in detail.
   \end{enumerate}
   Our job is to explicitly study the classical solution and its quantum effects from these mentioned three regions separately in the following subsections of this paper.
    \subsubsection{Region I: Slow Roll (SR) region}
    This region is exactly identical to the SRI region of Framework I. For this reason, the final solution and its physical interpretation are not going to change in this part. Though for completeness we will write down the final expression for the scalar mode function, which is given by the following expression:
   \bea
 \zeta_{\bf k}(\tau)&=&\left(\frac{ic_sH}{2M_{\rm pl}\sqrt{\epsilon}}\right)\frac{1}{(c_sk)^{3/2}}\left(1+ikc_s\tau\right)\; e^{-ikc_s\tau}.\quad
 \eea
 This solution is valid in the region $\tau\leq\tau_s$, which is here described as the SR phase.
 Here to write down the above expression we have chosen the Bunch Davies initial vacuum state which in turn fixes the structure of the Bogoliubov coefficients.

 \subsubsection{Region II: Ultra Slow Roll (USR) region}
 \label{s3e}

 This region is exactly similar to the USR phase as described in the case of Framework I, except for one fact. In the present framework, the end of the USR phase and the end of inflation coincides. On the other hand, in Framework I after ending the USR phase it transits to a SRII phase, and after a while inflation ends. However, the mathematical structure of the MS equation and the Bogoliubov coefficients are exactly the same as described in the case of Framework I. In the present case USR period is valid within the window, $\tau_s\leq \tau\leq \tau_e$, where $\tau_s$ and $\tau_e$ are physically interpreted as the SR to USR sharp transition scale and end of inflation with USR, respectively. The final solution of the scalar mode function is given by the following expression:
 \bea
 \zeta_{\bf k}(\tau)&=&\left(\frac{ic_sH}{2M_{ pl}\sqrt{\epsilon}}\right)\left(\frac{\tau_s}{\tau}\right)^{3}\frac{1}{(c_sk)^{3/2}}\times\bigg[\alpha_{\bf k}\left(1+ikc_s\tau\right)\; e^{-ikc_s\tau}-\beta_{\bf k}\left(1-ikc_s\tau\right)\; e^{ikc_s\tau}\bigg],
 \eea
 where the Bogoliubov coefficients $\alpha_{\bf k}$ and $\beta_{\bf k}$ in the USR region:
\bea \label{b2a2}\alpha_{\bf k}&=&1-\frac{3}{2ik^{3}c^{3}_s\tau^{3}_s}\left(1+k^{2}c^{2}_s\tau^{2}_s\right),\\
\label{b2b2}\beta_{\bf k}&=&-\frac{3}{2ik^{3}c^{3}_s\tau^{3}_s}\left(1+ikc_s\tau_s\right)^{2}\; e^{-2ikc_s\tau_s}.\eea
It is to be noted that, due to having the above mentioned non-trivial structure of the Bogoliubov coefficients, corresponding structure of the vacuum state is also modified compared to the Bunch Davies state as appeariing in SR region.
 This information will going to be extremely useful for the analysis performed in the rest of the paper.

\subsection{Quantization of classical scalar modes}
\label{s3f}

We must explicitly quantize the scalar modes appropriately in order to compute the expression for the two-point correlation function and the related power spectrum in Fourier space, which is needed to calculate the cosmological correlations. To accomplish this, we must first construct the creation operator, $\hat{a}^{\dagger}_{\bf k}$ and annihilation operator $\hat{a}_{\bf k}$, which will, respectively, produce an excited state from the initial Bunch Davies state and destroy it. To serve the rest of the purpose we now recognize $|0\rangle$ as Bunch Davies initial state, which must adhere to the ensuing restriction, 
\be \hat{a}_{\bf k}|0\rangle=0\quad\forall {\bf k}.\ee
Now the equal time commutation relations must be satisfied by the  the scalar perturbed mode and its corresponding canonically conjugate momenta, 
\bea &&\left[\hat{\zeta}_{\bf k}(\tau),\hat{\zeta}^{'}_{{\bf k}^{'}}(\tau)\right]_{\rm ETCR}=i\;\delta^{3}\left({\bf k}+{\bf k}^{'}\right),\\
&&\left[\hat{\zeta}_{\bf k}(\tau),\hat{\zeta}_{{\bf k}^{'}}(\tau)\right]_{\rm ETCR}=0,\\
&&\left[\hat{\zeta}^{'}_{\bf k}(\tau),\hat{\zeta}^{'}_{{\bf k}^{'}}(\tau)\right]_{\rm ETCR}=0.\eea
Here the following expressions represents the appropriate quantum mechanical operators for the scalar mode and its conjugate momenta:
\bea \hat{\zeta}_{\bf k}(\tau)&=&\bigg[{\zeta}_{\bf k}(\tau)\hat{a}_{\bf k}+{\zeta}^{*}_{\bf k}(\tau)\hat{a}^{\dagger}_{-{\bf k}}\bigg]=\frac{c_s}{a\sqrt{2\epsilon} M_{pl}}\bigg[v_{\bf k}(\tau)\hat{a}_{\bf k}+v^{*}_{\bf k}(\tau)\hat{a}^{\dagger}_{-{\bf k}}\bigg],\\
\hat{\zeta}^{'}_{\bf k}(\tau)&=&\bigg[{\zeta}^{'}_{\bf k}(\tau)\hat{a}_{\bf k}+{\zeta}^{*'}_{\bf k}(\tau)\hat{a}^{\dagger}_{-{\bf k}}\bigg]=\frac{c_s}{a\sqrt{2\epsilon} M_{pl}}\bigg[v^{'}_{\bf k}(\tau)\hat{a}_{\bf k}+v^{*'}_{\bf k}(\tau)\hat{a}^{\dagger}_{-{\bf k}}\bigg]-\frac{c^{2}_s}{2\epsilon a^{2} M_{pl}}\bigg(\frac{a\sqrt{2\epsilon}}{c_s}\bigg)^{'}\bigg[v_{\bf k}(\tau)\hat{a}_{\bf k}+v^{*}_{\bf k}(\tau)\hat{a}^{\dagger}_{-{\bf k}}\bigg].\quad\quad \eea
 which will going to be extremely useful when we perform the one loop computation using the in-in formalism in the next section.
 
This can also be expressed in terms of all conceivable commutation relations between the creation and annihilation operators as indicated above:
\bea \left[\hat{a}_{\bf k},\hat{a}^{\dagger}_{{\bf k}^{'}}\right]&=&\delta^{3}\left({\bf k}+{\bf k}^{'}\right),\\
 \left[\hat{a}_{\bf k},\hat{a}_{{\bf k}^{'}}\right]&=&0=\left[\hat{a}^{\dagger}_{\bf k},\hat{a}^{\dagger}_{{\bf k}^{'}}\right].\eea

  \subsection{Tree level primordial power spectrum from scalar modes}
  \label{s3g}
  
 We know that the co-moving curvature perturbation occurs at the late time scale,  $\tau\rightarrow 0$, the relevant tree-level contribution to the two-point cosmological correlation function for the scalar comoving curvature perturbation can be represented as follows:
\bea \langle \hat{\zeta}_{\bf k}\hat{\zeta}_{{\bf k}^{'}}\rangle_{{\bf Tree}} &=&(2\pi)^{3}\;\delta^{3}\left({\bf k}+{\bf k}^{'}\right)P^{\bf Tree}_{\zeta}(k),\quad\eea
where $P^{\bf Tree}_{\zeta}(k)$ represents the dimensionful power spectrum in the Fourier space, quantified by the following simplified expression:
\bea P^{\bf Tree}_{\zeta}(k)=\langle \hat{\zeta}_{\bf k}\hat{\zeta}_{-{\bf k}}\rangle_{(0,0)}=\left[{\zeta}_{\bf k}(\tau){\zeta}_{-{\bf k}}(\tau)\right]_{\tau\rightarrow 0}=|{\zeta}_{\bf k}(\tau)|^{2}_{\tau\rightarrow 0}.\quad\eea
However, for the practical purpose and to connect with cosmological observation it is always relevant to deal with a dimensionless form of the power spectrum in Fourier space, which is given by the following expression:
\bea \Delta^{2}_{\zeta,{\bf Tree}}(k)=\frac{k^{3}}{2\pi^{2}}P^{\bf Tree}_{\zeta}(k)=\frac{k^{3}}{2\pi^{2}}|{\zeta}_{\bf k}(\tau)|^{2}_{\tau\rightarrow 0}.\eea
  \subsubsection{Result from Framework I}

Now from the present analysis we already know that the solution of modes for the scalar cosmological perturbations become different in the first SR (SRI), USR, and in the second SR (SRII) region, which we have explicitly calculated in this paper.
Here at the tree level, the dimensionless power spectrum can be computed with the help of the computed scalar modes from the first SR (SRI), USR, and in the second SR (SRII) region as:  
\bea  \Delta^{2}_{\zeta,{\bf Tree}}(k)
&=& \left(\frac{H^{2}}{8\pi^{2}M^{2}_{ pl}\epsilon c_s}\right)_* \nonumber\\
&&\times
\left\{
	\begin{array}{ll}
		\displaystyle\left(1+k^{2}c^{2}_s\tau^{2}\right)& \mbox{when}\quad  k\ll k_s  \;(\rm SRI)  \\  
			\displaystyle 
			\displaystyle\left(\frac{k_e}{k_s }\right)^{6}\times\left|\alpha^{(2)}_{\bf k}\left(1+ikc_s\tau\right)\; e^{-ikc_s\tau}-\beta^{(2)}_{\bf k}\left(1-ikc_s\tau\right)\; e^{ikc_s\tau}\right|^{2} & \mbox{when }  k_s\leq k\leq k_e  \;(\rm USR)\\ 
   \displaystyle 
			\displaystyle\left(\frac{k_e}{k_s }\right)^{6}\times\left|\alpha^{(3)}_{\bf k}\left(1+ikc_s\tau\right)\; e^{-ikc_s\tau}-\beta^{(3)}_{\bf k}\left(1-ikc_s\tau\right)\; e^{ikc_s\tau}\right|^{2} & \mbox{when }  k_e\leq k\leq k_{\rm end}  \;(\rm SRII) 
	\end{array}
\right. .\eea

It is important to note that the above-mentioned expression is valid in the sub-horizon ($-kc_s\tau\gg 1$), horizon re-entry  ($-kc_s\tau= 1$), and super-horizon ($-kc_s\tau\ll 1$) scales. Here the Bogoliubov coefficients for the USR region, $(\alpha^{(2)}_{\bf k},\beta^{(2)}_{\bf k})$ and the second SR region (SRII), $(\alpha^{(3)}_{\bf k},\beta^{(3)}_{\bf k})$ are explicitly derived in equation (\ref{b2a}), equation (\ref{b2b}), equation (\ref{b3a}) and equation (\ref{b3b}) respectively. For the first SR region (SRI) the Bogoliubov coefficients, $(\alpha^{(1)}_{\bf k},\beta^{(1)}_{\bf k})$ are fixed by the Bunch Davies initial condition, which appears in equation (\ref{b1a}) and equation (\ref{b1b}).  The wave numbers $k_e$ and $k_ s$ that correspond to the time scales, $\tau_e$, and $\tau_s$ should also be taken into consideration for the present computation.  The symbol $*$ corresponds to the pivot scale relevant to CMB. From the above structure, one can distinguish the explicit contributions coming from the first SR region (SRI), USR region, and second SR region (SRII) in this present context of the discussion. 

\textcolor{black}{In the limit, $\tau\to 0$ of
interest, corresponding to super-hubble scales, $-kc_s\tau\ll 1$, the individual contribution of the tree-level scalar power spectrum in the three consecutive phases, SRI, USR, and SRII, can be further simplified as:
\bea  \Delta^{2}_{\zeta,{\bf Tree}}(k)
&=& \left(\frac{H^{2}}{8\pi^{2}M^{2}_{ pl}\epsilon c_s}\right)_* \times
\left\{
	\begin{array}{ll}
		\displaystyle 1& \mbox{when}\quad  k\ll k_s  \;(\rm SRI)  \\  
			\displaystyle 
			\displaystyle\left(\frac{k_e}{k_s }\right)^{6}\times\left|\alpha^{(2)}_{\bf k}-\beta^{(2)}_{\bf k}\right|^{2} & \mbox{when }  k_s\leq k\leq k_e  \;(\rm USR)\\ 
   \displaystyle 
			\displaystyle\left(\frac{k_e}{k_s }\right)^{6}\times\left|\alpha^{(3)}_{\bf k}-\beta^{(3)}_{\bf k}\right|^{2} & \mbox{when }  k_e\leq k\leq k_{\rm end}  \;(\rm SRII) 
	\end{array}
\right. \eea  
It may be noted that initially the scalar modes leave the horizon and go to the super-hubble regime, and further re-enter the horizon after a while. Once the modes leave the horizon it becomes classical and the corresponding amplitude freezes. 
For this reason, the contributions computed for the super-horizon scale are further used to give an estimate of the amplitude of the scalar power spectrum at the horizon re-entry.}

\textcolor{black}{Consequently, in the super-horizon scale ($-kc_s\tau\ll 1$) adding the contribution from SRI, USR, and SRII phases together in the presence of sharp transition the resulting expression for the tree-level amplitude of the primordial scalar power spectrum can be described by the following expression} \footnote{\textcolor{black}{Here it is important to note that, in the equation (\ref{totree}) first term corresponds to the contribution from the SRI phase. If other phases are absent for Framework I, then this particular contribution will remain the same from the small scale to the large scale. However, in the present context, we are interested in describing the production of PBHs which can be done by the insertion of a USR phase followed by another SRII phase. In both of these mentioned phases, enhancement of the scalar power spectrum is required to produce PBHs with the correct amplitude. For this reason in the equation (\ref{totree}) we have two additional terms apart from the contribution from the SRI phase.}}:
\textcolor{black}{\bea \label{totree}\Delta^{2}_{\zeta,{\bf Tree,Total}}(k)=\bigg[\Delta^{2}_{\zeta,{\bf Tree}}(k)\bigg]_{\bf SR}\times\Bigg\{1+\left(\frac{k_e}{k_s }\right)^{6}\times\bigg[\Theta(k-k_s)\left|\alpha^{(2)}_{\bf k}-\beta^{(2)}_{\bf k}\right|^{2}+\Theta(k-k_e)\left|\alpha^{(3)}_{\bf k}-\beta^{(3)}_{\bf k}\right|^{2}\bigg]\Bigg\},\eea
where the subscript {\bf SR} represents the amplitude of the scalar power spectrum in the SRI phase and this appears as an overall common factor which in the super-horizon scale is given by the following expression:
\bea \label{sramp}\bigg[\Delta^{2}_{\zeta,{\bf Tree}}(k)\bigg]_{\bf SR}:=\left(\frac{H^{2}}{8\pi^{2}M^{2}_{ pl}\epsilon c_s}\right)_*.\eea
Here two Heaviside theta functions are inserted to correctly implement the requirement of sharp transitions at the scales, $k_s$ and $k_e$ respectively.
This formula for the amplitude of the scalar power spectrum for the slow-roll phase (SRI) is consistent with the results obtained in the context of canonical ($c_s=1$) \cite{Baumann:2009ds,Chen:2010xka,Baumann:2018muz} and non-canonical ($c_s\neq 1$) \cite{Chen:2006nt,Chen:2010xka} models of single-field inflation} \footnote{\textcolor{black}{In the mentioned refs. \cite{Chen:2006nt,Chen:2010xka} the authors have used the fact that $M_{pl}=1$. In this Planckian unit, the power spectrum is itself dimensionless and described by the formula, $\bigg[\Delta^{2}_{\zeta,{\bf Tree}}(k)\bigg]_{\bf SR}:=\displaystyle\left(\frac{H^{2}}{8\pi^{2}\epsilon c_s}\right)_*$. However, in our computation, we have kept the factor $M_{pl}$ everywhere and for this particular case $H/M_{pl}$ is a dimensionless factor which ultimately makes the numerical estimate of the scalar power spectrum dimensionless and consistent with the observed value, $\bigg[\Delta^{2}_{\zeta,{\bf Tree}}(k)\bigg]_{\bf SR}\sim 2.2\times 10^{-9}$. To avoid any further confusion we have decided to include this clarification which we believe will be helpful to explore the underlying physics of this portion.} }.\textcolor{black}{Within the framework of single field EFT framework, one can club them together and write them in terms of a unified formula. For further clarification purposes to justify the correctness of the tree-level result, it is to be noted that the obtained results in this section are completely consistent with the recently studied various related issues and have been pointed out in the refs. \cite{Kristiano:2022maq,Riotto:2023hoz,Kristiano:2023scm,Riotto:2023gpm,Choudhury:2023hvf,Choudhury:2023kdb,Choudhury:2023hfm,Bhattacharya:2023ysp,Motohashi:2023syh,Firouzjahi:2023ahg,Franciolini:2023lgy,Firouzjahi:2023aum,Cheng:2023ikq,Tasinato:2023ukp}.}

\subsubsection{Result from Framework II}
Now from the present analysis we already know that the solution of modes for the scalar cosmological perturbations become different in the first SR (SRI), USR, and in the second SR (SRII) region, which we have explicitly calculated in this paper.
Here at the tree level, the dimensionless power spectrum can be computed with the help of the computed scalar modes from the first SR (SRI), USR, and in the second SR (SRII) region as:  
\bea  \Delta^{2}_{\zeta,{\bf Tree}}(k)
&=& \left(\frac{H^{2}}{8\pi^{2}M^{2}_{ pl}\epsilon c_s}\right)_* \nonumber\\
&&\times
\left\{
	\begin{array}{ll}
		\displaystyle\left(1+k^{2}c^{2}_s\tau^{2}\right)& \mbox{when}\quad  k\ll k_s  \;(\rm SR)  \\ 
			\displaystyle 
			\displaystyle\left(\frac{k_e}{k_s }\right)^{6}\times\left|\alpha_{\bf k}\left(1+ikc_s\tau\right)\; e^{-ikc_s\tau}-\beta_{\bf k}\left(1-ikc_s\tau\right)\; e^{ikc_s\tau}\right|^{2} & \mbox{when }  k_s\leq k\leq k_e  \;(\rm USR)
	\end{array}
\right.. \eea

It is important to note that the above-mentioned expression is valid in the sub-horizon ($-kc_s\tau\gg 1$), horizon re-entry  ($-kc_s\tau= 1$), and super-horizon ($-kc_s\tau\ll 1$) scales. Here the Bogoliubov coefficients for the USR region, $(\alpha_{\bf k},\beta_{\bf k})$ are explicitly mentioned in equation (\ref{b2a2}) and equation (\ref{b2b2}) respectively. For the SR region, the Bogoliubov coefficients are fixed by the Bunch Davies initial condition. The wave numbers $k_e$ and $k_s$ that correspond to the time scales, $\tau_e$, and $\tau_s$. Only here $\tau_e$ corresponds to both the end of inflation and the USR phase. The symbol $*$ corresponds to the pivot scale. From the above structure, one can distinguish the explicit contributions coming from the SR and USR regions in this present context of the discussion.

\textcolor{black}{Now since from the observational perspective, we can only probe the the super-horizon scale, where we have $-kc_s\tau\ll 1$, the individual contribution of the tree-level scalar power spectrum in the three consecutive phases, SRI, USR and SRII can be further simplified as:
\bea  \Delta^{2}_{\zeta,{\bf Tree}}(k)
&=& \left(\frac{H^{2}}{8\pi^{2}M^{2}_{ pl}\epsilon c_s}\right)_* \times
\left\{
	\begin{array}{ll}
		\displaystyle 1& \mbox{when}\quad  k\ll k_s  \;(\rm SRI)  \\  
			\displaystyle 
			\displaystyle\left(\frac{k_e}{k_s }\right)^{6}\times\left|\alpha^{(2)}_{\bf k}-\beta^{(2)}_{\bf k}\right|^{2} & \mbox{when }  k_s\leq k\leq k_e  \;(\rm USR)
	\end{array}
\right. \eea  
Additionally, it is important to note that the contributions computed for the super-horizon scale are further used to give an estimate of the amplitude of the scalar power spectrum at the horizon re-entry which happens when $-kc_s\tau=1$.}

\textcolor{black}{Consequently, in the super-horizon scale ($-kc_s\tau\ll 1$) adding the contribution from SR and USR phases together in the presence of sharp transition the total contribution to the tree-level amplitude of the primordial scalar power spectrum can be cast as follows} \footnote{\textcolor{black}{Here it is important to note that, in the equation (\ref{totree1}) first term corresponds to the contribution from the SR phase. If another USR phase is absent for Framework II, then this particular contribution will remain the same from the small scale to the large scale. However, in the present context, we are interested in describing the production of PBHs which can be done by the insertion of a USR phase followed by the end of inflation in this particular framework. In this USR phase for Framework II, enhancement of the scalar power spectrum is required to produce PBHs with the correct amplitude. For this reason in the equation (\ref{totree1}) we have an additional term apart from the contribution from the SR phase which will give rise to the correct estimate in the end.}}:
\textcolor{black}{\bea \label{totree1}\Delta^{2}_{\zeta,{\bf Tree,Total}}(k)=\bigg[\Delta^{2}_{\zeta,{\bf Tree}}(k)\bigg]_{\bf SR}\times\Bigg\{1+\left(\frac{k_e}{k_s }\right)^{6}\Theta(k-k_s)\left|\alpha_{\bf k}-\beta_{\bf k}\right|^{2}\Bigg\},\eea
where the subscript {\bf SR} represents the amplitude of the scalar power spectrum in the SR phase and this appears as an overall common factor which in the super-horizon scale is given by the following expression:
\bea \label{srampa}\bigg[\Delta^{2}_{\zeta,{\bf Tree}}(k)\bigg]_{\bf SR}:=\left(\frac{H^{2}}{8\pi^{2}M^{2}_{ pl}\epsilon c_s}\right)_*,\eea
which is exactly the same as we have obtained for Framework I.
Here a single Heaviside theta function is inserted to correctly implement the requirement of sharp transitions at the scale, $k_s$ where the PBHs formation occurs for the Framewok II.
Just like the Framework I in the present context this formula for the amplitude of the scalar power spectrum for the slow-roll phase (SR) is consistent with the results obtained in the context of canonical ($c_s=1$) \cite{Baumann:2009ds,Chen:2010xka,Baumann:2018muz} and non-canonical ($c_s\neq 1$) \cite{Chen:2006nt,Chen:2010xka} models of single-field inflation. Within the framework of single field EFT framework, one can club them together and write them in terms of a unified formula as we have done it for the Framework II here. For further clarification purposes to justify the correctness of the tree-level result, it is to be noted that the obtained results in this section are completely consistent with the recently studied various related issues and have been pointed out in the refs. \cite{Kristiano:2022maq,Riotto:2023hoz,Kristiano:2023scm,Riotto:2023gpm,Choudhury:2023hvf,Choudhury:2023kdb,Choudhury:2023hfm,Bhattacharya:2023ysp,Motohashi:2023syh,Firouzjahi:2023ahg,Franciolini:2023lgy,Firouzjahi:2023aum,Cheng:2023ikq,Tasinato:2023ukp}. This also justifies the correctness and the vast applications of the derived result in the present context of discussion.}

\section{Cut-off regulated one-loop scalar power spectrum from EFT}
\label{s4}

\subsection{Third order perturbation from scalar mode}
\label{s4a}

The next stage in our analysis will be to directly calculate the effect of the one-loop correction to the power spectrum from the scalar modes of the perturbation. The following calculation will be performed presuming that the typical EFT action is expanded in third order by the curvature perturbation \footnote{Here it is important to note that, if we set $M_2=0$, then we have $c_s=1$. Additionally, if we set $M_3=0$, $\bar{M}_1=0$, then we get back the third order action for the canonical single field slow roll model. Since the contributions having the EFT coefficients apart from $M_2$ are highly suppressed in the SR region in the one-loop contribution it will not make any significant change the final one-loop corrected expression for the scalar power spectrum. For the other classes of non-canonical or non-minimal $P(X,\phi)$ models of single field inflation these mentioned EFT coefficients might not be exactly zero. However, to hold the perturbation theory for the scalar modes perfectly during performing the one-loop computation in the present context we need to restrict us for the small values of these parameters. We will discuss about this with proper numerical examples in the later half of this paper.}:
\bea &&S^{(3)}_{\zeta}=\int d\tau\;  d^3x\;  M^2_{ pl}a^2\; \bigg[\left(3\left(c^2_s-1\right)\epsilon+\epsilon^2-\frac{1}{2}\epsilon^3\right)\zeta^{'2}\zeta+\frac{\epsilon}{c^2_s}\bigg(\epsilon-2s+1-c^2_s\bigg)\left(\partial_i\zeta\right)^2\zeta\nonumber\\ 
&&\quad\quad\quad\quad\quad\quad\quad\quad\quad\quad\quad-\frac{2\epsilon}{c^2_s}\zeta^{'}\left(\partial_i\zeta\right)\left(\partial_i\partial^{-2}\left(\frac{\epsilon\zeta^{'}}{c^2_s}\right)\right)-\frac{1}{aH}\left(1-\frac{1}{c^2_{s}}\right)\epsilon \bigg(\zeta^{'3}+\zeta^{'}(\partial_{i}\zeta)^2\bigg)
     \nonumber\\
&& \quad\quad\quad\quad\quad\quad\quad\quad\quad\quad\quad+\frac{1}{2}\epsilon\zeta\left(\partial_i\partial_j\partial^{-2}\left(\frac{\epsilon\zeta^{'}}{c^2_s}\right)\right)^2
+\boxed{\underbrace{\frac{1}{2c^2_s}\epsilon\partial_{\tau}\left(\frac{\eta}{c^2_s}\right)\zeta^{'}\zeta^{2}}_{\bf Most~dominant ~term~in~USR}}\nonumber\\
&&+\underbrace{\frac{3}{2}\frac{1}{aH}\frac{\bar{M}^3_1}{ HM^2_{ pl}}
	  \zeta^{'}(\partial_{i}\zeta)^2+\frac{9}{2}\frac{\bar{M}^3_1}{ HM^2_{ pl}}\zeta \zeta^{'2}+\bigg(\frac{3}{2}\frac{\bar{M}^3_1}{ HM^2_{ pl}}-\frac{4}{3}\frac{M^4_3}{H^2M^2_{ pl}}\bigg)\zeta^{'3}-\frac{3}{2}\frac{1}{aH}\frac{\bar{M}^3_1}{ HM^2_{ pl}} \zeta\partial_{\tau}\left(\partial_{i}\zeta\right)^2}_{\bf Suppressed~contributions
	  ~in~the~one-loop~corrected~power~spectrum~from~SR~region~only}
  \bigg],\quad\quad\eea
We also introduce another slow roll parameter, $s$,  which is appearing due to having effective sound speed $c_s$ and its slow variation with respect to the underlying conformal time scale in the present EFT framework of cosmological perturbation, and given by the following expression:
\bea s=\frac{\dot{c}_s}{Hc_s}=\frac{1}{aH}\frac{c^{'}_s}{c_s}=\frac{1}{{\cal H}}\frac{c^{'}_s}{c_s}.\eea
We will utilize all of the above-mentioned contributions explicitly to extract the correction from the one-loop quantum effect. The most significant contribution is appearing in the highlighted box which contributes as ${\cal O}(\epsilon^{3})$, ${\cal O}(\epsilon^{3})$ and ${\cal O}(\epsilon)$, in the first SR (SRI) region, second SR (SRII) region and USR region respectively for the Framework I. On the other hand, for Framework II the highlighted contribution contributes as ${\cal O}(\epsilon^{3})$ and ${\cal O}(\epsilon)$, in the SR region and USR region respectively. The prime reason for such drastic change appearing in the overall contribution for this particularly highlighted term from SRI and SRII regime to the USR regime is because of the second slow roll parameter changes from $\eta\sim 0$ (SRI) to $\eta\sim -6$ (USR) during such mentioned sharp transition. Then it becomes $\eta\sim 0$ (SRII) in the second sharp transition where the USR to SRII sharp transition takes place and persists up to the end of the inflation in the case of Framework I. In the case of Framework II second slow roll parameter changes from $\eta\sim 0$ (SR) to $\eta\sim -6$ (USR) and then inflation ends.
This particular highlighted term corresponds to the leading cubic self-interaction, which is all that is of particular relevance to us for the present computation. The five contributions appearing in the last line of the above expression will be going to contribute only to the one-loop corrected expression for the scalar power spectrum only in the SR region. In the USR region, these terms have more suppression than the SR region in the mentioned one-loop correction. For this reason in the final expression, these terms will not going to change the overall amplitude of the enhanced one-loop corrected scalar power spectrum. Again, the rest of the contributions have more suppression in the USR region compared to the SR region in the corresponding one-loop result. Additionally, for more completeness, it is important to note that, if we compare the strength of all the terms including boxed one then one can easily visualize that all the five contributions appearing in the last line of the third order action give extremely small correction to the one-loop correction to hold the perturbative approximations in the present context. This means that, if we increase the strength of these mentioned five contributions by a considerable amount then the perturbative approximation breaks down in the SRI, SRII, and USR region, which will give rise to the wrong enhanced amplitude of the scalar power spectrum during PBH formation for the Framework I. The exact argument holds good for Framework II, where only the SRII phase is absent and inflation ends at the end of USR phase.

\subsection{In-In formalism and one-loop computation of primordial power spectrum for scalar modes}
\label{s4b}

In rest half of the paper we are going to explicitly analyze all the terms included the highlighted most significant boxed term appearing as the bi-product of the EFT framework that we have adopted for our study. In order to achieve this, we employ the well-known in-in formalism.  According to this, the two-point correlation function of the following quantum operator at the late time scale, $\tau\rightarrow 0$, can be written as follows:
    \bea \label{Hamx}\langle\hat{\zeta}_{\bf p}\hat{\zeta}_{-{\bf p}}\rangle:&=&\langle\hat{\zeta}_{\bf p}(\tau)\hat{\zeta}_{-{\bf p}}(\tau)\rangle_{\tau\rightarrow 0}\nonumber\\
    &=&\left\langle\bigg[\overline{T}\exp\bigg(i\int^{\tau}_{-\infty(1-i\epsilon)}d\tau^{'}\;H_{\rm int}(\tau^{'})\bigg)\bigg]\;\;\hat{\zeta}_{\bf p}(\tau)\hat{\zeta}_{-{\bf p}}(\tau)
\;\;\bigg[{T}\exp\bigg(-i\int^{\tau}_{-\infty(1+i\epsilon)}d\tau^{''}\;H_{\rm int}(\tau^{''})\bigg)\bigg]\right\rangle_{\tau\rightarrow 0},\quad\quad \eea
where $\overline{T}$ and $T$, respectively, stand for the anti-time and time ordering operation, $H_{\rm int}(\tau)$ represents the interaction Hamiltonian in the given context and can be computed from the third order EFT action as:
\bea && H_{\rm int}(\tau)=-\int d^3x\;  M^2_{ pl}a^2\; \bigg[\left(3\left(c^2_s-1\right)\epsilon+\epsilon^2-\frac{1}{2}\epsilon^3\right)\zeta^{'2}\zeta+\frac{\epsilon}{c^2_s}\bigg(\epsilon-2s+1-c^2_s\bigg)\left(\partial_i\zeta\right)^2\zeta\nonumber\\ 
&&\quad\quad\quad\quad\quad\quad\quad\quad\quad\quad\quad-\frac{2\epsilon}{c^2_s}\zeta^{'}\left(\partial_i\zeta\right)\left(\partial_i\partial^{-2}\left(\frac{\epsilon\zeta^{'}}{c^2_s}\right)\right)-\frac{1}{aH}\left(1-\frac{1}{c^2_{s}}\right)\epsilon \bigg(\zeta^{'3}+\zeta^{'}(\partial_{i}\zeta)^2\bigg)
     \nonumber\\
&& \quad\quad\quad\quad\quad\quad\quad\quad\quad\quad\quad+\frac{1}{2}\epsilon\zeta\left(\partial_i\partial_j\partial^{-2}\left(\frac{\epsilon\zeta^{'}}{c^2_s}\right)\right)^2
+\boxed{\underbrace{\frac{1}{2c^2_s}\epsilon\partial_{\tau}\left(\frac{\eta}{c^2_s}\right)\zeta^{'}\zeta^{2}}_{\bf Most~dominant ~term~in~USR}}\nonumber\\
&&+\underbrace{\frac{3}{2}\frac{1}{aH}\frac{\bar{M}^3_1}{ HM^2_{ pl}}
	  \zeta^{'}(\partial_{i}\zeta)^2+\frac{9}{2}\frac{\bar{M}^3_1}{ HM^2_{ pl}}\zeta \zeta^{'2}+\bigg(\frac{3}{2}\frac{\bar{M}^3_1}{ HM^2_{ pl}}-\frac{4}{3}\frac{M^4_3}{H^2M^2_{ pl}}\bigg)\zeta^{'3}-\frac{3}{2}\frac{1}{aH}\frac{\bar{M}^3_1}{ HM^2_{ pl}} \zeta\partial_{\tau}\left(\partial_{i}\zeta\right)^2}_{\bf Suppressed~contributions
	  ~in~the~one-loop~corrected~power~spectrum~from~SR~region~only}
  \bigg],\quad\quad\eea
Further using equation (\ref{Hamx}) and considering the contribution up to the one-loop correction in the Dyson Swinger series, we get the following expression for the two-point correlation function for the scalar modes:
    \bea  &&\label{g}\langle\hat{\zeta}_{\bf p}\hat{\zeta}_{-{\bf p}}\rangle= \underbrace{\langle\hat{\zeta}_{\bf p}\hat{\zeta}_{-{\bf p}}\rangle_{(0,0)}}_{\bf Tree\;level\;result}+\underbrace{\langle\hat{\zeta}_{\bf p}\hat{\zeta}_{-{\bf p}}\rangle_{(0,1)}+\langle\hat{\zeta}_{\bf p}\hat{\zeta}_{-{\bf p}}\rangle^{\dagger}_{(0,1)}+\langle\hat{\zeta}_{\bf p}\hat{\zeta}_{-{\bf p}}\rangle_{(0,2)}+\langle\hat{\zeta}_{\bf p}\hat{\zeta}_{-{\bf p}}\rangle^{\dagger}_{(0,2)}+\langle\hat{\zeta}_{\bf p}\hat{\zeta}_{-{\bf p}}\rangle_{(1,1)}}_{\bf One-loop\;level\;result},
\eea
where each of the contributions can be written as:
\bea
     &&\label{g0}\langle\hat{\zeta}_{\bf p}\hat{\zeta}_{-{\bf p}}\rangle_{(0,0)}=\left[\langle \hat{\zeta}_{\bf p}(\tau)\hat{\zeta}_{-{\bf p}}(\tau)\rangle\right]_{\tau\rightarrow 0},\\
    &&\label{g1}\langle\hat{\zeta}_{\bf p}\hat{\zeta}_{-{\bf p}}\rangle_{(0,1)}=\left[-i\int^{\tau}_{-\infty}d\tau_1\;\langle \hat{\zeta}_{\bf p}(\tau)\hat{\zeta}_{-{\bf p}}(\tau)H_{\rm int}(\tau_1)\rangle\right]_{\tau\rightarrow 0},\\
 &&\label{g2}\langle\hat{\zeta}_{\bf p}\hat{\zeta}_{-{\bf p}}\rangle^{\dagger}_{(0,1)}=\left[-i\int^{\tau}_{-\infty}d\tau_1\;\langle \hat{\zeta}_{\bf p}(\tau)\hat{\zeta}_{-{\bf p}}(\tau)H_{\rm int}(\tau_1)\rangle^{\dagger}\right]_{\tau\rightarrow 0},\\
 &&\label{g3}\langle\hat{\zeta}_{\bf p}\hat{\zeta}_{-{\bf p}}\rangle_{(0,2)}=\left[\int^{\tau}_{-\infty}d\tau_1\;\int^{\tau}_{-\infty}d\tau_1\;\langle \hat{\zeta}_{\bf p}(\tau)\hat{\zeta}_{-{\bf p}}(\tau)H_{\rm int}(\tau_1)H_{\rm int}(\tau_2)\rangle\right]_{\tau\rightarrow 0},\\
 &&\label{g4}\langle\hat{\zeta}_{\bf p}\hat{\zeta}_{-{\bf p}}\rangle^{\dagger}_{(0,2)}=\left[\int^{\tau}_{-\infty}d\tau_1\;\int^{\tau}_{-\infty}d\tau_1\;\langle \hat{\zeta}_{\bf p}(\tau)\hat{\zeta}_{-{\bf p}}(\tau)H_{\rm int}(\tau_1)H_{\rm int}(\tau_2)\rangle^{\dagger}\right]_{\tau\rightarrow 0},\\
  &&\label{g5}\langle\hat{\zeta}_{\bf p}\hat{\zeta}_{-{\bf p}}\rangle^{\dagger}_{(1,1)}=\left[\int^{\tau}_{-\infty}d\tau_1\;\int^{\tau}_{-\infty}d\tau_1\;\langle H_{\rm int}(\tau_1)\hat{\zeta}_{\bf p}(\tau)\hat{\zeta}_{-{\bf p}}(\tau)H_{\rm int}(\tau_2)\rangle^{\dagger}\right]_{\tau\rightarrow 0}.\eea
In this context, our job is to explicitly evaluate the equations (\ref{g1})-(\ref{g5}) both in the SR as well as in the USR region. The explicit contribution from equation (\ref{g0}) representing the tree level effect is already computed in the previous section explicitly.

Let us now look into the boxed highlighted cubic self interaction of the Hamiltonian, which will further contribute to the two-point correlation function of the scalar modes at the one-loop level in the USR period, quantified by the following expression:
     \bea   \langle\hat{\zeta}_{\bf p}\hat{\zeta}_{-{\bf p}}\rangle_{(0,1)}&=& -\frac{iM^2_{ pl}}{2}\int^{0}_{-\infty}d\tau\frac{a^2(\tau)}{c^2_s(\tau)}\epsilon(\tau)\partial_{\tau}\left(\frac{\eta(\tau)}{c^2_s(\tau)}\right)\nonumber\\
  &&\times\int \frac{d^{3}{\bf k}_1}{(2\pi)^3} \int \frac{d^{3}{\bf k}_2}{(2\pi)^3} \int \frac{d^{3}{\bf k}_3}{(2\pi)^3} \nonumber\\
  && \times\delta^3\bigg({\bf k}_1+{\bf k}_2+{\bf k}_3\bigg) \times \langle \hat{\zeta}_{\bf p}\hat{\zeta}_{-{\bf p}}\hat{\zeta}^{'}_{{\bf k}_1}(\tau)\hat{\zeta}_{{\bf k}_2}(\tau)\hat{\zeta}_{{\bf k}_3}(\tau)\rangle,\\
   \langle\hat{\zeta}_{\bf p}\hat{\zeta}_{-{\bf p}}\rangle_{(0,1)}&=& -\frac{iM^2_{ pl}}{2}\int^{0}_{-\infty}d\tau\frac{a^2(\tau)}{c^2_s(\tau)}\epsilon(\tau)\partial_{\tau}\left(\frac{\eta(\tau)}{c^2_s(\tau)}\right)\nonumber\\
  &&\times\int \frac{d^{3}{\bf k}_1}{(2\pi)^3} \int \frac{d^{3}{\bf k}_2}{(2\pi)^3} \int \frac{d^{3}{\bf k}_3}{(2\pi)^3} \nonumber\\
  && \times\delta^3\bigg({\bf k}_1+{\bf k}_2+{\bf k}_3\bigg) \times \langle \hat{\zeta}_{\bf p}\hat{\zeta}_{-{\bf p}}\hat{\zeta}^{'}_{{\bf k}_1}(\tau)\hat{\zeta}_{{\bf k}_2}(\tau)\hat{\zeta}_{{\bf k}_3}(\tau)\rangle^{\dagger},\\
     \langle\hat{\zeta}_{\bf p}\hat{\zeta}_{-{\bf p}}\rangle_{(0,2)}&=& -\frac{M^4_{ pl}}{4}\int^{0}_{-\infty}d\tau_1\frac{a^2(\tau_1)}{c^2_s(\tau_1)}\epsilon(\tau_1)\partial_{\tau_1}\left(\frac{\eta(\tau_1)}{c^2_s(\tau_1)}\right)\;\int^{0}_{-\infty}d\tau_2\;\frac{a^2(\tau_2)}{c^2_s(\tau_2)}\epsilon(\tau_2)\partial_{\tau_2}\left(\frac{\eta(\tau_2)}{c^2_s(\tau_2)}\right)\nonumber\\
  &&\times\int \frac{d^{3}{\bf k}_1}{(2\pi)^3} \int \frac{d^{3}{\bf k}_2}{(2\pi)^3} \int \frac{d^{3}{\bf k}_3}{(2\pi)^3} \int \frac{d^{3}{\bf k}_4}{(2\pi)^3} \int \frac{d^{3}{\bf k}_5}{(2\pi)^3} \int \frac{d^{3}{\bf k}_6}{(2\pi)^3}\nonumber\\
  &&\times \delta^3\bigg({\bf k}_1+{\bf k}_2+{\bf k}_3\bigg) \delta^3\bigg({\bf k}_4+{\bf k}_5+{\bf k}_6\bigg)\nonumber\\
  &&\times \langle \hat{\zeta}_{\bf p}\hat{\zeta}_{-{\bf p}}\hat{\zeta}^{'}_{{\bf k}_1}(\tau_1)\hat{\zeta}_{{\bf k}_2}(\tau_1)\hat{\zeta}_{{\bf k}_3}(\tau_1)\hat{\zeta}^{'}_{{\bf k}_4}(\tau_2)\hat{\zeta}_{{\bf k}_5}(\tau_2)\hat{\zeta}_{{\bf k}_6}(\tau_2)\rangle,\\
  \langle\hat{\zeta}_{\bf p}\hat{\zeta}_{-{\bf p}}\rangle^{\dagger}_{(0,2)}&=& -\frac{M^4_{ pl}}{4}\int^{0}_{-\infty}d\tau_1\frac{a^2(\tau_1)}{c^2_s(\tau_1)}\epsilon(\tau_1)\partial_{\tau_1}\left(\frac{\eta(\tau_1)}{c^2_s(\tau_1)}\right)\;\int^{0}_{-\infty}d\tau_2\;\frac{a^2(\tau_2)}{c^2_s(\tau_2)}\epsilon(\tau_2)\partial_{\tau_2}\left(\frac{\eta(\tau_2)}{c^2_s(\tau_2)}\right)\nonumber\\
  &&\times\int \frac{d^{3}{\bf k}_1}{(2\pi)^3} \int \frac{d^{3}{\bf k}_2}{(2\pi)^3} \int \frac{d^{3}{\bf k}_3}{(2\pi)^3} \int \frac{d^{3}{\bf k}_4}{(2\pi)^3} \int \frac{d^{3}{\bf k}_5}{(2\pi)^3} \int \frac{d^{3}{\bf k}_6}{(2\pi)^3}\nonumber\\
  &&\times \delta^3\bigg({\bf k}_1+{\bf k}_2+{\bf k}_3\bigg) \delta^3\bigg({\bf k}_4+{\bf k}_5+{\bf k}_6\bigg)\nonumber\\
  &&\times \langle \hat{\zeta}_{\bf p}\hat{\zeta}_{-{\bf p}}\hat{\zeta}^{'}_{{\bf k}_1}(\tau_1)\hat{\zeta}_{{\bf k}_2}(\tau_1)\hat{\zeta}_{{\bf k}_3}(\tau_1)\hat{\zeta}^{'}_{{\bf k}_4}(\tau_2)\hat{\zeta}_{{\bf k}_5}(\tau_2)\hat{\zeta}_{{\bf k}_6}(\tau_2)\rangle^{\dagger},\\
 \langle\hat{\zeta}_{\bf p}\hat{\zeta}_{-{\bf p}}\rangle_{(1,1)}&=& \frac{M^4_{ pl}}{4}\int^{0}_{-\infty}d\tau_1\frac{a^2(\tau_1)}{c^2_s(\tau_1)}\epsilon(\tau_1)\partial_{\tau_1}\left(\frac{\eta(\tau_1)}{c^2_s(\tau_1)}\right)\;\int^{0}_{-\infty}d\tau_2\;\frac{a^2(\tau_2)}{c^2_s(\tau_2)}\epsilon(\tau_2)\partial_{\tau_2}\left(\frac{\eta(\tau_2)}{c^2_s(\tau_2)}\right)\nonumber\\
  &&\times\int \frac{d^{3}{\bf k}_1}{(2\pi)^3} \int \frac{d^{3}{\bf k}_2}{(2\pi)^3} \int \frac{d^{3}{\bf k}_3}{(2\pi)^3} \int \frac{d^{3}{\bf k}_4}{(2\pi)^3} \int \frac{d^{3}{\bf k}_5}{(2\pi)^3} \int \frac{d^{3}{\bf k}_6}{(2\pi)^3}\nonumber\\
  &&\times \delta^3\bigg({\bf k}_1+{\bf k}_2+{\bf k}_3\bigg) \delta^3\bigg({\bf k}_4+{\bf k}_5+{\bf k}_6\bigg)\nonumber\\
  &&\times \langle \hat{\zeta}^{'}_{{\bf k}_1}(\tau_1)\hat{\zeta}_{{\bf k}_2}(\tau_1)\hat{\zeta}_{{\bf k}_3}(\tau_1)\hat{\zeta}_{\bf p}\hat{\zeta}_{-{\bf p}}\hat{\zeta}^{'}_{{\bf k}_4}(\tau_2)\hat{\zeta}_{{\bf k}_5}(\tau_2)\hat{\zeta}_{{\bf k}_6}(\tau_2)\rangle. \eea
 
Here it is important to note to compute the above-mentioned correlators explicitly we have to utilize the fact that the effective sound speed parameter for EFT $c_s$ and the second slow-roll parameter are taken to be almost constants in the SRI, SRII, and USR periods, except at $\tau=\tau_s$ and $\tau=\tau_e$ where we have considered sharp transitions from SRI to USR and USR to SRII respectively. For this reason, one can immediately write the following equation\footnote{During our computation we have considered sharp transitions at the conformal time scales, $\tau=\tau_s$ (SRII to USR) and $\tau=\tau_e$ (USR to SRII), where we define the following equation:
\bea \partial_{\tau}\left(\frac{\eta(\tau)}{c^2_s(\tau)}\right)\approx\frac{\Delta \eta(\tau)}{c^2_s(\tau)}\bigg(\underbrace{\delta(\tau-\tau_e)}_{\bf USR\rightarrow SRII}-\underbrace{\delta(\tau-\tau_s)}_{\bf SRI\rightarrow USR}\bigg).\eea}:
\be \displaystyle \partial_{\tau}\left(\frac{\eta(\tau)}{c^2_s(\tau)}\right)\approx 0,\ee 
which means that the above results perfectly hold good in the region $\tau<\tau_s$, within the conformal time window, $\tau_s<\tau<\tau_e$ and in the region, $\tau>\tau_e$. Outside the mentioned regions, let's say at the SRI to USR sharp transition point, $\tau=\tau_s$ and at the USR to SRII sharp transition point $\tau=\tau_e$ the above approximation can't be used. For this reason instead of computing the full integral over the conformal time scale within the range $-\infty<\tau<0$ we need to collect the contributions only from the boundaries of the USR region which are attached to the SRI region and SRII region, and appearing at the scale $\tau=\tau_s$ and $\tau=\tau_e$, as pointed before. To know about this particular crucial fact in detail and its impact on the present computation look at Appendix \ref{app:A}, where with proper physical justification we have established this fact and also discussed its direct consequence in the simplification of the final result of the one-loop integral. For a better understanding purpose let us further mention the time-dependent parametrization of the effective sound speed parameter in the present context of discussion. Throughout the evolution of the time scale, the effective sound speed parameter is fixed at the CMB pivot scale value, which is $c_s(\tau_*)=c_{s}$. However, at the sharp transition points, it is parameterized as, $c_s(\tau_e)=c_s(\tau_s)=\tilde{c}_s=1\pm \delta$, where $\delta$ represents a fine-tuning parameter which is fixed by maintaining $\delta\ll 1$. This additional information will be extremely helpful during the computation of the normalized and resumed contribution of the one-loop contribution to the momentum loop integrals computed in the different phases both for Framework I and II respectively.

\subsection{Computing one-loop corrected primordial power spectrum for scalar modes}
\label{s4c}

\subsubsection{Result for the Framework I}
After using every conceivable Wick contraction for the Framework I, the one-loop contribution to the two-point correlation function of the scalar perturbation can be further denoted by the following \footnote{Here it is important to note that, 
\be \langle\hat{\zeta}_{\bf p}\hat{\zeta}_{-{\bf p}}\rangle_{(0,1)}+\langle\hat{\zeta}_{\bf p}\hat{\zeta}_{-{\bf p}}\rangle^{\dagger}_{(0,1)}=2{\rm Re}\left[\langle\langle\hat{\zeta}_{\bf p}\hat{\zeta}_{-{\bf p}}\rangle\rangle_{(0,1)}\right]=0,\ee
by applying the Wick contraction. For this reason equation (\ref{g0}) and equation (\ref{g1}) will not finally contribute to the one-loop correction of the primordial power spectrum for the scalar perturbation.}:
 \bea \label{ff1}\langle\langle\hat{\zeta}_{\bf p}\hat{\zeta}_{-{\bf p}}\rangle\rangle_{\bf One-loop}&=&\langle\langle\hat{\zeta}_{\bf p}\hat{\zeta}_{-{\bf p}}\rangle\rangle_{(1,1)}+2{\rm Re}\bigg[\langle\langle\hat{\zeta}_{\bf p}\hat{\zeta}_{-{\bf p}}\rangle\rangle_{(0,2)}\bigg]\nonumber\\
 &\approx & \frac{M^4_{ pl}}{4}\Bigg(a^4(\tau_e)\epsilon^2(\tau_e)\frac{\left(\Delta\eta(\tau_e)\right)^2}{c^8_s} {\cal D}({\bf p},\tau_e)-a^4(\tau_s)\epsilon^2(\tau_s)\frac{\left(\Delta\eta(\tau_s)\right)^2}{c^8_s}{\cal D}({\bf p},\tau_s)\Bigg)\nonumber\\
 &&\quad\quad\quad\quad+\frac{M^2_{ pl}}{2} \Bigg(a^2(\tau_e)\epsilon(\tau_e)\frac{\left(\Delta\eta(\tau_e)\right)}{c^4_s} {\cal E}({\bf p},\tau_e)-a^2(\tau_s)\epsilon(\tau_s)\frac{\left(\Delta\eta(\tau_s)\right)}{c^4_s}{\cal E}({\bf p},\tau_s)\Bigg),\eea
Additionally it is important to note that, here we define a new function which is momentum and conformal time dependent functions ${\cal D}({\bf p},\tau)$ and ${\cal E}({\bf p},\tau)$, given by\footnote{After performing all possible  Wick contractions and performing all the momentum integrals in presence of Dirac Delta functions we finally left with equation (\ref{ff1}). Here we found that due to having symmetry we finally left with effectively $16$ contributions for ${\cal D}({\bf p},\tau)$, which turn out to be exactly identical and one additional contribution for ${\cal E}({\bf p},\tau)$. }:
 \bea {\cal D}({\bf p},\tau)&=&16\int \frac{d^{3}{\bf k}}{(2\pi)^3}\bigg[|\zeta_{\bf p}|^{2}|\zeta_{{\bf k}-{\bf p}}|^{2}{\rm Im}\bigg(\zeta^{'}_{{\bf p}}\zeta^{*}_{{\bf p}}\bigg){\rm Im}\bigg(\zeta^{'}_{{\bf k}}\zeta^{*}_{{\bf k}}\bigg)\bigg],\\
 {\cal E}({\bf p},\tau)&=&\int \frac{d^{3}{\bf k}}{(2\pi)^3}\bigg[|\zeta_{\bf p}|^{2}|\zeta_{{\bf k}-{\bf p}}|^{2}\frac{d\ln |\zeta_{{\bf k}-{\bf p}}|^{2} }{d\ln k}\bigg].\eea
Here it is important to note that, $k$ is the loop momenta which is very large compared to external momenta $p$, i.e. $k\gg p$. For this reason during performing the loop integral one 
 can neglect the contribution from the momenta $p$. Consequently, the integrals ${\cal D}({\bf p},\tau)$ and ${\cal E}({\bf p},\tau)$ can be further approximated by the following expressions:
 \bea {\cal D}({\bf p},\tau)&=&16|\zeta_{\bf p}|^{2}\int \frac{d^{3}{\bf k}}{(2\pi)^3}\bigg[|\zeta_{{\bf k}}|^{2}{\rm Im}\bigg(\zeta^{'}_{{\bf p}}\zeta^{*}_{{\bf p}}\bigg){\rm Im}\bigg(\zeta^{'}_{{\bf k}}\zeta^{*}_{{\bf k}}\bigg)\bigg],\\
 {\cal E}({\bf p},\tau)&=&|\zeta_{\bf p}|^{2}\int \frac{d^{3}{\bf k}}{(2\pi)^3}\bigg[|\zeta_{{\bf k}}|^{2}\frac{d\ln |\zeta_{{\bf k}}|^{2} }{d\ln k}\bigg].\eea
 
 Our prime objective is now to evaluate this integral within the USR region and try to find out the exact contribution of this cubic self-interaction part in the one-loop corrected primordial power spectrum for scalar modes.
 
 To evaluate this integral we need to use the following facts:
\bea  &&\bigg[{\rm Im}\bigg(\zeta^{'}_{{\bf k}}\zeta^{*}_{{\bf k}}\bigg)\bigg]_{\tau=\tau_e}=-\frac{1}{\left(k_e\right)^2}\left(\frac{k_e}{k_s}\right)^{6}\left(\frac{H^{2}}{4\pi^{2}M^{2}_{ pl}\epsilon c_s}\right)_*,\\
&&\bigg[{\rm Im}\bigg(\zeta^{'}_{{\bf k}}\zeta^{*}_{{\bf k}}\bigg)\bigg]_{\tau=\tau_s}=-\frac{1}{\left(k_s\right)^2}\left(\frac{k_e}{k_s}\right)^{6}\left(\frac{H^{2}}{4\pi^{2}M^{2}_{pl}\epsilon c_s}\right)_*.\quad\quad\eea
 Using these results we get the following simplified result: 
 \bea &&\bigg[\Delta^{2}_{\zeta, {\bf One-loop}}(p)\bigg]_{\bf USR\;on\;SR}=\bigg\{\bigg[\frac{1}{4}\bigg[\Delta^{2}_{\zeta,{\bf Tree}}(p)\bigg]^2_{\bf SR}\times\bigg(\frac{\left(\Delta\eta(\tau_e)\right)^2}{c^8_s} \left(\frac{k_e}{k_s}\right)^{6}{\bf I}(\tau_e)- \frac{\left(\Delta\eta(\tau_s)\right)^2}{c^8_s}{\bf I}(\tau_s)\bigg)\nonumber\\
&&\quad\quad\quad\quad\quad\quad\quad\quad\quad\quad\quad\quad+\frac{1}{2}\bigg[\Delta^{2}_{\zeta,{\bf Tree}}(p)\bigg]^2_{\bf SR}\times\bigg(\frac{\left(\Delta\eta(\tau_e)\right)}{c^4_s}\left(\frac{k_e}{k_s}\right)^{6}{\bf F}(\tau_e)-\frac{\left(\Delta\eta(\tau_s)\right)}{c^4_s}{\bf F}(\tau_s)\bigg)\bigg]-c_{\bf USR}\bigg\},\quad\eea
 where in the super-horizon scale of the USR region we have the following simplified result of the one-loop integral:
 \bea &&\boxed{{\bf I}(\tau_e)={\bf I}(\tau_s)
		={\bf I}\approx  \ln \left(\frac{k_e}{k_s}\right)},\\ 
  &&\boxed{{\bf F}(\tau_e)={\bf F}(\tau_s)
		={\bf F}\approx -{\cal O}(1)}.\eea
  The details of the computation are found in Appendix \ref{app:B}, where we have explicitly derived the result in great detail. \textcolor{black}{Additionally, it is important to note that the term $c_{\bf USR}$ mimics the role of counterterm while performing the renormalization under a specified scheme. In the latter half of the related discussion, we are going to discuss two equivalent schemes, which are late-time renormalization and adiabatic/wave function renormalization. Such schemes can able to fix the explicit form of this counter term of a given specified scheme. } 
 
Last but not least, the one-loop contribution to the power spectrum of the scalar perturbation resulting from the USR period on SR contribution can be stated on the super-horizon scale using the following shortened form:
\bea \boxed{\bigg[\Delta^{2}_{\zeta, {\bf One-loop}}(p)\bigg]_{\bf USR\;on\;SR}=\bigg[\Delta^{2}_{\zeta,{\bf Tree}}(p)\bigg]_{\bf SR}V}.\quad\quad\eea
Here the quantity $V$ is defined as\footnote{To know about the relative contributions from ${\bf Z}_1$ and ${\bf Z}_2$ see Appendix \ref{app:B}. }:
\bea V&=&\bigg({\bf Z}_1+{\bf Z}_2-c_{\bf USR}\bigg),\eea
where the factors ${\bf Z}_1$ and ${\bf Z}_2$ are defined by the following expressions:
\bea {\bf Z}_1:&=&\frac{1}{4}\bigg[\Delta^{2}_{\zeta,{\bf Tree}}(p)\bigg]_{\bf SR}\times \Bigg(\frac{\left(\Delta\eta(\tau_e)\right)^2}{c^8_s} \left(\frac{k_e}{k_s}\right)^{6}-\frac{\left(\Delta\eta(\tau_s)\right)^2}{c^8_s}\Bigg)\times{\bf I},\\
{\bf Z}_2:&=&\frac{1}{2}\bigg[\Delta^{2}_{\zeta,{\bf Tree}}(p)\bigg]_{\bf SR}\times\bigg(\frac{\left(\Delta\eta(\tau_e)\right)}{c^4_s}\left(\frac{k_e}{k_s}\right)^{6}-\frac{\left(\Delta\eta(\tau_s)\right)}{c^4_s}\bigg)\times{\bf F}.\eea

 Similarly, the one-loop contribution in the SRI and SRII period combiningly given by:
  \bea \boxed{\bigg[\Delta^{2}_{\zeta, {\bf One-loop}}(p)\bigg]_{\bf SR}=\bigg[\Delta^{2}_{\zeta,{\bf Tree}}(p)\bigg]_{\bf SR}U},\eea
  where the quantity $U$ is defined as:
  \bea U=U_{\bf SRI}+U_{\bf SRII},\eea 
  where each of the contributions from SRI and SRII is given by the following expressions:
\bea U_{\bf SRI}&=&\bigg[\Delta^{2}_{\zeta,{\bf Tree}}(p)\bigg]_{\bf SR}\times\Bigg(1+\frac{2}{15\pi^2}\frac{1}{c^2_{s}p^2_*}\bigg(-\left(1-\frac{1}{c^2_{s}}\right)\epsilon+6\frac{\bar{M}^3_1}{ HM^2_{ pl}}-\frac{4}{3}\frac{M^4_3}{H^2M^2_{ pl}}\bigg)\Bigg)\times\Bigg(c_{\bf SR_{I}}-\frac{4}{3}{\bf K}\Bigg),\\ 
U_{\bf SRII}&=&\bigg[\Delta^{2}_{\zeta,{\bf Tree}}(p)\bigg]_{\bf SR}\times\Bigg(1+\frac{2}{15\pi^2}\frac{1}{c^2_{s}p^2_*}\bigg(-\left(1-\frac{1}{c^2_{s}}\right)\epsilon+6\frac{\bar{M}^3_1}{ HM^2_{ pl}}-\frac{4}{3}\frac{M^4_3}{H^2M^2_{ pl}}\bigg)\Bigg)\times\Bigg(c_{\bf SR_{II}}+{\bf O}\Bigg),\quad\quad\eea
where the factors ${\bf K}$ and ${\bf O}$ are defined as\footnote{For the detailed computation of factors ${\bf K}$ and ${\bf O}$ see Appendix \ref{app:B}.}:
\bea &&\boxed{{\bf K}=\ln \left(\frac{k_e}{p_*}\right)},\\
&&\boxed{{\bf O}=-\ln \left(\frac{k_e}{k_{\rm end}}\right)-\frac{27}{32}\Bigg\{1-\left(\frac{k_e}{k_{\rm end}}\right)^{12}\Bigg\}}.\eea

\textcolor{black}{In this computation two types of divergences appear, which are UV and IR. The UV divergence appears as quadratic contributions and the IR one appears as logarithmic contributions as a byproduct of the cut-off regularization. The UV regularization
is the main issue compared to the IR. While performing the loop integrals we have explicitly used a fixed UV cutoff scale while carrying out this time-dependent calculation. As an outcome, one can immediately think that the momentum integration at a given time may include irrelevant momentum modes and such overcounting of modes may lead to a wrong result. However, such a notion is not correct which we will explain now in detail. To demonstrate our argument let us first start with the following representative momentum integral which appears in the SRI, USR, and SRII phases explicitly:
\bea {\cal I}_{\bf rep}:=\int^{k_{\rm UV}}_{k_{\rm IR}}\frac{dk}{k}\;\left(A+Bc^2_sk^2\tau^2\right)+C=\bigg[A\ln\left(\frac{k_{\rm UV}}{k_{\rm IR}}\right)+\frac{1}{2}B\tau^2c^2_s\left(k^2_{\rm UV}-k^2_{\rm IR}\right)+C\bigg],\eea 
where $k_{\rm UV}$ and $k_{\rm IR}$ represent the UV and IR momentum cut-offs, which are different for the mentioned three phases. Here $c_s$ is the value of the effective sound speed at the CMB pivot scale, $k=k_*$, which is considered to be a constant in this computation. Also, it is important to note that, constant factors $A$ and $B$ are also different for SRI, USR, and SRII regions. Here $C$ mimics the role of counter terms as appearing during performing renormalization discussed in the next section of this paper. During this demonstration, it is important to note that we have not fixed the value of the UV and IR momentum cut-offs. Further considering the super-horizon limit in the above-mentioned derived result we get the following simplified answer from the above-mentioned formula for the momentum loop-integral:
\bea \label{df} {\cal I}_{\bf rep}(\tau\rightarrow 0):=\bigg[\lim_{\tau\rightarrow 0}\bigg\{A\ln\left(\frac{k_{\rm UV}}{k_{\rm IR}}\right)+\frac{1}{2}B\tau^2c^2_s\left(k^2_{\rm UV}-k^2_{\rm IR}\right)\bigg\}+C\bigg]=A\ln\left(\frac{k_{\rm UV}}{k_{\rm IR}}\right),\eea
where the counter term $C$ is fixed at the following value so that the contribution from the quadratic UV divergence is completely removed from the underlying theoretical setup under consideration:
\bea \label{plwq} C=-\lim_{\tau\rightarrow 0}\frac{1}{2}B\tau^2c^2_s\left(k^2_{\rm UV}-k^2_{\rm IR}\right)=0.\eea
Here we have explicitly used the fact that $k_{\rm UV}\neq k_{\rm IR}$ and $k_{\rm UV}\gg k_{\rm IR}$ during fixing the counter term $C$.
In the present context, IR divergence is not harmful, and employing the prescribed counter-term constraint condition in the super-horizon limit one cannot able to completely remove such contribution from the IR end. Only this prescribed constraint condition helps us to soften the behaviour of IR logarithmic divergence in the present context of the computation. This derived result is generic and appears to be completely consistent with the analysis performed in the refs. \cite{Kristiano:2022maq,Kristiano:2023scm,Choudhury:2023hvf,Choudhury:2023kdb,Choudhury:2023hfm,Bhattacharya:2023ysp,Motohashi:2023syh,Firouzjahi:2023ahg,Franciolini:2023lgy,Firouzjahi:2023aum,Cheng:2023ikq,Tasinato:2023ukp}, where no over counting of momentum modes is appearing. At the super-horizon scale and during the time of re-entry through the horizon we can clearly observe that all such additional irrelevant momentum modes which mainly capture the information of the sub-horizon quantum fluctuations after performing momentum integration and taking the late time limit (which is extremely relevant from the observational perspective) completely washed out from the computation and finally give rise to a smoother version of IR softened result.}

\textcolor{black}{Though we have not fixed the value of the UV and IR cut-offs while performing the loop integrals, to give a full-proof technical justification and to explicitly show that no overcounting of momentum modes will appear during performing such computation we re-evaluate the previously mentioned integral considering the following trick:
\bea \int^{k_{\rm UV}}_{k_{\rm IR}}:=\bigg(\int^{k_{\rm INT}}_{k_{\rm IR}}+\int^{k_{\rm UV}=\Lambda_{\rm UV} a(\tau)/c_s}_{k_{\rm INT}}\bigg),\eea
 Since the associated wave number $k_{\rm UV}$ is expressed in the comoving scale and all the results we are computing in the super-horizon scale at the time of horizon re-entry it is quite physically consistent to express the UV limit of the integration in terms of the conformal time dependence. Here we have incorporated the information of the time dependence in the UV limit of the integration $k_{\rm UV}=\Lambda_{\rm UV} a(\tau)/c_s$ for the better understanding purpose explicitly. Here $\Lambda_{\rm UV}$ is the contribution of the UV cutoff which is introduced after extracting out the conformal time dependence in the present context. Additionally, it is important to note that we have decomposed the limits of the momentum integral into two parts. The first part basically takes care of the finite contribution of the integral and the second part is taking care of the quadratic UV divergence in the presence of the time-dependence in the upper limit of the momentum-dependent loop integration. After using the above-mentioned decomposition in the limits of the integration we can write down the following simplified expression:
\bea \label{yh} {\cal I}_{\bf rep}:&=&\bigg\{\bigg(\int^{k_{\rm INT}}_{k_{\rm IR}}+\int^{k_{\rm UV}=\Lambda_{\rm UV} a(\tau)/c_s}_{k_{\rm INT}}\bigg)\frac{dk}{k}\;\left(A+Bc^2_sk^2\tau^2\right)\bigg\}+C\nonumber\\
&=&\bigg[A\bigg\{\ln\left(\frac{k_{\rm INT}}{k_{\rm IR}}\right)+\ln\left(\frac{\Lambda_{\rm UV}}{H}\right)\bigg\}+\frac{1}{2}B\bigg\{\left(\frac{\Lambda_{\rm UV}}{H}\right)^2-1\bigg\}+C\bigg],\eea  
where during the computation of the last step of the loop integration we have utilized the fact that $a(\tau)/c_sk_{\rm INT}=H^{-1}$ and in the super-horizon scale $Bc^2_s\tau^2(k^2_{\rm INT}-k^2_{\rm IR})/2$ vanishes respecting the constraint $k_{\rm INT}\gg k_{\rm IR}$. Further imposing the constraint from wave function/adiabatic renormalization at the pivot scale $k=k_*$ one can further determine the explicit expression for the counter term $C$ in the present context of discussion and this is given by the following simplified formula:
  \bea C(\mu,\Lambda_{\rm UV})=\bigg[A\bigg\{\ln\left(\frac{\mu}{H}\right)-\ln\left(\frac{\Lambda_{\rm UV}}{H}\right)\bigg\}+\frac{1}{2}B\bigg\{\left(\frac{\mu}{H}\right)^2-\left(\frac{\Lambda_{\rm UV}}{H}\right)^2\bigg\}\bigg].\eea
  Here $\mu$ represents the renormalization scale associated with the adiabatic renormalization scheme.
  Further substituting the above-mentioned result in the equation (\ref{yh}) we get the following answer from the one-loop contribution, which is valid in the SRI, USR, and SRII phases separately:
\bea \label{yh1} {\cal I}_{\bf rep}(\mu):
&=&\bigg[A\bigg\{\ln\left(\frac{k_{\rm INT}}{k_{\rm IR}}\right)+\ln\left(\frac{\Lambda_{\rm UV}}{H}\right)\bigg\}+\frac{1}{2}B\bigg\{\left(\frac{\Lambda_{\rm UV}}{H}\right)^2-1\bigg\}\bigg]\nonumber\\
&&\quad\quad\quad\quad\quad\quad+\bigg[A\bigg\{\ln\left(\frac{\mu}{H}\right)-\ln\left(\frac{\Lambda_{\rm UV}}{H}\right)\bigg\}+\frac{1}{2}B\bigg\{\left(\frac{\mu}{H}\right)^2-\left(\frac{\Lambda_{\rm UV}}{H}\right)^2\bigg\}\bigg]\nonumber\\
&=&\bigg[A\bigg\{\ln\left(\frac{k_{\rm INT}}{k_{\rm IR}}\right)+\ln\left(\frac{\mu}{H}\right)\bigg\}+\frac{1}{2}B\bigg\{\left(\frac{\mu}{H}\right)^2-1\bigg\}\bigg].\eea  
Next, considering the fact that the renormalization scale is associated with an adiabatic renormalization scheme at $\mu=H$ we get the following simplified answer for the loop-integral:
\bea \label{yh2} {\cal I}_{\bf rep}(\mu=H)
&=&A\ln\left(\frac{k_{\rm INT}}{k_{\rm IR}}\right),\eea 
which is exactly the similar result that we have derived in the equation (\ref{df}). Here it is important to note that the final derived result is completely independent of the UV cut-off $\Lambda_{\rm UV}$, which we have introduced during the inclusion of conformal time dependence in the upper limit of the momentum loop integration. This comparative analysis not only helps us to establish the equivalence between the mentioned two procedures, late time renormalization and adiabatic/ wave function renormalization but also confirms the correctness of the computations performed in this paper. Most importantly, this technical justification allows us to confirm that during performing the computation no irrelevant momentum modes are overpowering the final result. As a consequence are completely on the safe side as overcounting does not appear in the present context of discussion and as an immediate outcome computations performed in this paper are completely trustworthy. With the help of any of the renormalization procedures stated before one can able to completely remove the contribution of the dangerous quadratic UV divergence term from the one-loop momentum integration. However, the logarithmic IR divergent contribution cannot able to removed by implementing any of the previously mentioned renormalization schemes. However, such IR divergence is not at all harmful, and the corresponding contribution can be further smoothened by implementing the power spectrum renormalization, which serves the purpose of course graining as discussed in the next section of this paper in great detail. The same discussions will also hold for the computations performed for Framework II where the only difference is the SRII phase is absent and just after the USR phase inflation ends.}

Here in the SRI phase, $p_*$ represents the pivot scale which is expected to be $p_*\ll k_s$ in the present framework under consideration\footnote{\bf {It is necessary to have the required number of e-foldings in the present set up in the first slow roll regime.}}. Additionally, it is important to note that, in the SRII phase $k_e\ll k_{\rm end}$, for this reason, $k_e/k_{\rm end}\ll 1$ and such contributions are highly suppressed. For the numerical purpose we choose, $k_e=10^{22}{\rm Mpc}^{-1}$ and $k_{\rm end}=10^{24}{\rm Mpc}^{-1}$, for which we have $k_e/k_{\rm end}=10^{-2}\ll 1$. Also, it is important to note that, $c_{\bf SR_I}$ and  $c_{\bf SR_{II}}$ are two parameters from the SRI and SRII phases which depend on any renormalization method \footnote{\textcolor{black}{For the numerical estimation purpose during the implementation of the underlying physical principles of late time renomrlaziartion scheme in the super-horizon limit we fix, $c_{\bf SR_I}=0$ and  $c_{\bf SR_{II}}=0$, without losing any generality of the underlying problem under consideration. For a detailed clarification of this logical argument see equation (\ref{plwq}) which correctly justifies the necessity of this requirement. }} and appears after the contributions from UV divergences in the underlying theory have been canceled during the SRI and SRII phases. The computation details are found in Appendix \ref{app:B}.

Let's now write down the complete equation for the one loop adjusted scalar power spectrum, which will be useful for the subsequent discussions, before moving on to any other topics. The following is the equivalent expression for the scalar power spectrum during the PBH creation in the USR period \footnote{To avoid any further confusion here it is important to note that, for the Framework I when we have written SR in the final expression for the one-loop corrected power spectrum for scalar modes it means particularly the SRI phase, not the SRII phase. This is a very crucial point to technically write and physically interpret the obtained result in the context of Framework II.}:
\bea \boxed{\Delta^{2}_{\zeta, {\bf EFT}}(p)
=\bigg[\Delta^{2}_{\zeta,{\bf Tree}}(p)\bigg]_{\bf SR}\Bigg(1+U+V\Bigg)}.\eea
\textcolor{black}{Here the scalar power spectrum's slow-roll (SRI) contribution to the Framework I can be written as:
\bea \bigg[\Delta^{2}_{\zeta,{\bf Tree}}(p)\bigg]_{\bf SR}&=&\left(\frac{H^{2}}{8\pi^{2}M^{2}_{ pl}\epsilon c_s}\right)_{*}\left(1+\left(\frac{p}{k_s}\right)^2\right)\xrightarrow{\rm Super-hozon\, scale\, p\ll k_s}\left(\frac{H^{2}}{8\pi^{2}M^{2}_{ pl}\epsilon c_s}\right)_{*},\quad\quad\eea
which is perfectly consistent with the results we have derived for the Framework I in the super-horizon scale as stated in equation (\ref{sramp}). This clarification helps us to understand the correctness and applicability of the derived result in the present context of discussion associated with Framework I. }

\subsubsection{Result for the Framework II}

The contribution from the USR for this framework is exactly same as computed for the Framework I. For the completeness we write the one-loop correction directly, which is given by the following expression: 
\bea \boxed{\bigg[\Delta^{2}_{\zeta, {\bf One-loop}}(p)\bigg]_{\bf USR\;on\;SR}=\bigg[\Delta^{2}_{\zeta,{\bf Tree}}(p)\bigg]_{\bf SR}V}.\quad\quad\eea
Here the quantity $V$ is defined previously for Framework I, which is exactly same for our Framework II. The only difference is in the expressions, we need to interpret now $k_e$ as the wave number associated with end of inflation as well as the end of USR period.

 Similarly, the one-loop contribution in the single SR period given by:
  \bea \boxed{\bigg[\Delta^{2}_{\zeta, {\bf One-loop}}(p)\bigg]_{\bf SR}=\bigg[\Delta^{2}_{\zeta,{\bf Tree}}(p)\bigg]_{\bf SR}U},\eea
  where the quantity $U$ is defined as:
\bea U&=&\bigg[\Delta^{2}_{\zeta,{\bf Tree}}(p)\bigg]_{\bf SR}\times\Bigg(1+\frac{2}{15\pi^2}\frac{1}{c^2_{s}p^2_*}\bigg(-\left(1-\frac{1}{c^2_{s}}\right)\epsilon+6\frac{\bar{M}^3_1}{ HM^2_{ pl}}-\frac{4}{3}\frac{M^4_3}{H^2M^2_{ pl}}\bigg)\Bigg)\times\Bigg(c_{\bf SR_{I}}-\frac{4}{3}{\bf K}\Bigg),\eea
where the factors ${\bf K}$ is defined as:
\bea &&\boxed{{\bf K}=\ln \left(\frac{k_e}{p_*}\right)}.\eea
Here in SRI phase $p_*$ represents the pivot scale which is fixed at $p_*=0.02{\rm Mpc}^{-1}$ for all the numerical purposes. Also it is important to note that, $c_{\bf SR}$  depend on any renormalization method \footnote{For the numerical estimation purpose we fix, $c_{\bf SR}=0$, without loosing any generality of the underlying problem under consideration.} and appears after the contributions from UV divergences in the underlying theory have been cancelled during the SR phase.

  The final expression for the one-loop corrected power spectrum exactly looking same like for Framework I. The only difference are the function $U$ is made of with a single SR one-loop contribution and in the function $V$ the UV momenta is, $k_e$, which is interpreted as the wave number at the end of inflation as well as at the end of the mentioned USR phase. For completeness again we write down the total contribution of the one-loop corrected spectrum for the scalar modes as obtained from Framework II, which is given by the following expression:
\bea \boxed{\Delta^{2}_{\zeta, {\bf EFT}}(p)
=\bigg[\Delta^{2}_{\zeta,{\bf Tree}}(p)\bigg]_{\bf SR}\Bigg(1+U+V\Bigg)},\eea
\textcolor{black}{where the single slow-roll (SR) contribution in the spectrum for the Framework II can be written as:
\bea \bigg[\Delta^{2}_{\zeta,{\bf Tree}}(p)\bigg]_{\bf SR}&=&\left(\frac{H^{2}}{8\pi^{2}M^{2}_{ pl}\epsilon c_s}\right)_{*}\left(1+\left(\frac{p}{k_s}\right)^2\right)\xrightarrow{\rm Super-hozon\, scale\, p\ll k_s}\left(\frac{H^{2}}{8\pi^{2}M^{2}_{ pl}\epsilon c_s}\right)_{*},\quad\quad\eea
which is perfectly consistent with the results we have derived for Framework II in the super-horizon scale as stated in equation (\ref{sramp}). This clarification helps us to understand the correctness and applicability of the derived result in the present context of discussion associated with Framework II. }

\section{Renormalized one-loop scalar power spectrum  from EFT}  
\label{s5}

Before performing the renormalization of the computed one-loop corrected scalar power spectrum as computed in the previous section it is important to note that the renormalization procedure is exactly same for the Framework I and Framework II. The only difference appears in the structure of the SR phase dependent function $U$ in both the cases. Also, though the structure of the other USR phase function $V$ is exactly same in both of these mentioned frameworks, the physical interpretation is different due to having two different ending mechanism for inflation. Now we develop the present formalism in such a fashion that the explicit structure of the functions $U$ and $V$ will not required to construct a renormalized version of one-loop corrected power spectrum for the scalar modes. For this specific reason during performing renormalization the explicit details of the Framework I and Framework II will be not required and the final derived result in this section will hold good perfectly for both of the mentioned frameworks.

Now it is important to note that the following renormalized power spectrum for the scalar perturbation for the recommended EFT set up is necessarily needed in order to totally eliminate the effects of one-loop logarithmic divergences from- (1) SRI, SRII as well as the USR effects on SRI period for the Framework I and, (2) single SR as well as the USR effects on SR phase for the Framework II. It is technically demonstrated by the following expression for both of the mentioned frameworks:
\bea  \overline{\Delta^{2}_{\zeta,{\bf EFT}}(p)}&=&{\cal Z}_{\zeta,{\bf EFT}}\times\Delta^{2}_{\zeta,{\bf EFT}}(p),\eea
where the renormalization factor, also called the counter-term, is determined by the explicit renormalization condition and is expressed by the quantity  ${\cal Z}_{\zeta,{\bf EFT}}$, which we need to be determined in this paper correctly.  Any renormalization strategy necessitates deriving the counter term from the background theoretical framework. Renormalization condition is fixed at the pivot scale $p_*$, which technically expressed as:
\bea \overline{\Delta^{2}_{\zeta,{\bf EFT}}(p_*)}&=& \bigg[\bigg[\Delta^{2}_{\zeta,{\bf Tree}}(p)\bigg]_{\bf SR}\bigg]_{p=p_*}=\bigg[\Delta^{2}_{\zeta,{\bf Tree}}(p_*)\bigg]_{\bf SR},\eea
Hence the counter term can be calculated from the above mentioned boundary condition as:
\bea  \boxed{{\cal Z}_{\zeta,{\bf EFT}}=\frac{ \overline{\Delta^{2}_{\zeta,{\bf EFT}}(p_*)}}{\Delta^{2}_{\zeta,{\bf EFT}}(p_*)}
=\frac{\bigg[\Delta^{2}_{\zeta,{\bf Tree}}(p_*)\bigg]_{\bf SR}}{\Delta^{2}_{\zeta,{\bf EFT}}(p_*)}=\frac{1}{\left(1+U_*+V_*\right)}\approx \left(1-U_*-V_*+\cdots\right)}.\quad\quad\eea
Here the $U_*$ and $V_*$ both are logarithm dependent terms. In the series expansion of $\frac{1}{\left(1+U_*+V_*\right)}$ we have truncated this in the first order and neglected all the higher order contributions in the expansion.
Following that, using this counter term, the associated one-loop corrected renormalized power spectrum for the scalar modes can be written as:
\bea  \boxed{\overline{\Delta^{2}_{\zeta,{\bf EFT}}(p)}=\bigg[\Delta^{2}_{\zeta,{\bf Tree}}(p)\bigg]_{\bf SR}\Bigg(1+{\cal Q}_{\bf EFT}\Bigg)},\eea
where the EFT dependent function ${\cal Q}_{\bf EFT}$ after renormalization and performing proper truncation can be computed as:
\bea 
{\cal Q}_{\bf EFT}&=&\sum^{3}_{j=1}{\cal Q}_j,\eea
where the explicit form of the functions ${\cal Q}_1$, ${\cal Q}_2$, and ${\cal Q}_3$ can be expressed by the following expression:
\bea 
\label{e1} {\cal Q}_1&=&\Bigg(\frac{\bigg[\Delta^{2}_{\zeta,{\bf Tree}}(p)\bigg]_{\bf SR}}{\bigg[\Delta^{2}_{\zeta,{\bf Tree}}(p_*)\bigg]_{\bf SR}}-1\Bigg)U_*,\\
\label{e2} {\cal Q}_2&=&\Bigg(\frac{\bigg[\Delta^{2}_{\zeta,{\bf Tree}}(p)\bigg]_{\bf SR}}{\bigg[\Delta^{2}_{\zeta,{\bf Tree}}(p_*)\bigg]_{\bf SR}}-1\Bigg)V_*,\\
\label{e3} {\cal Q}_3&=&-\frac{\bigg[\Delta^{2}_{\zeta,{\bf Tree}}(p)\bigg]_{\bf SR}}{\bigg[\Delta^{2}_{\zeta,{\bf Tree}}(p_*)\bigg]_{\bf SR}}\Bigg(U^{2}_*+V^{2}_*+\cdots\Bigg).\eea
Now from the observational perspective we use the following approximation for the equation (\ref{e1}) and equation (\ref{e2}), which is given by:
\bea \frac{\bigg[\Delta^{2}_{\zeta,{\bf Tree}}(p)\bigg]_{\bf SR}}{\bigg[\Delta^{2}_{\zeta,{\bf Tree}}(p_*)\bigg]_{\bf SR}}&=&\frac{\displaystyle\left(1+\left(\frac{p}{k_s}\right)^2\right)}{\displaystyle\left(1+\left(\frac{p_*}{k_s}\right)^2\right)}\approx\left(1+\left(\frac{p}{k_s}\right)^2\right)\left(1-\left(\frac{p_*}{k_s}\right)^2\right)\approx\left(1+\frac{p^2-p^2_*}{k^2_s}+\cdots\right)\approx 1,\eea 
where in the last line of the above computation we have taken the approximation that $p$ is very near to $p_*$, which is from observational perspective quite natural because we try to probe the outcome at the pivot scale, $p_*$. For this specific reason, we have the following simplified results:
\bea && Q_1\approx 0,\\
&& Q_2\approx 0.\eea
This will further give rise to the following result where the function ${\cal Q}_{\bf EFT}$ is solely fixed in terms of the function ${\cal Q}_3$, i.e.\footnote{During representing the contribution coming from the function ${\cal Q}_3$ we have kept the distinction between the momentum $p$ and $p_*$, even though the difference between them are very small. However, if we even take they are same there will be no change in the overall final contribution in the one-loop corrected renormalized power spectrum of scalar modes. The prime reason for such fact is that the ratio $\displaystyle \frac{\bigg[\Delta^{2}_{\zeta,{\bf Tree}}(p)\bigg]_{\bf SR}}{\bigg[\Delta^{2}_{\zeta,{\bf Tree}}(p_*)\bigg]_{\bf SR}}\sim 1$, in the present set-up. The small corrections in the ratio we kept intact, though it will not change the final conclusion as pointed out before. \textcolor{black}{Careful observation pointing towards to fact that in the super-horizon limit, when $p\ll k_s$ and $p_*\ll k_s$ we have exactly, $\bigg[\Delta^{2}_{\zeta,{\bf Tree}}(p)\bigg]_{\bf SR}=\bigg[\Delta^{2}_{\zeta,{\bf Tree}}(p_*)\bigg]_{\bf SR}$, which justifies complete correctness of this approximation in the present context of discussion where we are implementing the constraints for power spectrum renormalization. For this reason, henceforth we can completely rely on the derived results which we have projected as the prime findings of this paper.}}
\bea {\cal Q}_{\bf EFT}\approx {\cal Q}_3=-\frac{\bigg[\Delta^{2}_{\zeta,{\bf Tree}}(p)\bigg]_{\bf SR}}{\bigg[\Delta^{2}_{\zeta,{\bf Tree}}(p_*)\bigg]_{\bf SR}}\Bigg(U^{2}_*+V^{2}_*+\cdots\Bigg).\quad\quad\eea
Here to comment on the ratio we have considered the fact that the wave number $p$ is not very far from the pivot scale $p_*$ which is explicitly used in the present computation. Now one can ask an immediate question that if the wave number $p$ is very far from the pivot scale, say at the sharp transition scale or at the end of inflation does our computation holds well perfectly within the present EFT setup? The answer is yes. The appearance of this ration will induce two factors ${\cal Q}_1$ and ${\cal Q}_3$. In that case to quantify the function ${\cal Q}_{\bf EFT}$ instead of only ${\cal Q}_3$ one need to take care of the summation of ${\cal Q}_1$, ${\cal Q}_2$ and ${\cal Q}_3$ together. However, this will not be going to change the final conclusion of the paper and will not be going to show any strong effect from ${\cal Q}_1$ and ${\cal Q}_3$ in ${\cal Q}_3$ during resummation, which we explicitly perform in the next section. The prime reason for this fact is in the USR period the logarithm square term becomes larger in ${\cal Q}_3$ compared to the linear logarithmic divergent term as appearing from the SR and USR phases in ${\cal Q}_1$ and ${\cal Q}_2$. We found that ${\cal Q}_3$ becomes extremely dominant compared to ${\cal Q}_1$ and ${\cal Q}_2$, though the perturbation theory will not break during this computation and resummed result will give consistent result in the present context, which we are going to explicitly show in the next section. Hence quantifying the function ${\cal Q}_{\bf EFT}$ solely in terms of ${\cal Q}_3$ is completely physically justifiable in the present context of the discussion. However, since for the effective sound speed $c_s>1.5$ the perturbative approximation does not hold well perfectly, in that case, the present approximation as well as the computation will not work perfectly. For this particular reason, resummed version of the power spectrum is more trustworthy than the renormalized power spectrum computed in this section. We will be going to elaborately discuss this issue in the next section to justify our computed result in this paper.

As an immediate consequence of the present computation the scalar spectral tilt, running and running of the running is going to be completely independent of the quantum loop effects at the pivot scale $p_*$. This can be equivalently interpreted as cosmological $\beta$ functions 
\footnote{All of these cosmological $\beta$ functions are connected via the flow equation, using which one can able to find connection among themselves. } small values of which determine the shape of the primordial scalar power spectrum and confirms the small deviation from exact scale invariance for the same. Technically this statement can be written as:
\bea &&\overline{n_{\zeta,{\bf EFT}}(p_*)}-1= n_{\zeta,{\bf SR}}(p_*)-1,\\
&&\overline{\alpha_{\zeta,{\bf EFT}}(p_*)}= \alpha_{\zeta,{\bf SR}}(p_*),\\
&&\overline{\beta_{\zeta,{\bf EFT}}(p_*)}= \beta_{\zeta,{\bf SR}}(p_*).\quad\quad \quad\eea
Not only the results show independence of the loop effects, but also it clearly explain everything in terms of the results obtained in the SR limit, at the pivot scale $p_*$ where the observational probes try to check the validity of the underlying theoretical framework. 
This is the result of a very deft yet very reasonable selection of the renormalization condition for determining the correct counter-term.  Since there should not be any influence from the quantum loops at all in these results at the pivot scale $p _*$, it becomes justified. Otherwise, the observational investigations that have been made so far must directly detect and test the loop effects in the shape of power spectrum. Here it is important to note that, the cosmological $\beta$ functions computed from the the one-loop corrected power spectrum for the scalar modes from the given Goldstone EFT set up is not equal to the SR contribution (SRI for the Framework I and single SR for the Framework II) in all the momentum scales, except the pivot scale, which we fix at $p_*=0.02{\rm Mpc}^{-1}$ for the numerical purpose in this paper. This is not at all an odd fact and quite reasonably expected from the underlying theoretical set up under consideration. The reason is as follows, if the cosmological $\beta$ functions computed from the present EFT set up become same with the SR contributions in all scale then the enhancement of the spectrum will not able to achieve to a certain required value at the peak, which is necessarily required to produce PBH in the present context of discussion. Fortunately, the obtained result suggest towards the fact that though large mass PBH can't be produced but for the range of the sound speed $1<c_s<1.5$ power spectrum enhancement from its SR counterpart becomes significantly dominant due to having large USR contribution in the one-loop correction. Beyond $c_s>1.5$ the perturbation theory breaks due to having greater one-loop contribution from the USR period compared to the tree level contribution. This fact is extremely strange but true for our set up, which we have found from the numerical analysis. Detailed outcome can be visalized in the numerical plots that we have obtained from both the Framework I and Framework II.

\section{Dynamical Renormalization Group (DRG) resummed scalar power spectrum from EFT}
\label{s6}

Before going to the technical details of the computation in this section let us clearly mention that the present computation is not based on the explicit structure of the prime component of the one-loop correction, which is identified as ${\cal Q}_{\bf EFT}$ from the Framework I and Framework II. To perform the resummation the overall magnitude of this factor has be maintained within the perturbative limit so that we can able to get the finite result in the end. We quote the result in such a fashion that for both of the mentioned frameworks this technique works well.

In this section, our prime objective is to describe the Dynamical Renormalization Group (DRG) approach \cite{Boyanovsky:1998aa,Boyanovsky:2001ty,Boyanovsky:2003ui,Burgess:2009bs,Dias:2012qy,Chen:2016nrs,Baumann:2019ghk,Burgess:2009bs,Chaykov:2022zro,Chaykov:2022pwd}, which enables us to resum over all of the contributions that are logarithmically divergent in the present computation. More precisely this resummed result is valid in all loop order of the perturbative computation which can able to correctly capture the quantum effects. However, this is only feasible when the resummed infinite series follow the strict convergent criteria at horizon crossing and super-horizon scales. These terms in the mentioned convergent series are all byproducts of the cosmological perturbation theory of scalar modes in every conceivable loop order. DRG is typically thought as as the natural process that allows the legitimacy of secular  momentum dependent contributions to the convergent infinite series at horizon crossing and super-horizon scales, which we are employing within the context of primordial EFT driven cosmological set-up, to be easily justified. Instead of knowing the complete behaviour from the series expansion term by term, after executing the resummation process, this technique allows to correctly know the behaviour at the horizon crossing and super-horizon scales. Initially the concept of Renomrmalization Group (RG) resummation technique came into the picture,  which involves momentum dependent contributions into the equations for scale dependent running couplings of the underlying theory in terms of $\beta$ functions. DRG resummation technique is the more refined version of the previously mentioned RG resummation. This result can be applied in a certain observationally viable broader range of running of momentum scales in the tiny coupling domain where the perturbative approximation perfectly holds good within the framework of underlying EFT set-up. In the context of Cosmological DGR resummation at the late time scale, instead of studying the behaviour of running coupling with respect to underlying scale we study the shape and features of the primordial power spectrum and these features are characterized by the spectral tilt, running and running of the running of spectral tilt with respect to the momentum scale. All these characteristic physical quantities are interpreted as the cosmological $\beta$ functions,  existence of which confirms the small deviation from the exact scale invariant feature of the primordial power spectrum. In the corresponding context of discussion, DRG resummation is often referred as the resummation under the influence of exponentiation.   

 Using the DRG method, the final form of the resummed dimensionless primordial power spectrum for the scalar modes can be expressed as follows:
\bea \overline{\overline{\Delta^{2}_{\zeta,{\bf EFT}}(p)}}
&=&\bigg[\Delta^{2}_{\zeta,{\bf Tree}}(p)\bigg]_{\bf SR}\times \Bigg\{1+{\cal Q}_{\bf EFT}+\frac{1}{2!}{\cal Q}^2_{\bf EFT}+\cdots\Bigg\}\times\bigg\{1+{\cal O}\bigg(\bigg[\Delta^{2}_{\zeta,{\bf Tree}}(p_*)\bigg]^2_{\bf SR}\bigg)\bigg\}\nonumber\\
&\approx&\bigg[\Delta^{2}_{\zeta,{\bf Tree}}(p)\bigg]_{\bf SR}\times \Bigg(\sum^{\infty}_{n=0}\frac{{\cal Q}^n_{\bf EFT}}{n!}\Bigg)\times\bigg\{1+{\cal O}\bigg(\bigg[\Delta^{2}_{\zeta,{\bf Tree}}(p_*)\bigg]^2_{\bf SR}\bigg)\bigg\}\nonumber\\
&=&\bigg[\Delta^{2}_{\zeta,{\bf Tree}}(p)\bigg]_{\bf SR}\exp\bigg({\cal Q}_{\bf EFT}\bigg)\times\bigg\{1+{\cal O}\bigg(\bigg[\Delta^{2}_{\zeta,{\bf Tree}}(p_*)\bigg]^2_{\bf SR}\bigg)\bigg\},\eea
which actually describes the large scale behaviour after clubbing the contributions from the secular terms\footnote{In general the uncontrollable growth of the secular terms becomes very problematic as it invalidate the perturbative computation at the late time scale, particularly in the horizon crossing and super-horizon scale. DRG method helps us to resolve this issue in primordial cosmology.}. Here the DRG resummed version of the result holds good for all order of the function ${\cal Q}_{\bf EFT}$. Here it is to be noted that the convergence criteria strictly demands, $|{\cal Q}_{\bf EFT}|\ll 1$, which is perfectly satisfied in the present context of discussion. Most significant outcome of DRG resummed version of one-loop corrected primordial power spectrum for the scalar modes is it produces a controlled version of the dimensionless primordial power spectrum where the behaviour of the logarithimically divergent contribution becomes more softened compared to the renormalized version of the one-loop power spectrum explicitly derived in the previous section. Though the explicit details of the Feynman diagrams and the sub graphs are not needed during performing the DRG resummation method, but it is important to mention that the leading order logarithmically divergent contributions are appearing from chain diagrams which continually add cubic self-energy in the present context. The applicability of the DRG resummation not needed the domination of all of the chain diagrams compared to the other possible diagrams appearing in the computation, but in the leading contributions coming from these logarithmic dependent factors it will surely contribute. In the above mentioned result the higher order convergent terms in the infinite series exactly mimics the role of higher-loop contributions in the present context of discussion. It is really a remarkable fact, because without explicitly performing the higher-loop correction to the primordial power spectrum for the scalar modes one can study the behaviour of each correction terms in all-loop order, which means we can study the non-perturbative though convergent behaviour of the spectrum as the sum over all-loop contribution becomes finite and expressed in terms of an exponential function in this context. 

Now to understand and clearly visualize the softening the behaviour of the logarithimic divergence we utilize the fact that the contribution coming from the function $U$, which is the byproduct of the SR phase, is extremely small compared to the USR contribution $V$ as it contains a factor $(k_e/k_s)^6\ln (k_e/k_s)$. For this reason one can for the time being can neglect the function $U$ in the factor ${\cal Q}_{\bf EFT}$. As a consequence, we can write the following approximated result:
\bea {\cal Q}_{\bf EFT}&\approx& -{\bf R}(p)\Bigg(V^{2}_*+\cdots\Bigg)= -{\bf R}(p)\Bigg(\xi_{*} {\bf Z}^2_{1,*}+\cdots\Bigg),\eea
where we define the ratio of the tree level SR contributions ${\bf R}(p)$, the function ${\bf Z}_1$ and the co-efficient $\xi_{*}$ at the pivot scale as:
\bea {\bf R}(p)&=&\frac{\bigg[\Delta^{2}_{\zeta,{\bf Tree}}(p)\bigg]_{\bf SR}}{\bigg[\Delta^{2}_{\zeta,{\bf Tree}}(p_*)\bigg]_{\bf SR}},\\
{\bf Z}_{1,*}:&\approx&\frac{1}{4}\bigg[\Delta^{2}_{\zeta,{\bf Tree}}(p_*)\bigg]_{\bf SR}\times \frac{\left(\Delta\eta(\tau_e)\right)^2}{c^8_s} \left(\frac{k_e}{k_s}\right)^{6}\times\ln\left(\frac{k_e}{k_s}\right),\\ 
\xi_{*}&=&\Bigg(1+\frac{{\bf Z}_{2,*}}{{\bf Z}_{1,*}}\Bigg).\eea
Here we have used the following fact for the simplification purpose:
\bea \left(\Delta\eta(\tau_e)\right)^2\left(\frac{k_e}{k_s}\right)^{6}\gg\left(\Delta\eta(\tau_s)\right)^2.\eea
Here for the numerical purpose we choose, $\Delta\eta(\tau_e)=1$, $k_e=10^{22}{\rm Mpc}^{-1}$ and $k_s=10^{21}{\rm Mpc}^{-1}$. Additionally we consider the sound speed will lie within the widow, $0.6<c_s<1.5$. Then the numerical value of the co-efficient $\xi_{*}$ at the pivot scale is estimated as:
\bea &&\underline{{\bf For}\;\; 0.6< c_s< 1 :}\quad\quad 1.12<\xi_{*}<1.86,\\
&&\underline{{\bf For}\;\; c_s= 1 :}\quad\quad\quad\quad\quad\quad\quad\;\;\;\xi_{*}\sim 1.86,\\
&&\underline{{\bf For}\;\; 1< c_s< 1.5 :}\quad\quad 1.86<\xi_{*}<5.35.\eea
Hence the factor ${\cal Q}_{\bf EFT}$ can be recast in the following form after performing some simplification:
\bea {\cal Q}_{\bf EFT}&\approx&  -\delta_*\ln^2\left(\frac{k_e}{k_s}\right),\eea
where the factor $\delta$ is defined as:
\bea \delta_*=\frac{\xi_*}{4}\bigg[\Delta^{2}_{\zeta,{\bf Tree}}(p_*)\bigg]_{\bf SR}\times \frac{\left(\Delta\eta(\tau_e)\right)^2}{c^8_s} \left(\frac{k_e}{k_s}\right)^{6}\ll 1.\eea
Then the DRG resummed dimensionless primordial power spectrum for the scalar modes can be simplified as follows:
\bea \boxed{\overline{\overline{\Delta^{2}_{\zeta,{\bf EFT}}(p)}}
\approx\bigg[\Delta^{2}_{\zeta,{\bf Tree}}(p)\bigg]_{\bf SR}\left(\frac{k_e}{k_s}\right)^{-2.3\delta_*}\times\bigg\{1+{\cal O}\bigg(\bigg[\Delta^{2}_{\zeta,{\bf Tree}}(p_*)\bigg]^2_{\bf SR}\bigg)\bigg\}}.\eea
The above mentioned result clearly implies after exponenciation due to having logarithimic divergent contribution the behaviour of the divergences softened after performing DRG resummation. Additionally it is important to note that, apart from logarithimic divergence if we have some other type of divergences then after DRG resummation the softening of the behaviour of the divergences might not be always possible. Fortunately, in the super-horizon scale we don't have any other type of divergences in the present set up which will spoil the scenario and using the DRG method in the end we get a controlled version of the primordial power spectrum for the scalar modes.

\section{Numerical results: Studying further constraints on PBH mass from causality from EFT framework}\label{s7}  

\subsection{Framework I}\label{s7a}
    \begin{figure*}[htb!]
    	\centering
    	\subfigure[For $c_s=0.6(<1)$  with $M^4_2/\dot{H}M^2_p\sim -0.89$ (non-canonical and causal).]{
      	\includegraphics[width=8cm,height=8cm] {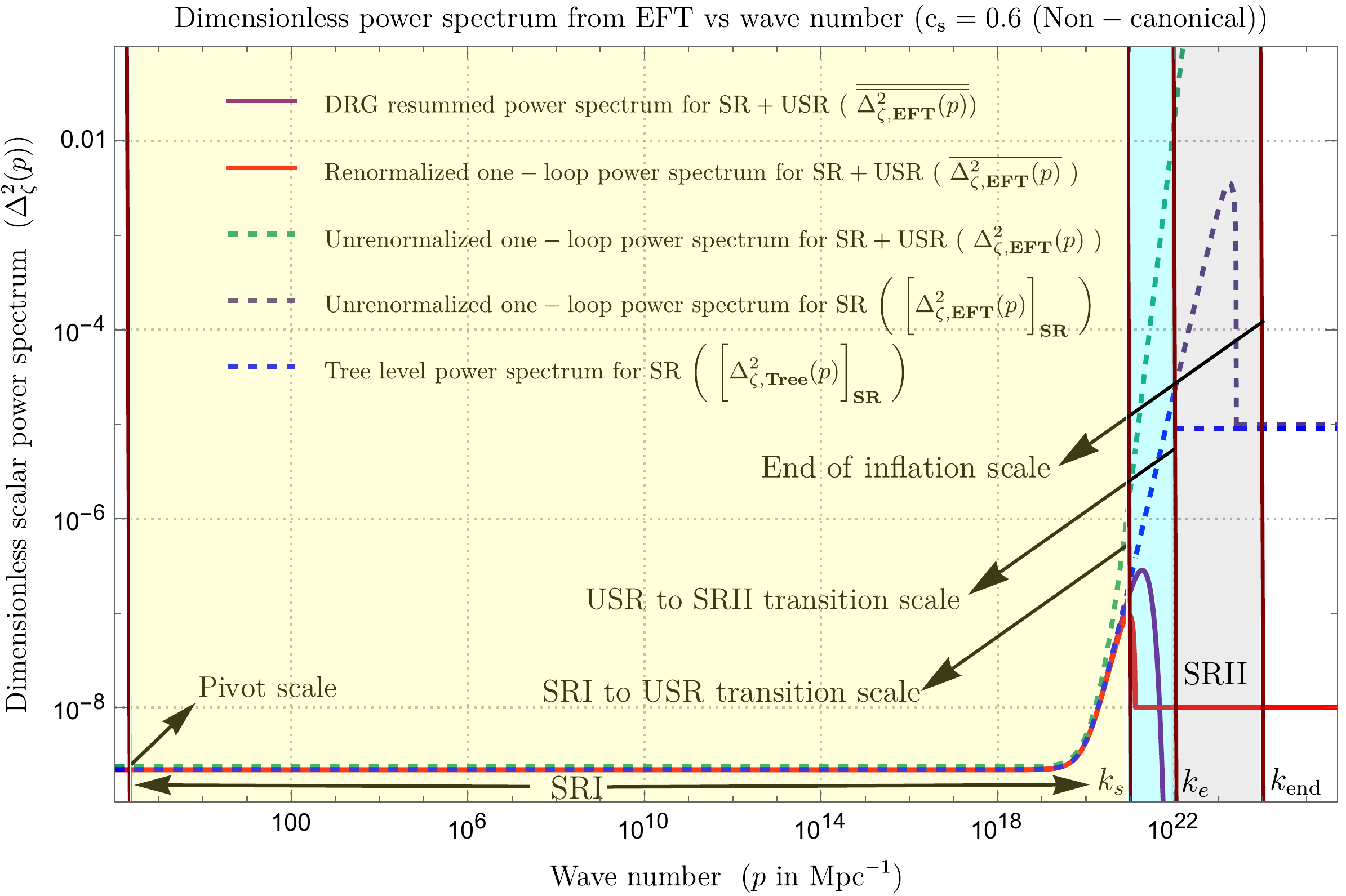}
        \label{G1}
    }
    \subfigure[For $c_s=1$  with $M^4_2/\dot{H}M^2_p\sim 0$ (canonical and causal).]{
       \includegraphics[width=8cm,height=8cm] {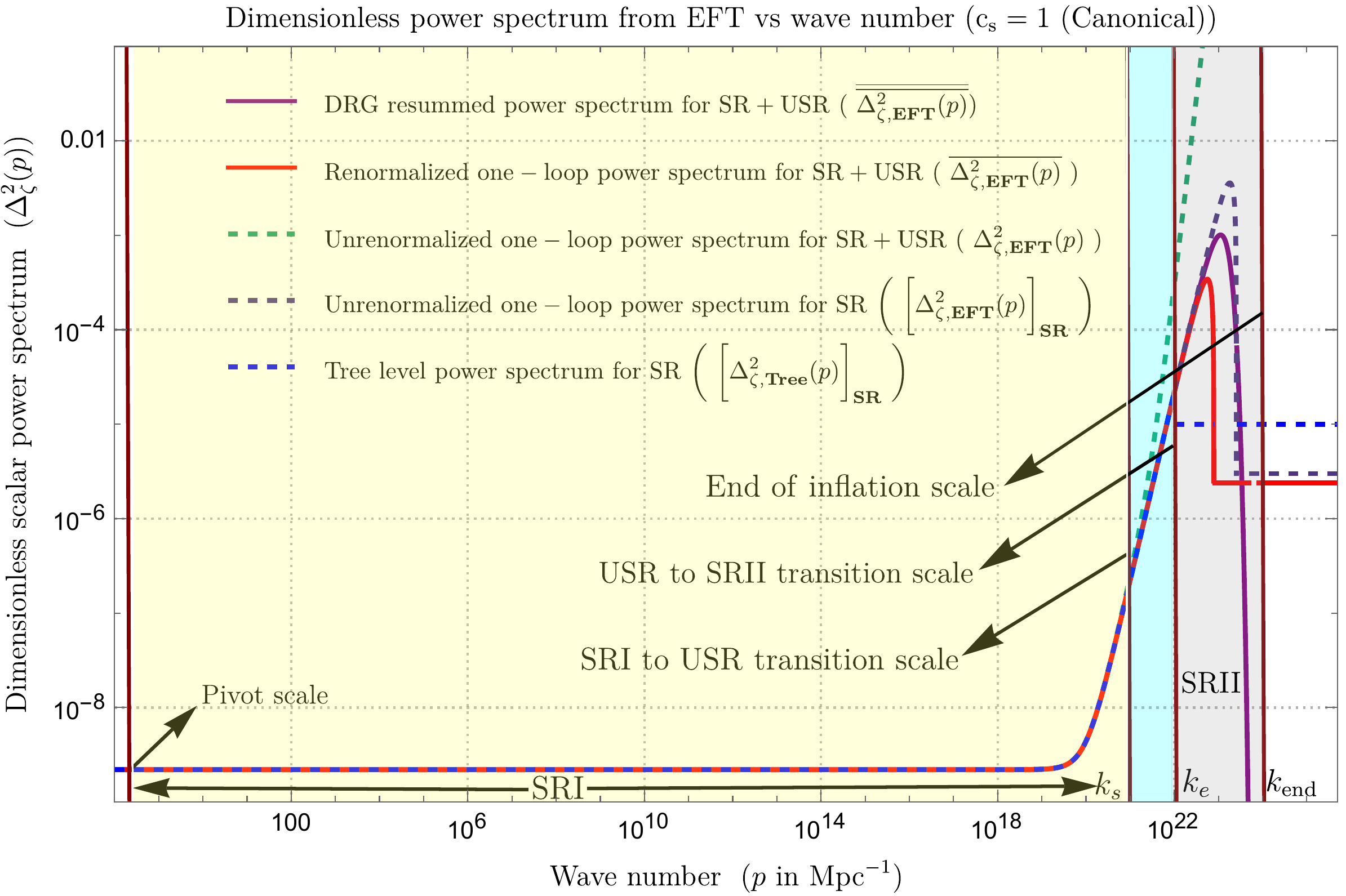}
        \label{G2}
       }
        \subfigure[For $c_s=1.17(>1)$  with $M^4_2/\dot{H}M^2_p\sim 0.13$ (non-canonical and a-causal).]{
       \includegraphics[width=8cm,height=8cm] {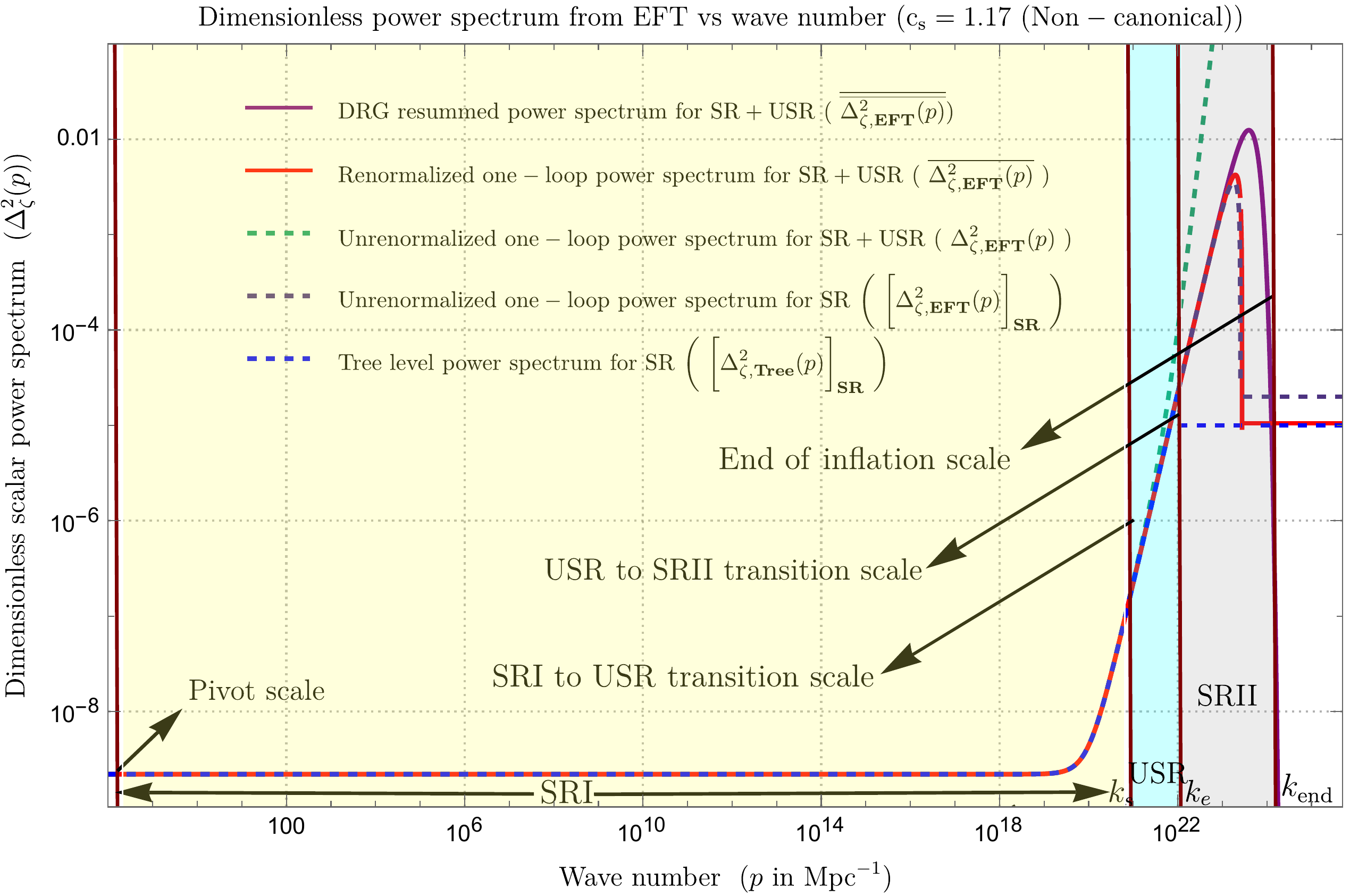}
        \label{G3}
       }
       \subfigure[For $c_s=1.5(>1)$  with $M^4_2/\dot{H}M^2_p\sim 0.28$ (non-canonical and a-causal).]{
       \includegraphics[width=8cm,height=8cm] {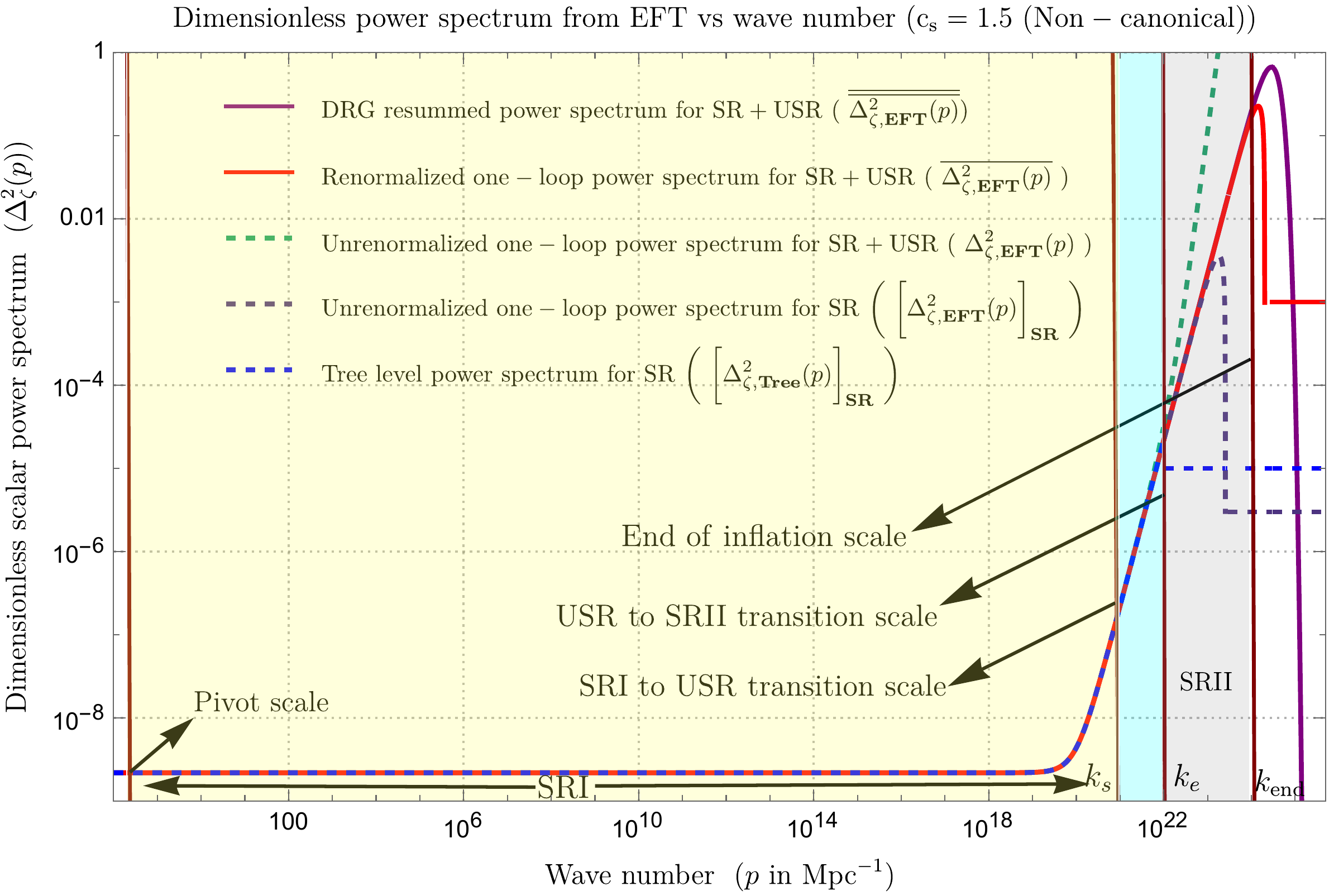}
        \label{G4}
       }
    	\caption[Optional caption for list of figures]{Behaviour of the dimensionless primordial power spectrum for scalar modes with respect to the wave number from the Framework I. In these representative plots, we have studied the behaviour for different values of the sound speed of EFT, $c_s=0.6,1,1.17,1.5$. Here we fix the pivot scale at $p_*=0.02\;{\rm Mpc}^{-1}$, SR to USR sharp transition scale at $k_s=10^{21}\;{\rm Mpc}^{-1}$, end of USR scale at $k_e=10^{22}\;{\rm Mpc}^{-1}$, end of inflation scale at $k_{\rm end}=10^{24}\;{\rm Mpc}^{-1}$, the renomalization parameter $c_{\bf SR}=0$, $\Delta\eta(\tau_e)=1$ and $\Delta\eta(\tau_s)=-6$. In this plot we have found that, $k_{\rm UV}/k_{\rm IR}=k_e/k_s\approx{\cal O}(10)$ and $k_{\rm end}/k_e\approx{\cal O}(100)$. Plots show that $c_s\gtrsim 1$ is the allowed range of effective sound speed, out of which $c_s=1.17$ gives the best outcome to have ${\cal O}(10^{-2})$ amplitude of the corresponding spectrum. At $c_s=1.5$ the amplitude reaches at ${\cal O}(1)$, at which the perturbation theory breaks. } 
    	\label{Spectrum1}
    \end{figure*}
    \begin{figure*}[htb!]
    	\centering
    	\subfigure[For $c_s=0.6(<1)$  with $M^4_2/\dot{H}M^2_p\sim -0.89$ (non-canonical and causal).]{
      	\includegraphics[width=8cm,height=8cm] {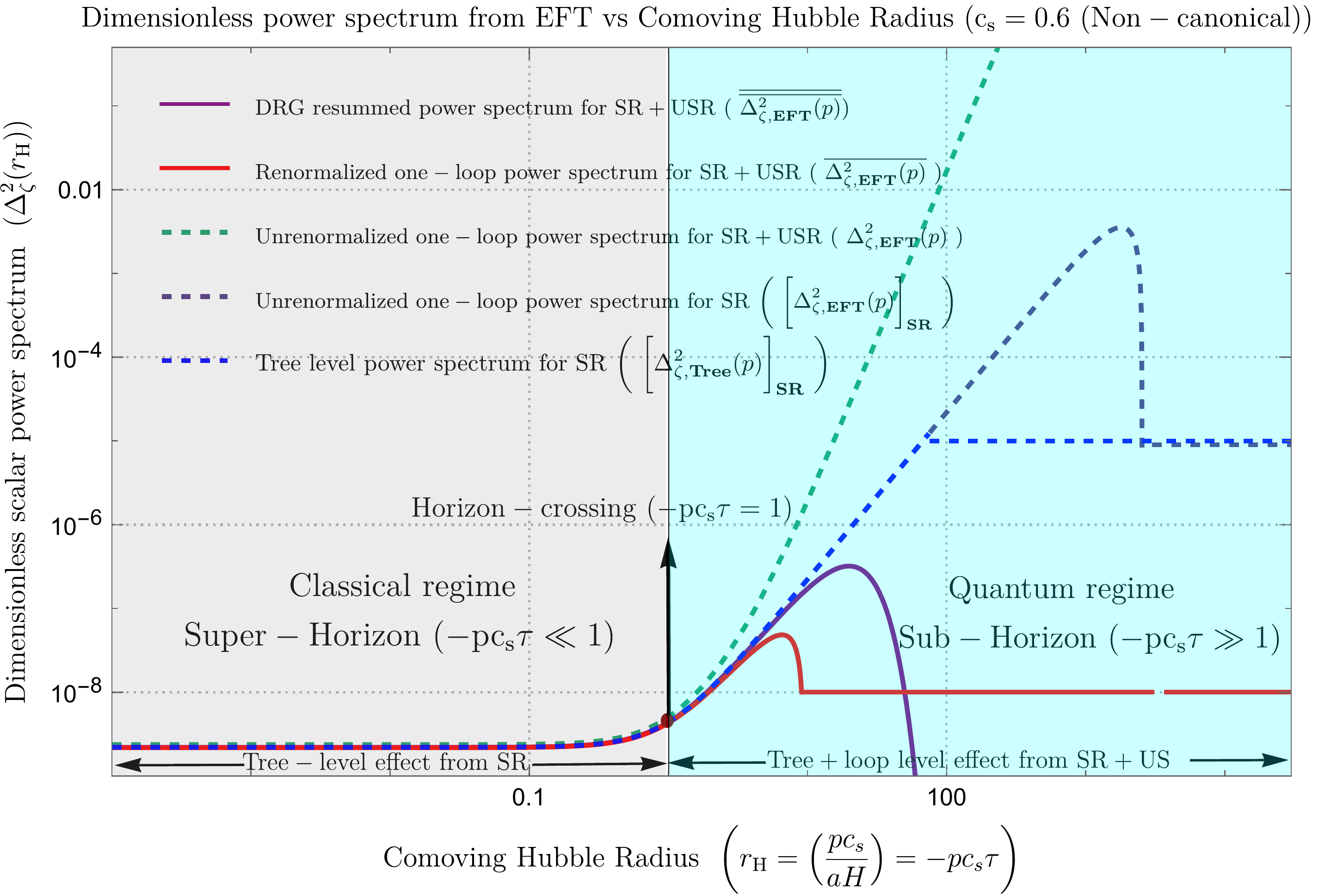}
        \label{H1}
    }
    \subfigure[For $c_s=1$  with $M^4_2/\dot{H}M^2_p\sim 0$ (canonical and causal).]{
       \includegraphics[width=8cm,height=8cm] {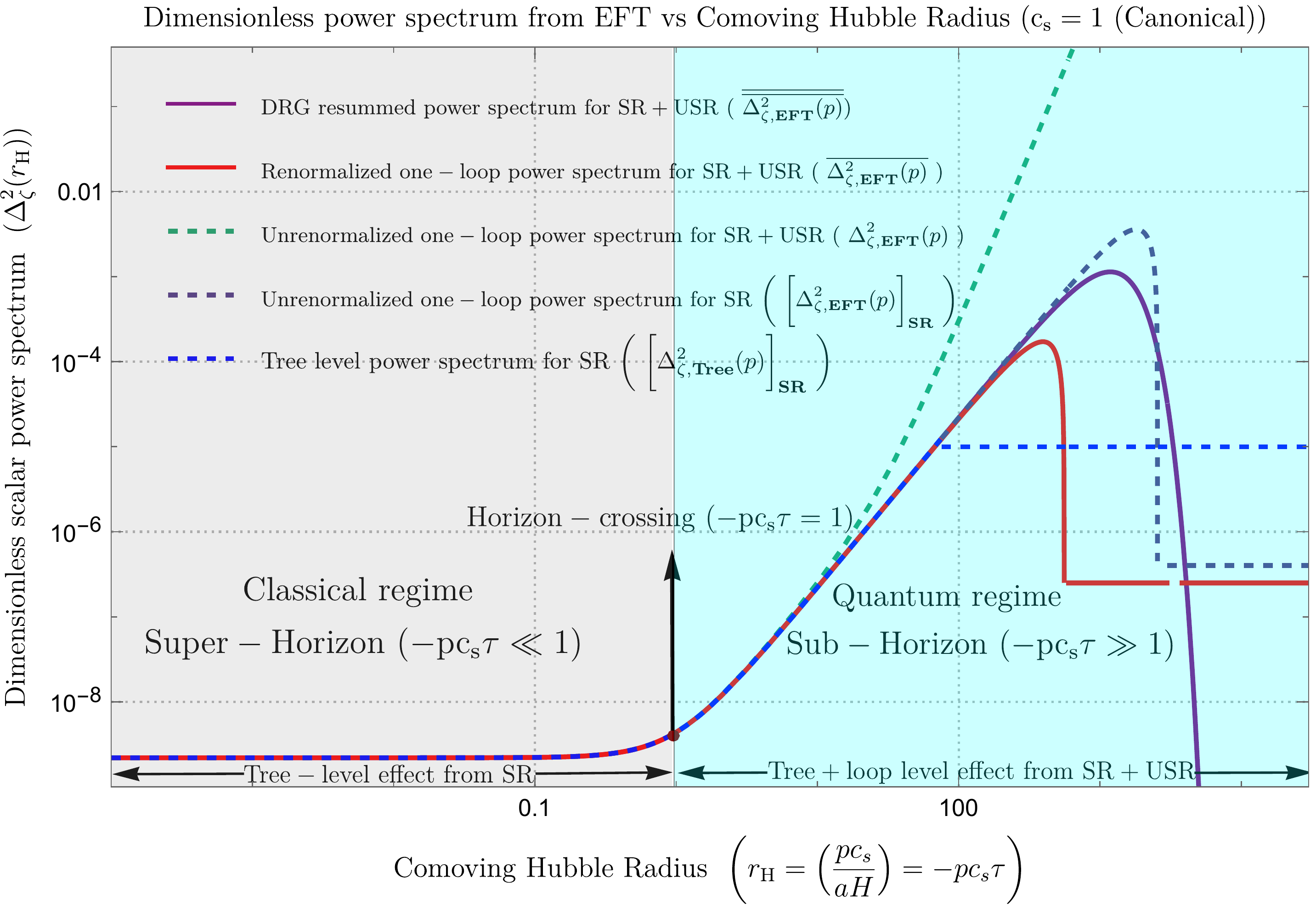}
        \label{H2}
       }
        \subfigure[For $c_s=1.17(>1)$  with $M^4_2/\dot{H}M^2_p\sim 0.13$ (non-canonical and a-causal).]{
       \includegraphics[width=8cm,height=8cm] {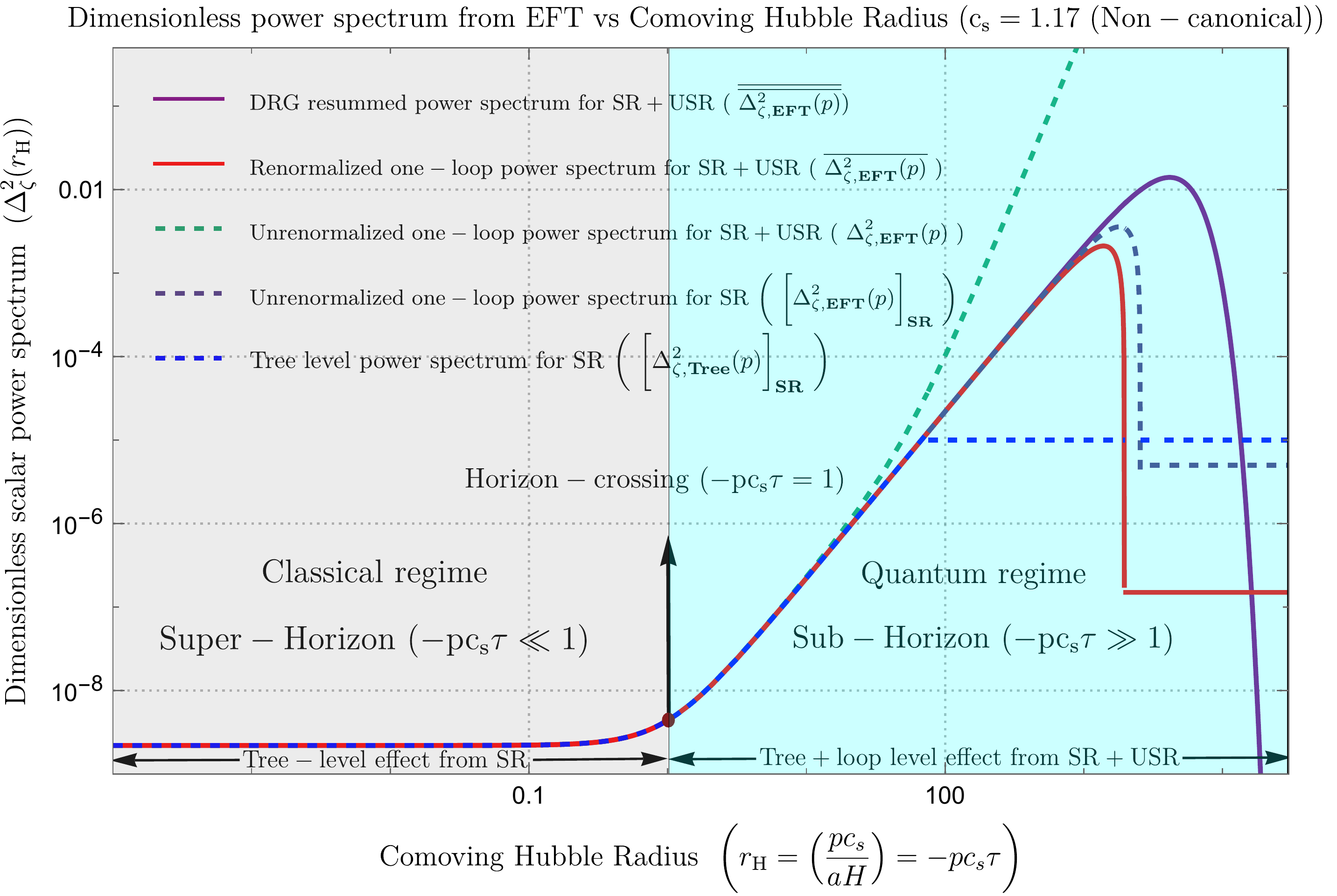}
        \label{H3}
       }
        \subfigure[For $c_s=1.5(>1)$  with $M^4_2/\dot{H}M^2_p\sim 0.28$ (non-canonical and a-causal).]{
       \includegraphics[width=8cm,height=8cm] {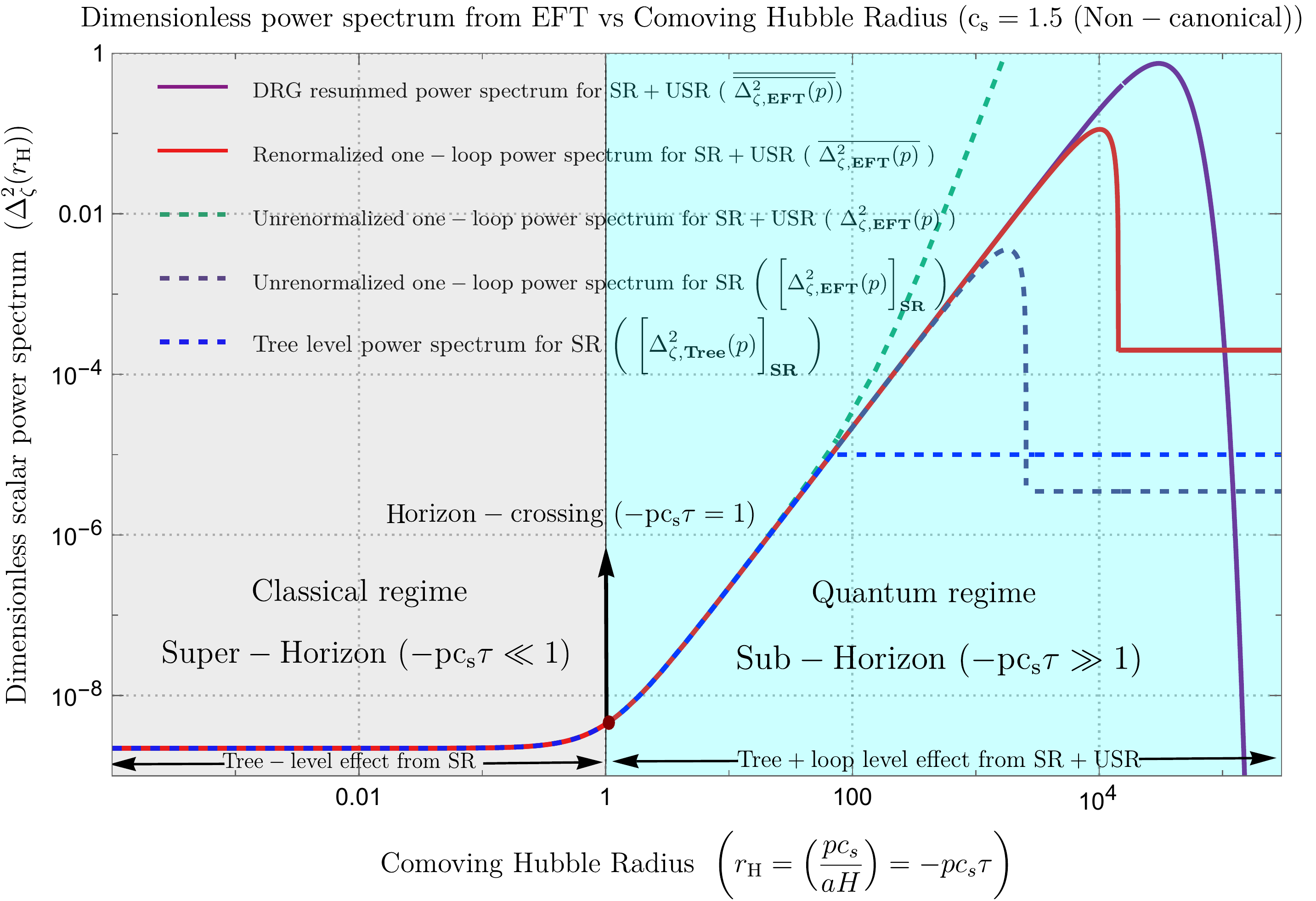}
        \label{H4}
       }
    	\caption[Optional caption for list of figures]{Behaviour of the dimensionless power spectrum for scalar modes with respect to the Comoving Hubble Radius from the Framework I. In these representative plots, we have studied the behaviour for different values of the sound speed of EFT, $c_s=0.6,1,1.17,1.5$. Here we fix the pivot scale at $p_*=0.02\;{\rm Mpc}^{-1}$, SR to USR sharp transition scale at $k_s=10^{21}\;{\rm Mpc}^{-1}$, end of USR scale at $k_e=10^{22}\;{\rm Mpc}^{-1}$, end of inflation scale at $k_{\rm end}=10^{24}\;{\rm Mpc}^{-1}$, the renomalization parameter $c_{\bf SR}=0$, $\Delta\eta(\tau_e)=1$ and $\Delta\eta(\tau_s)=-6$. In this plot we have found that, $k_{\rm UV}/k_{\rm IR}=k_e/k_s\approx{\cal O}(10)$ and $k_{\rm end}/k_e\approx{\cal O}(100)$. Plots show that $c_s\gtrsim 1$ is the allowed range of effective sound speed, out of which $c_s=1.17$ gives the best outcome to have ${\cal O}(10^{-2})$ amplitude of the corresponding spectrum. At $c_s=1.5$ the amplitude reaches at ${\cal O}(1)$, at which the perturbation theory breaks. } 
    	\label{Spectrum2}
    \end{figure*}

    \begin{figure*}[htb!]
    	\centering
    \subfigure[For $c_s=0.6(<1)$  with $M^4_2/\dot{H}M^2_p\sim -0.89$ (non-canonical and causal).]{
      	\includegraphics[width=8cm,height=8cm] {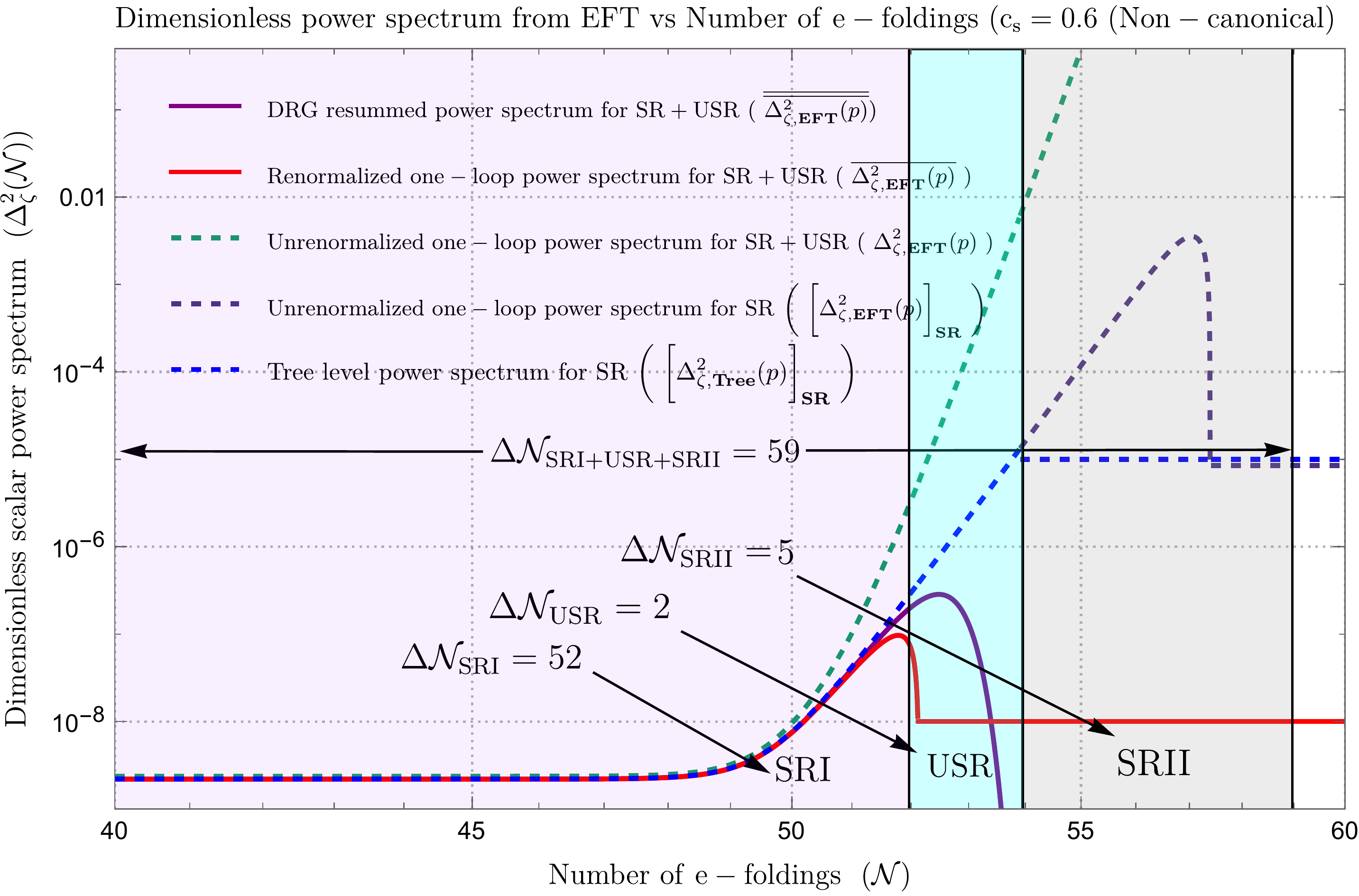}
        \label{I1}
    }
    \subfigure[For $c_s=1$  with $M^4_2/\dot{H}M^2_p\sim 0$ (canonical and causal).]{
       \includegraphics[width=8cm,height=8cm] {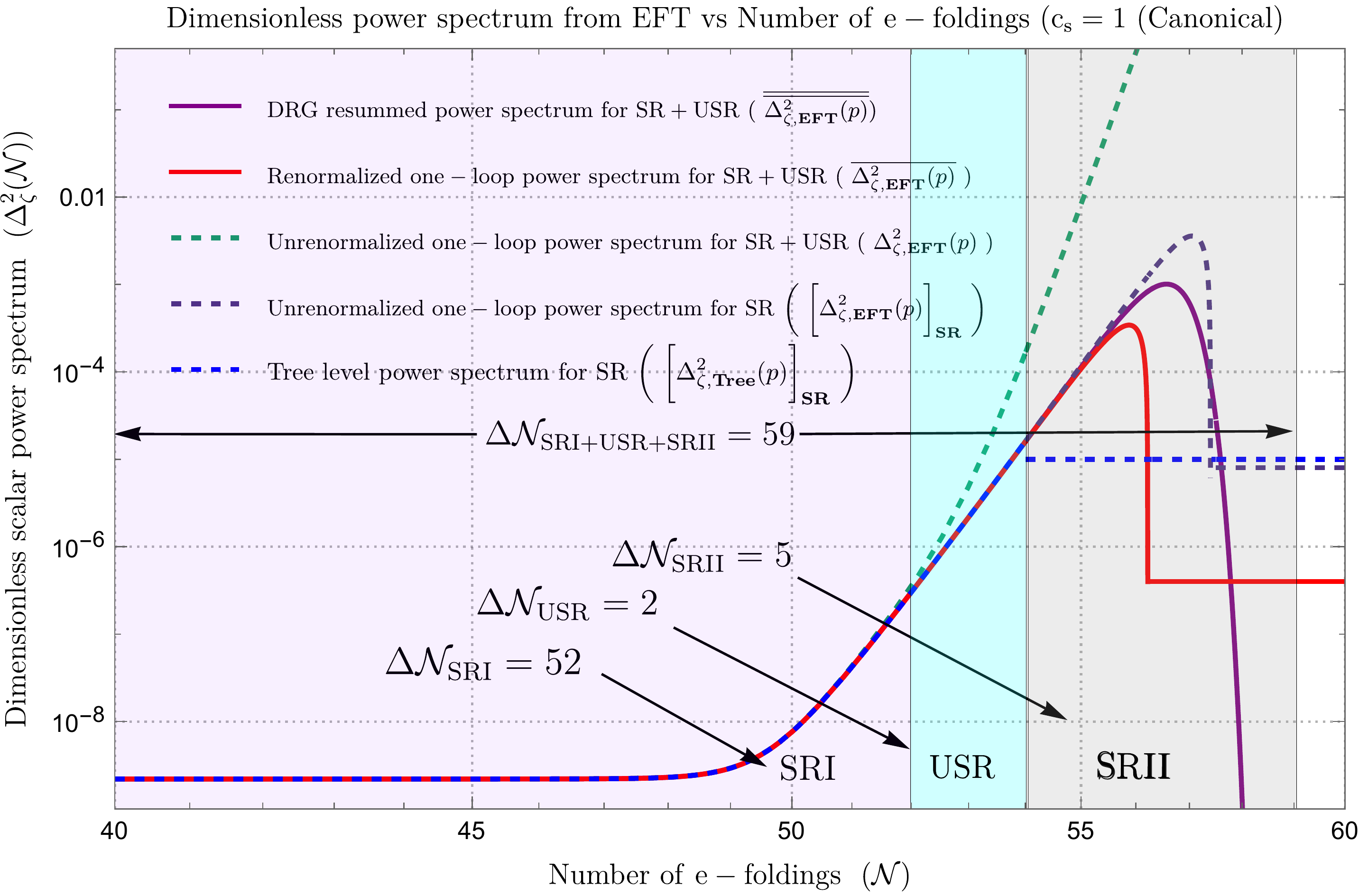}
        \label{I2}
       }
        \subfigure[For $c_s=1.17(>1)$  with $M^4_2/\dot{H}M^2_p\sim 0.13$ (non-canonical and a-causal).]{
       \includegraphics[width=8cm,height=8cm] {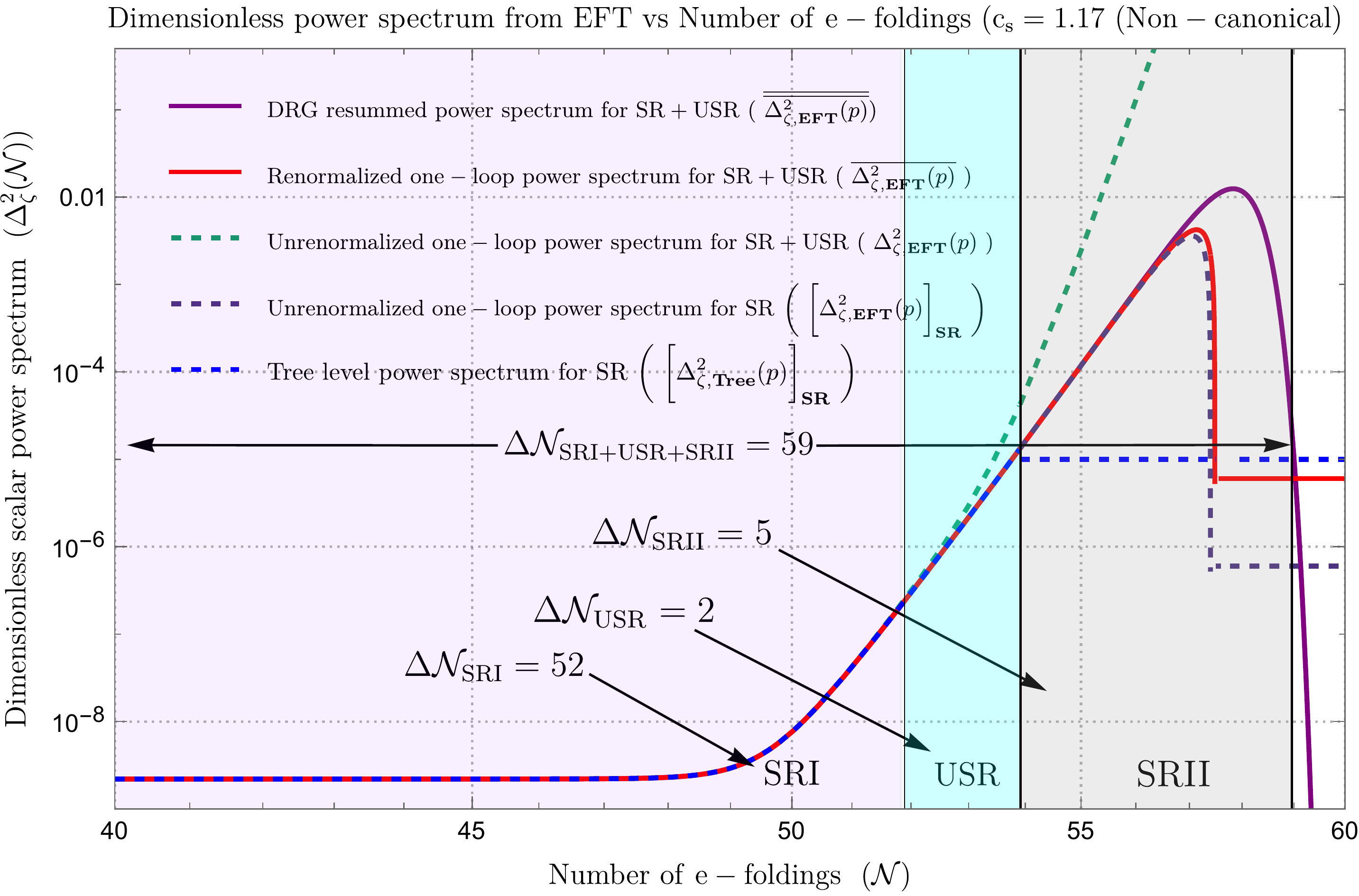}
        \label{I3}
       }
        \subfigure[For $c_s=1.5(>1)$  with $M^4_2/\dot{H}M^2_p\sim 0.28$ (non-canonical and a-causal).]{
       \includegraphics[width=8cm,height=8cm] {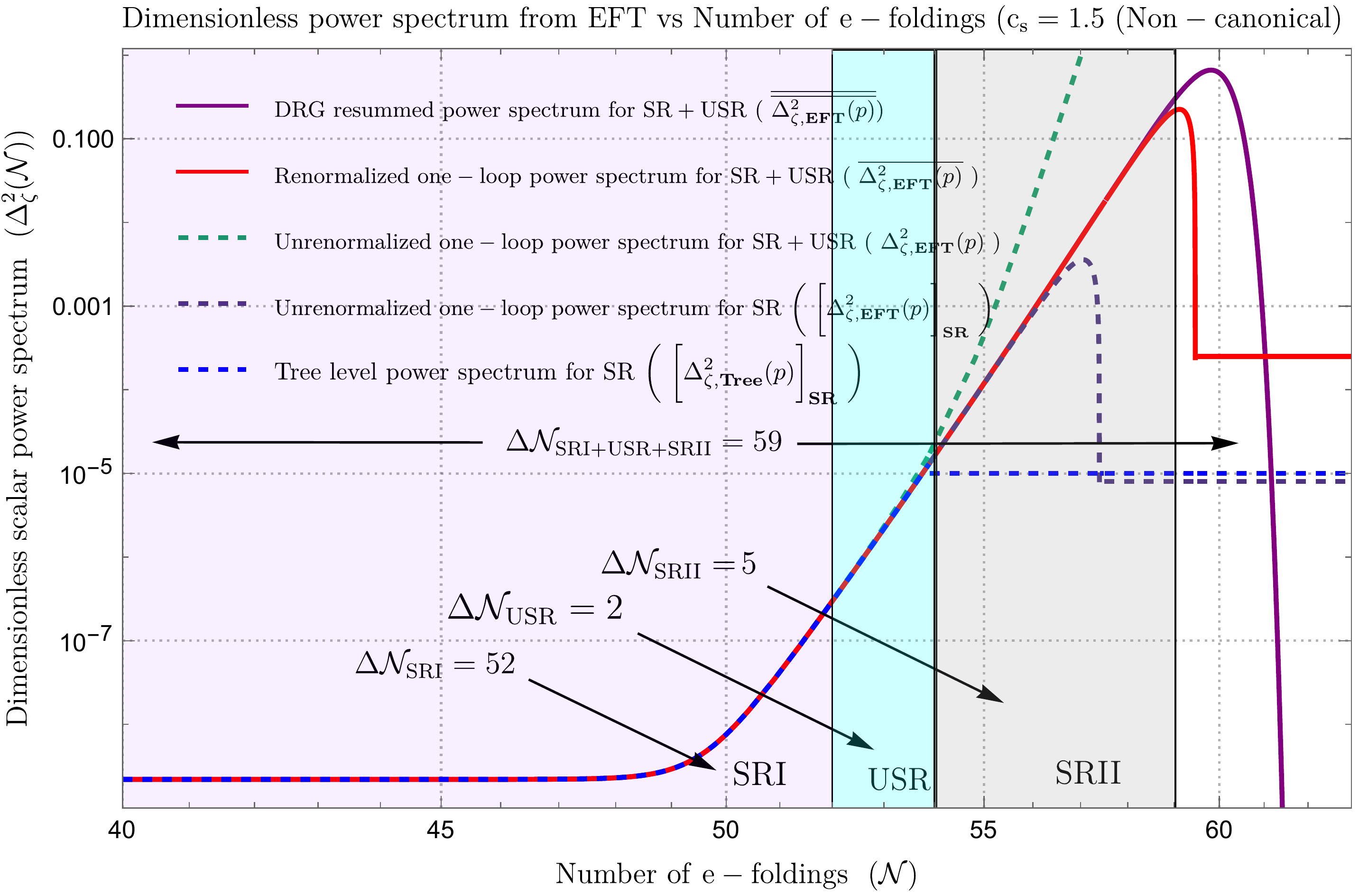}
        \label{I4}
       }
    	\caption[Optional caption for list of figures]{Behaviour of the dimensionless power spectrum for scalar modes with respect to the number of e-foldings from the Framework I. In these representative plots, we have studied the behaviour for different values of the sound speed of EFT, $c_s=0.6,1,1.17,1.5$. Here we fix the pivot scale at $p_*=0.02\;{\rm Mpc}^{-1}$, SR to USR sharp transition scale at $k_s=10^{21}\;{\rm Mpc}^{-1}$, end of USR scale at $k_e=10^{22}\;{\rm Mpc}^{-1}$, end of inflation scale at $k_{\rm end}=10^{24}\;{\rm Mpc}^{-1}$, the renomalization parameter $c_{\bf SR}=0$, $\Delta\eta(\tau_e)=1$ and $\Delta\eta(\tau_s)=-6$. In this plot we have found that, $k_{\rm UV}/k_{\rm IR}=k_e/k_s\approx{\cal O}(10)$ and $k_{\rm end}/k_e\approx{\cal O}(100)$. Plots show that $c_s\gtrsim 1$ is the allowed range of effective sound speed, out of which $c_s=1.17$ gives the best outcome to have ${\cal O}(10^{-2})$ amplitude of the corresponding spectrum. At $c_s=1.5$ the amplitude reaches at ${\cal O}(1)$, at which the perturbation theory breaks.  } 
    	\label{Spectrum3}
    \end{figure*}
    
In Figure (\ref{G1}), figure(\ref{G2}), figure(\ref{G3}) and Figure (\ref{G4}), we have depicted the behaviour of the dimensionless power spectrum for scalar modes with respect to the wave number for different values of the effective sound speed $c_s=0.6$, $c_s=1$, $c_s=1.17$ and $c_S=1.5$ for the Framework I. In each of the plots we have shown the contribution coming from tree level, one-loop corrected unrenormalized, renormalized one-loop corrected part and DRG resummed one-loop corrected part.
We found from each of the plots that the DRG resummed spectrum gives the best interpretation for the present scenario where the spectrum falls very fast at the end of the USR phase. Most importantly, the resummed spectrum provides the distinguishable features which help to differentiate among all of these mentioned contributions. For the computational purpose we fix the sharp transition scale at $k_s=10^{21}\;{\rm Mpc}^{-1}$ (where we fix the IR cut-off) and the end of USR at $k_e=10^{22}\;{\rm Mpc}^{-1}$ (where we fix the UV cut-off), end of inflation scale at $k_{\rm end}=10^{24}\;{\rm Mpc}^{-1}$, the renomalization parameter $c_{\bf SR}=0$, $\Delta\eta(\tau_e)=1$ and $\Delta\eta(\tau_s)=-6$. We have maintained a restriction, $k_{\rm UV}/k_{\rm IR}=k_e/k_s\approx{\cal O}(10)$\footnote{It is important to note that this fact was first pointed out in ref. \cite{Choudhury:2023vuj} for canonical single field models of inflation and then further used in ref \cite{Choudhury:2023jlt} for Effective Field Theory of Single Field Inflation.} and $k_{\rm end}/k_e\approx{\cal O}(100)$\footnote{This is the completely new finding in this paper due to having an additional SRII phase for the Framework I.}, which is the key finding of this calculation. All of these plots show that $c_s\gtrsim 1$ is the allowed range of effective sound speed, out of which $c_s=1.17$ gives the best outcome to have highest amplitude, ${\cal O}(10^{-2})$ of the corresponding spectrum necessarily needed to form PBHs from the present set up. At $c_s=1.5$ the amplitude reaches at ${\cal O}(1)$, at which the perturbation theory breaks. For this reason we can't do our analysis beyond $c_s=1.5$. More precisely the allowed window of the effective sound speed is found within the range, $1<c_S<1.17$.

In figure(\ref{H1}), figure(\ref{H2}), figure(\ref{H3}) and figure(\ref{H4}), we have depicted the behaviour of the dimensionless power spectrum for scalar modes with respect to the Comoving Hubble Radius for different values of the effective sound speed $c_s=0.6$, $c_s=1$, $c_s=1.17$ and $c_S=1.5$ for the Framework I. In each of the plots we have shown the size of the super-horizon and sub-horizon region and horizon crossing point where we found that the classical, quantum effects and  semi-classical effects are prominent. All of these plots show that $1<c_S<1.17$ is the allowed range of effective sound speed, out of which $c_s=1.17$ gives the best outcome to have the highest amplitude, ${\cal O}(10^{-2})$ of the corresponding spectrum necessarily needed to form PBHs. At $c_s=1.5$ the amplitude reaches at ${\cal O}(1)$, at which the perturbation theory breaks.

Finally, in Figure (\ref{I1}), figure(\ref{I2}), figure(\ref{I3}), and Figure (\ref{I4}), we have depicted the behaviour of the dimensionless power spectrum for scalar modes with respect to the number of e-foldings for the Framework I. From this plot, we have found that for the Framework I:
\bea \Delta {\cal N}_{\rm USR}=\ln(k_e/k_s)\approx\ln(10)\approx 2,\eea 
which implies only approximately $2$ e-folds are allowed in the USR period for the PBH formation in the present setup. 

In this framework, the allowed e-folds for the SRI and SRII periods are given by:
\bea &&\Delta {\cal N}_{\rm SRI}=\ln(k_s/p_*)\approx 52,\\
&&\Delta {\cal N}_{\rm SRII}=\ln(k_{\rm end}/k_e)\approx 2\ln(10)\approx 5.\eea 
The restriction for the SRII phase comes to perform DRG resummation within the perturbative regime. As a consequence, the total number of e-foldings allowed by Framework I is given by the following expression:
\bea \Delta {\cal N}_{\rm Total}=\Delta {\cal N}_{\rm SRI}+\Delta {\cal N}_{\rm USR}+\Delta {\cal N}_{\rm SRII}\sim 52+2+5=59,\eea 
provided the sharp transition scale, the end of USR period and inflation are fixed at, $k_s=10^{21}\;{\rm Mpc}^{-1}$, $k_e=10^{22}\;{\rm Mpc}^{-1}$ and $k_{\rm end}=10^{24}\;{\rm Mpc}^{-1}$. We will show in the next section that this possibility will lead to a small mass PBH formation having a sufficient number of e-folds for inflation in the present context.

Now let us consider a situation where we shift the sharp transition scale, the end of the USR period, and inflation at the scale, $k_s=10^{6}\;{\rm Mpc}^{-1}$, $k_e=10^{7}\;{\rm Mpc}^{-1}$ and $k_{\rm end}=10^{9}\;{\rm Mpc}^{-1}$. In this case, the allowed number of e-folds for SRI can be computed as:
\bea &&\Delta {\cal N}_{\rm SRI}=\ln(k_s/p_*)\approx 18.\eea
For SRII and USR the results will be unchanged. This implies in that mentioned scenario, the total number of e-foldings allowed by Framework I is given by the following expression:
\bea \Delta {\cal N}_{\rm Total}=\Delta {\cal N}_{\rm SRI}+\Delta {\cal N}_{\rm USR}+\Delta {\cal N}_{\rm SRII}\sim 18+2+5=25.\eea
We will show in the next section that this possibility will lead to a large mass PBH formation having an insufficient number of e-folds for inflation. For this reason, this possibility can be immediately discarded.

\subsection{Framework II}\label{s7b}
    \begin{figure*}[htb!]
    	\centering
    	\subfigure[For $c_s=0.6(<1)$  with $M^4_2/\dot{H}M^2_p\sim -0.89$ (non-canonical and causal).]{
      	\includegraphics[width=8cm,height=8cm] {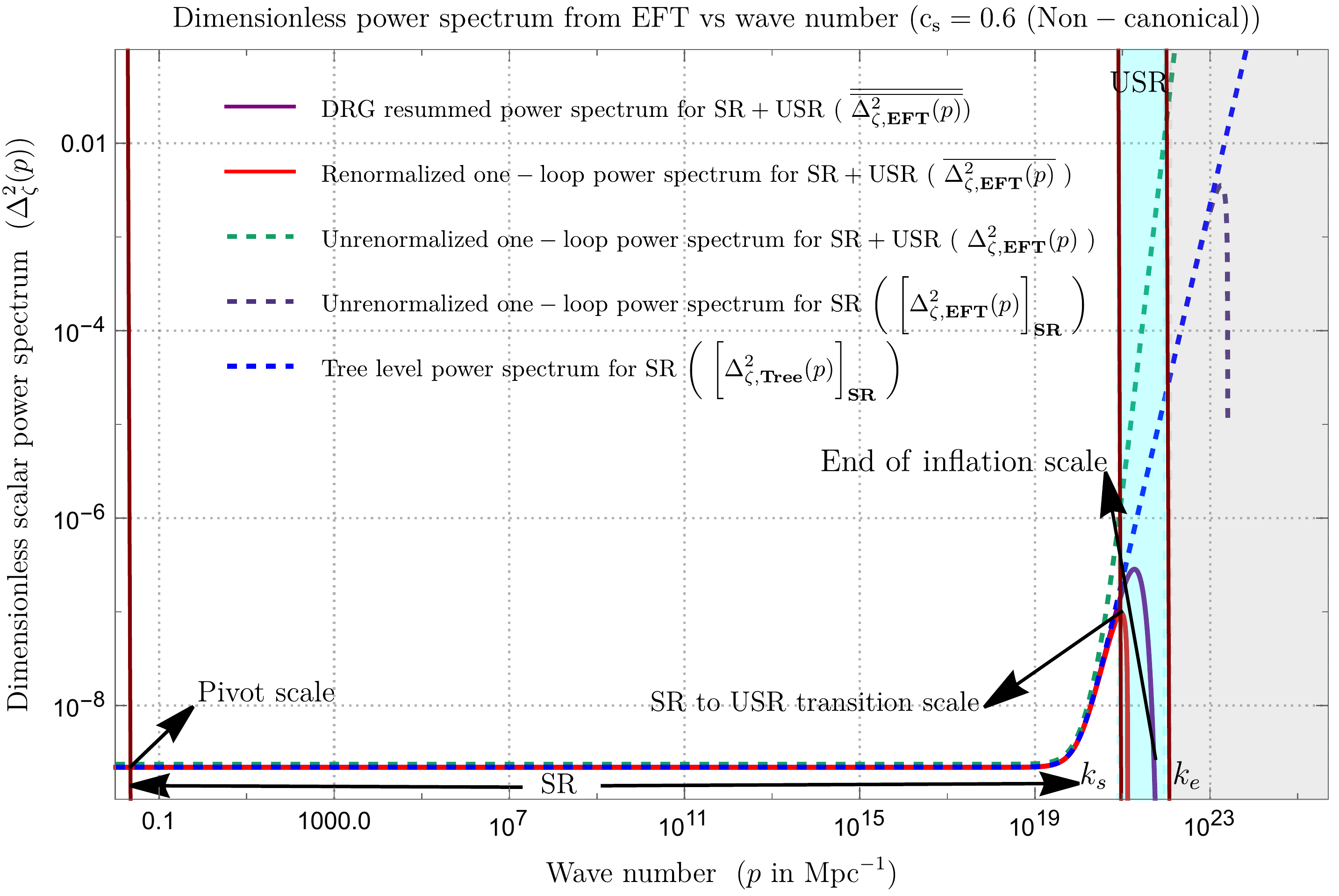}
        \label{Gv1}
    }
    \subfigure[For $c_s=1$  with $M^4_2/\dot{H}M^2_p\sim 0$ (canonical and causal).]{
       \includegraphics[width=8cm,height=8cm] {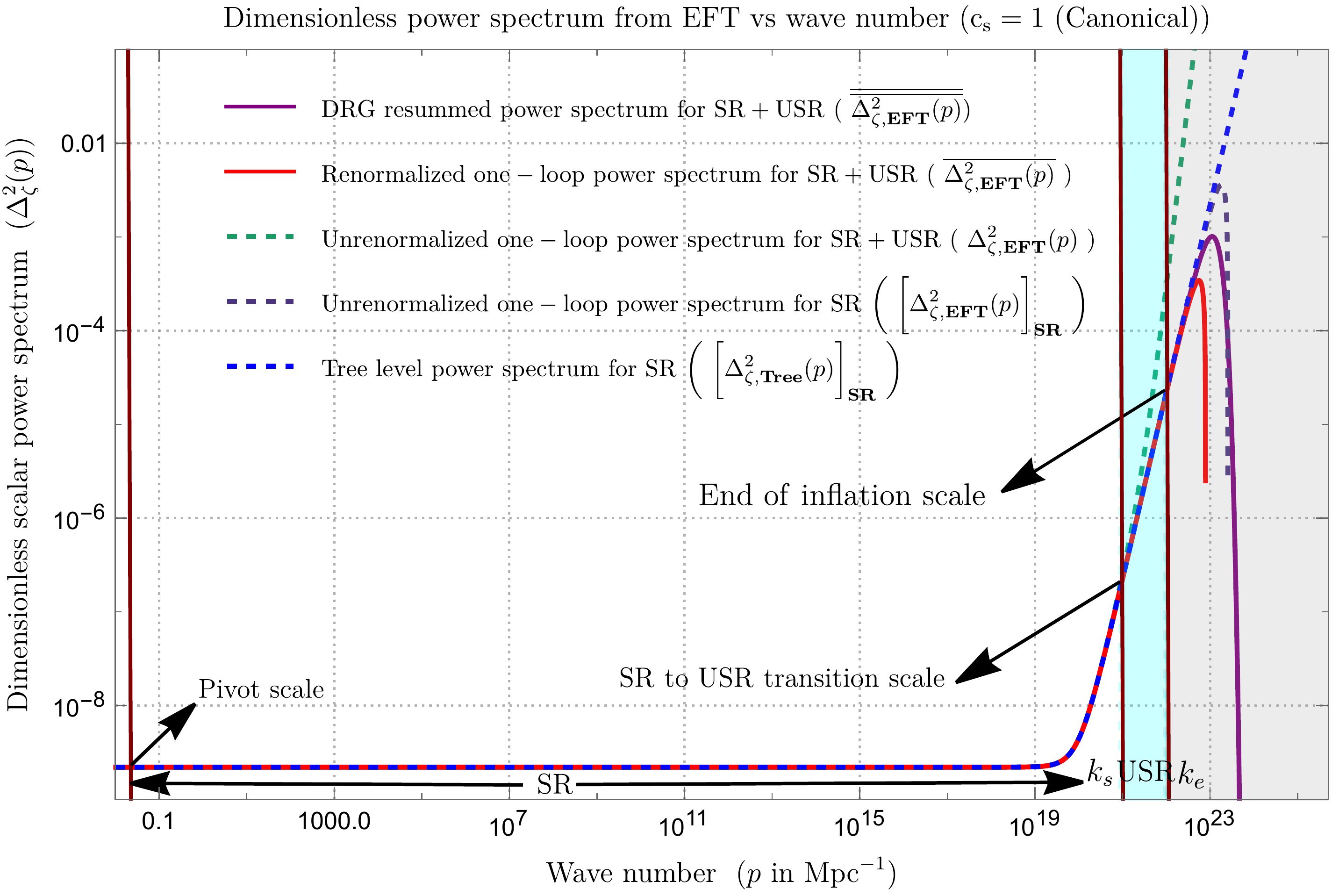}
        \label{Gv2}
       }
        \subfigure[For $c_s=1.17(>1)$  with $M^4_2/\dot{H}M^2_p\sim 0.13$ (non-canonical and a-causal).]{
       \includegraphics[width=8cm,height=8cm] {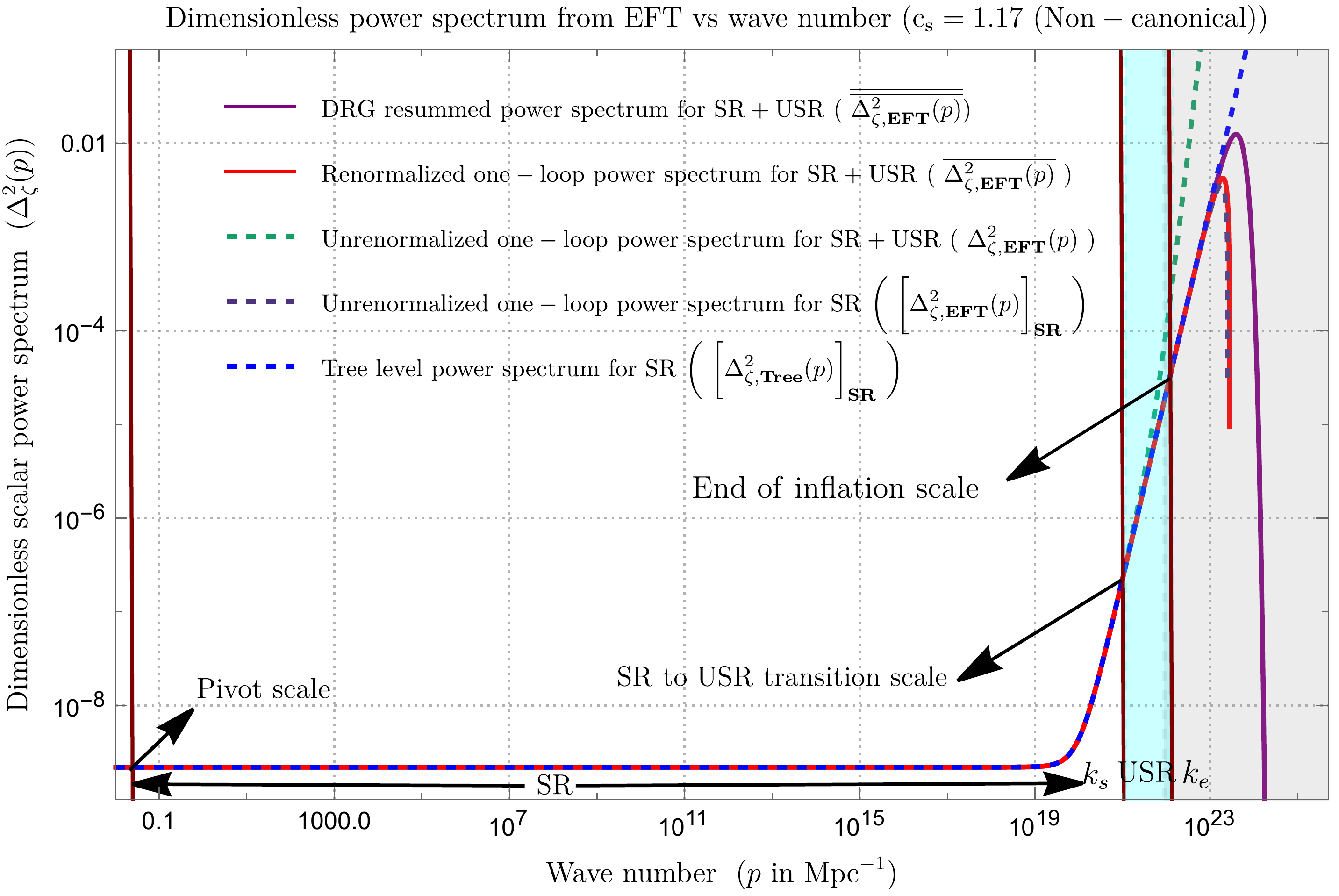}
        \label{Gv3}
       }
       \subfigure[For $c_s=1.5(>1)$  with $M^4_2/\dot{H}M^2_p\sim 0.28$ (non-canonical and a-causal).]{
       \includegraphics[width=8cm,height=8cm] {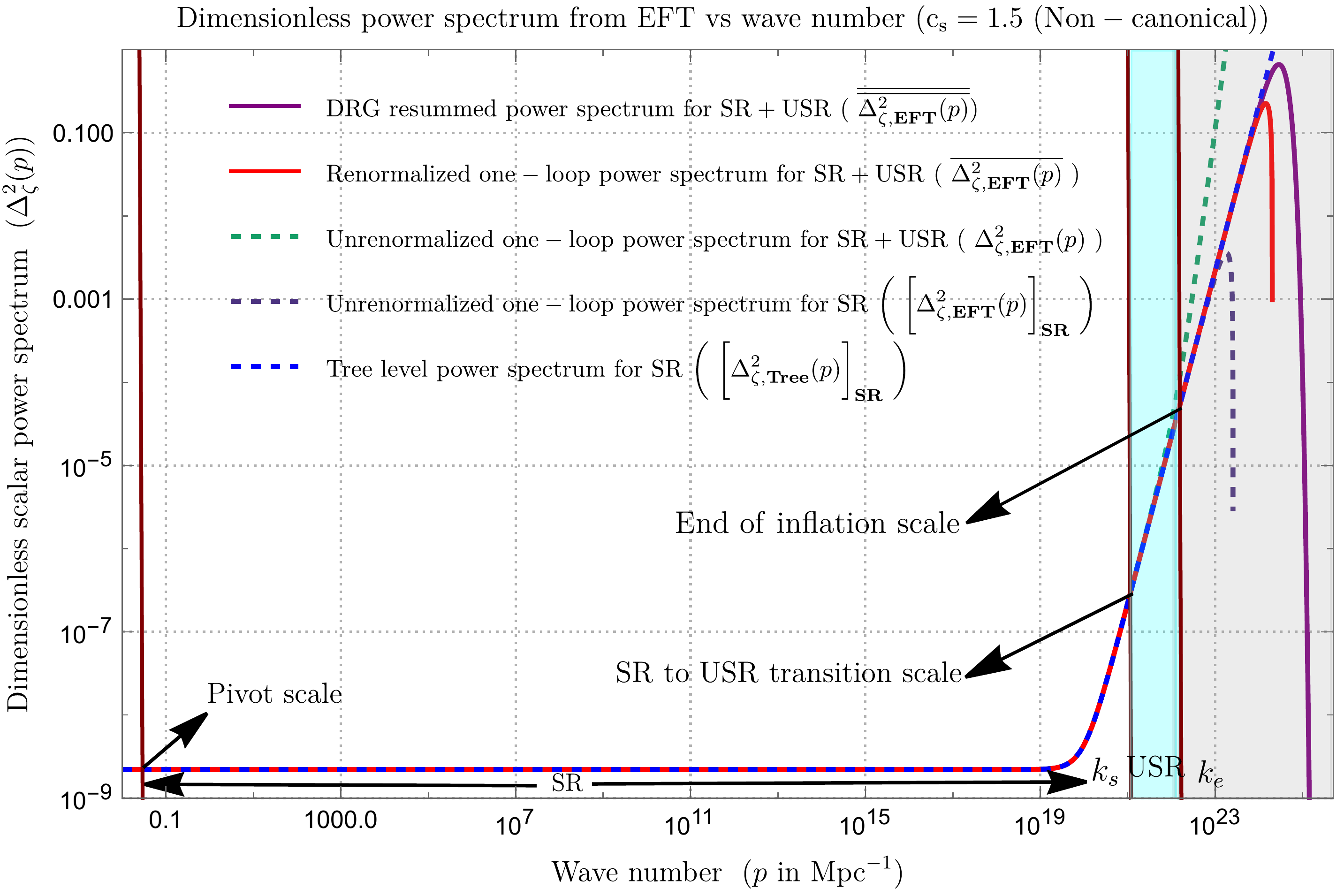}
        \label{Gv4}
       }
    	\caption[Optional caption for list of figures]{Behaviour of the dimensionless primordial power spectrum for scalar modes with respect to the wave number from Framework II. In these representative plots, we have studied the behaviour for different values of the sound speed of EFT, $c_s=0.6,1,1.17,1.5$. Here we fix the pivot scale at $p_*=0.02\;{\rm Mpc}^{-1}$, SR to USR sharp transition scale at $k_s=10^{21}\;{\rm Mpc}^{-1}$ and the end of inflation scale at $k_e=10^{22}\;{\rm Mpc}^{-1}$, the renomalization parameter $c_{\bf SR}=0$, $\Delta\eta(\tau_e)=1$ and $\Delta\eta(\tau_s)=-6$. In this plot we have found that, $k_{\rm UV}/k_{\rm IR}=k_e/k_s\approx{\cal O}(10)$. Plots show that $c_s\gtrsim 1$ is the allowed range of effective sound speed, out of which $c_s=1.17$ gives the best outcome to have ${\cal O}(10^{-2})$ amplitude of the corresponding spectrum. At $c_s=1.5$ the amplitude reaches at ${\cal O}(1)$, at which the perturbation theory breaks. } 
    	\label{Spectrumv1}
    \end{figure*}
    \begin{figure*}[htb!]
    	\centering
    \subfigure[For $c_s=0.6(<1)$  with $M^4_2/\dot{H}M^2_p\sim -0.89$ (non-canonical and causal).]{
      	\includegraphics[width=8cm,height=8cm] {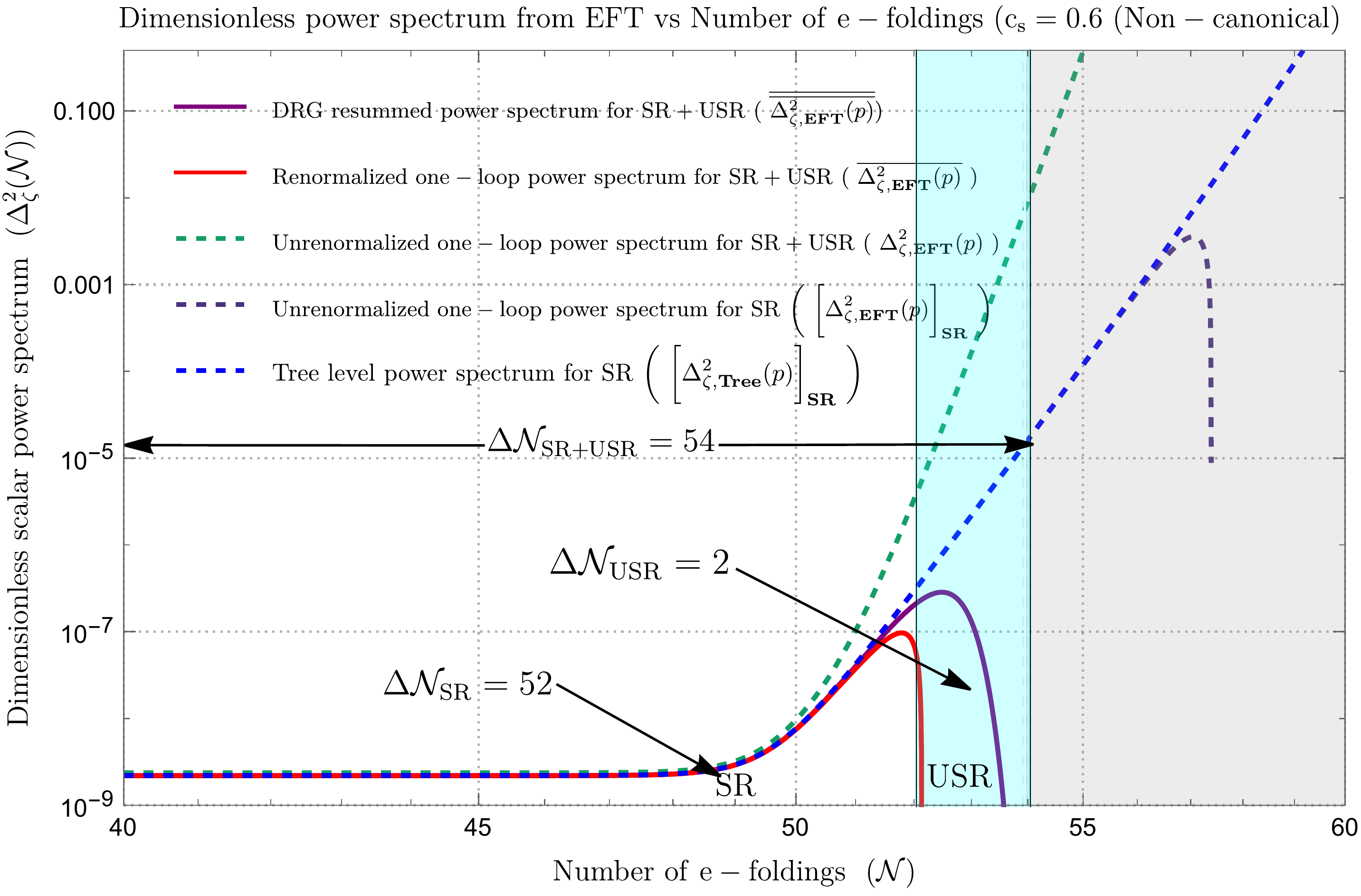}
        \label{Iv1}
    }
    \subfigure[For $c_s=1$  with $M^4_2/\dot{H}M^2_p\sim 0$ (canonical and causal).]{
       \includegraphics[width=8cm,height=8cm] {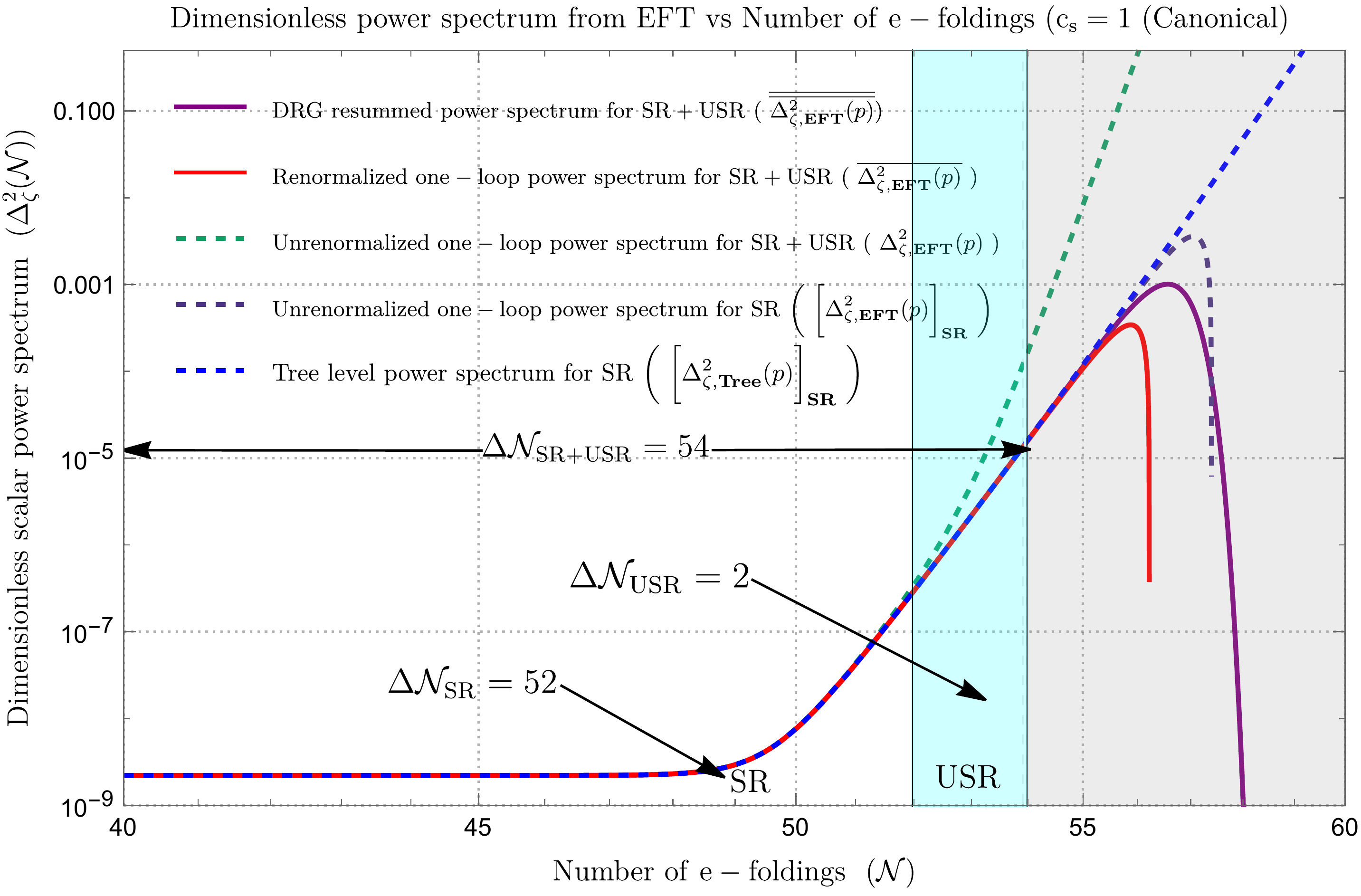}
        \label{Iv2}
       }
        \subfigure[For $c_s=1.17(>1)$  with $M^4_2/\dot{H}M^2_p\sim 0.13$ (non-canonical and a-causal).]{
       \includegraphics[width=8cm,height=8cm] {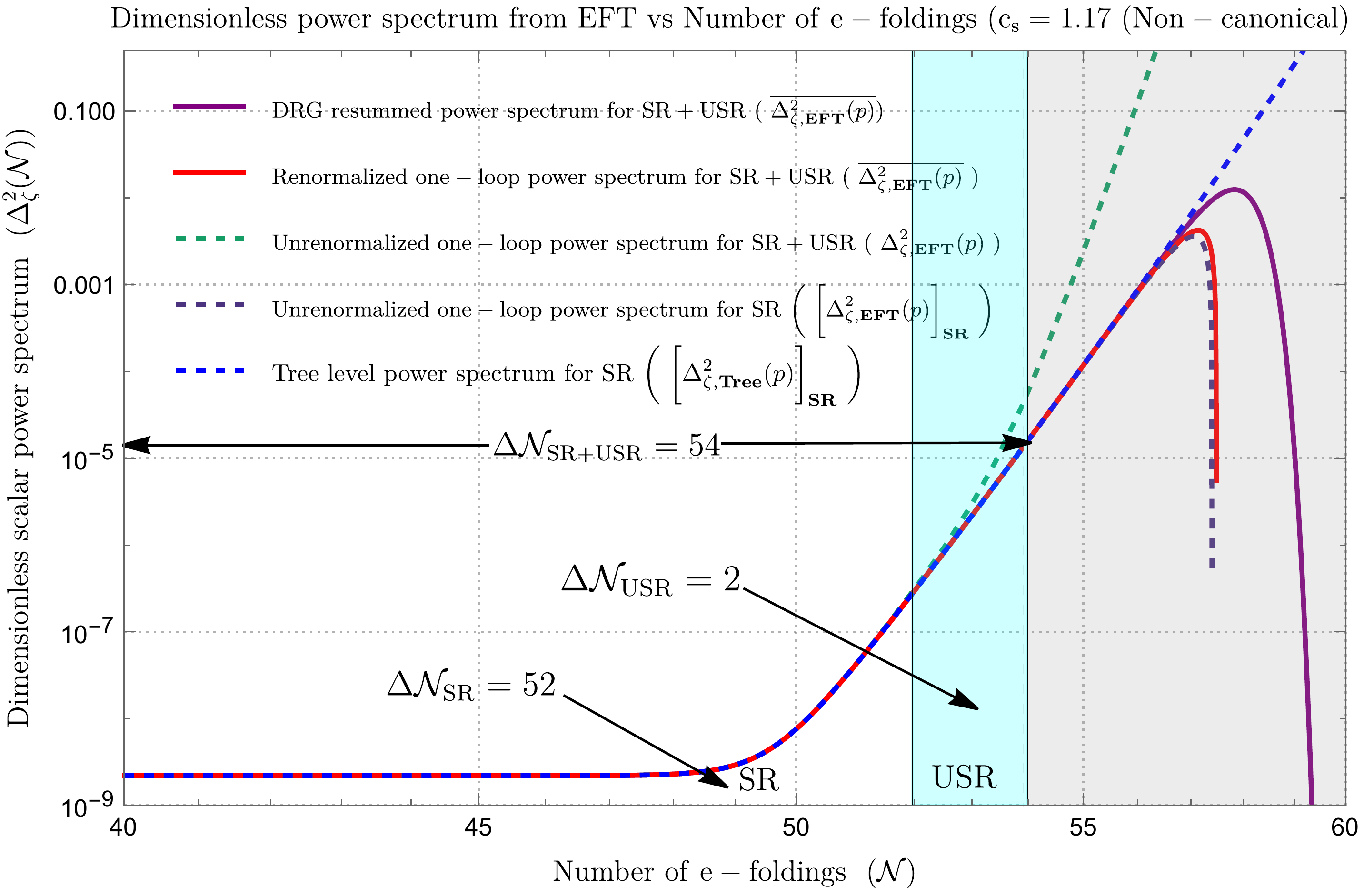}
        \label{Iv3}
       }
        \subfigure[For $c_s=1.5(>1)$  with $M^4_2/\dot{H}M^2_p\sim 0.28$ (non-canonical and a-causal).]{
       \includegraphics[width=8cm,height=8cm] {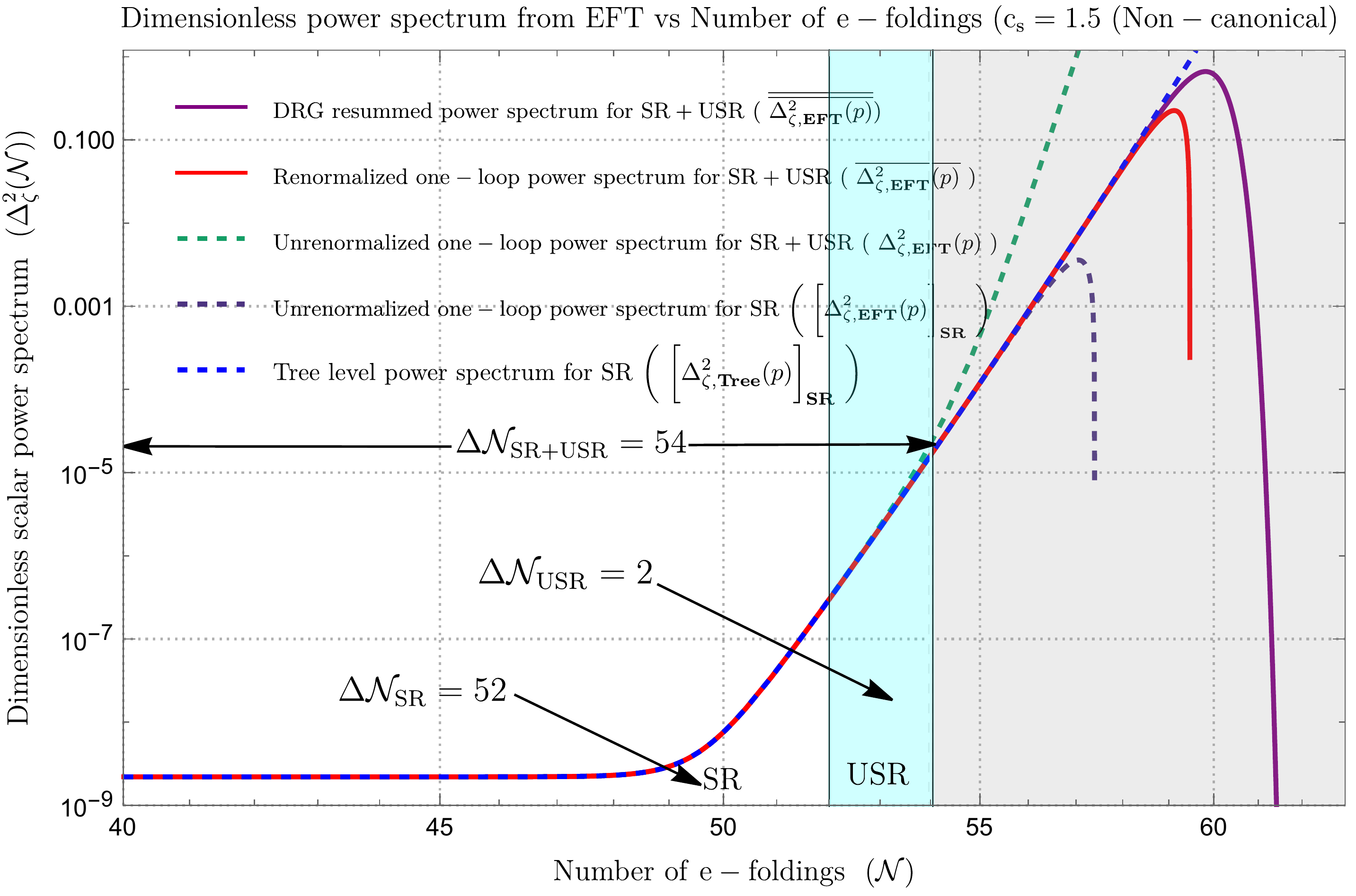}
        \label{Iv4}
       }
    	\caption[Optional caption for list of figures]{Behaviour of the dimensionless power spectrum for scalar modes with respect to the number of e-foldings from Framework II. In these representative plots, we have studied the behaviour for different values of the sound speed of EFT, $c_s=0.6,1,1.17,1.5$. Here we fix the pivot scale at $p_*=0.02\;{\rm Mpc}^{-1}$, SR to USR sharp transition scale at $k_s=10^{21}\;{\rm Mpc}^{-1}$ and the end of inflation scale at $k_e=10^{22}\;{\rm Mpc}^{-1}$, the renomalization parameter $c_{\bf SR}=0$, $\Delta\eta(\tau_e)=1$ and $\Delta\eta(\tau_s)=-6$. In this plot we have found that, $k_{\rm UV}/k_{\rm IR}=k_e/k_s\approx{\cal O}(10)$. Plots show that $c_s\gtrsim 1$ is the allowed range of effective sound speed, out of which $c_s=1.17$ gives the best outcome to have ${\cal O}(10^{-2})$ amplitude of the corresponding spectrum. At $c_s=1.5$ the amplitude reaches at ${\cal O}(1)$, at which the perturbation theory breaks.  } 
    	\label{Spectrumv3}
    \end{figure*}

In figure(\ref{Gv1}), figure(\ref{Gv2}), figure(\ref{Gv3}) and figure(\ref{Gv4}), we have depicted the behaviour of the dimensionless power spectrum for scalar modes with respect to the wave number for different values of the effective sound speed $c_s=0.6$, $c_s=1$, $c_s=1.17$ and $c_S=1.5$ for the Framework II. In each of the plots we have shown the contribution coming from tree level, one-loop corrected unrenormalized, renormalized one-loop corrected part and DRG resummed one-loop corrected part.
We found from each of the plots that the DRG resummed spectrum gives the best interpretation for the present scenario where the spectrum falls very fast at the end of the USR phase. Most importantly, the resummed spectrum provides the distinguishable features which help to differentiate among all of these mentioned contributions. 
For the computational purpose we fix the sharp transition scale at $k_s=10^{21}\;{\rm Mpc}^{-1}$ (where we fix the IR cut-off) and the end of USR and inflation at $k_e=10^{22}\;{\rm Mpc}^{-1}$ (where we fix the UV cut-off), the renomalization parameter $c_{\bf SR}=0$, $\Delta\eta(\tau_e)=1$ and $\Delta\eta(\tau_s)=-6$. We have maintained a restriction, $k_{\rm UV}/k_{\rm IR}=k_e/k_s\approx{\cal O}(10)$, which is the key finding of this calculation\footnote{It is important to note that this fact was first pointed in ref. \cite{Choudhury:2023vuj} for canonical single field models of inflation and then further used in \cite{Choudhury:2023jlt} for Effective Field Theory of Single Field Inflation.}. All of these plots show that $1<c_S<1.17$ is the allowed range of effective sound speed, out of which $c_s=1.17$ gives the best outcome to have the highest amplitude, ${\cal O}(10^{-2})$ of the corresponding spectrum necessarily needed to form PBHs from the present set up. At $c_s=1.5$ the amplitude reaches at ${\cal O}(1)$, at which the perturbation theory breaks. For this reason, we can't do our analysis beyond $c_s=1.5$.

Finally, in Figure (\ref{Iv1}), figure(\ref{Iv2}), figure(\ref{Iv3}), and Figure (\ref{Iv4}), we have depicted the behaviour of the dimensionless power spectrum for scalar modes with respect to the number of e-foldings. From this plot we have found that for Framework II:
\bea \Delta {\cal N}_{\rm USR}=\ln(k_e/k_s)\approx\ln(10)\approx 2,\eea 
which implies only approximately $2$ e-folds are allowed in the USR period for the PBH formation in the present setup. 

In this framework, the allowed e-folds for the SR period are given by:
\bea &&\Delta {\cal N}_{\rm SR}=\ln(k_s/p_*)\approx 52.\eea 
 As a consequence, the total number of e-foldings allowed by Framework II is given by the following expression:
\bea \Delta {\cal N}_{\rm Total}=\Delta {\cal N}_{\rm SR}+\Delta {\cal N}_{\rm USR}\sim 52+2=54,\eea 
provided the sharp transition scale, the end of the USR period and inflation are fixed at, $k_s=10^{21}\;{\rm Mpc}^{-1}$, and $k_e=10^{22}\;{\rm Mpc}^{-1}$. We will show in the next section that this possibility will lead to a small mass PBH formation having a sufficient number of e-folds for inflation in the present context.

Now let us consider a situation where we shift the sharp transition scale, the end of the USR period and inflation at the scale, $k_s=10^{6}\;{\rm Mpc}^{-1}$, and $k_e=10^{7}\;{\rm Mpc}^{-1}$. In this case, the allowed number of e-folds for SR can be computed as:
\bea &&\Delta {\cal N}_{\rm SR}=\ln(k_s/p_*)\approx 18.\eea
For USR the result will be unchanged. This implies in that mentioned scenario, the total number of e-foldings allowed by Framework II is given by the following expression:
\bea \Delta {\cal N}_{\rm Total}=\Delta {\cal N}_{\rm SR}+\Delta {\cal N}_{\rm USR}\sim 18+2=20.\eea
We will show in the next section that this possibility will lead to a large mass PBH formation having an insufficient number of e-folds for inflation. For this reason, this possibility can be immediately discarded.

\section{Stringent constraint on the PBH mass}\label{s8}

    \begin{figure*}[htb!]
    	\centering
{
      	\includegraphics[width=14cm,height=9cm] {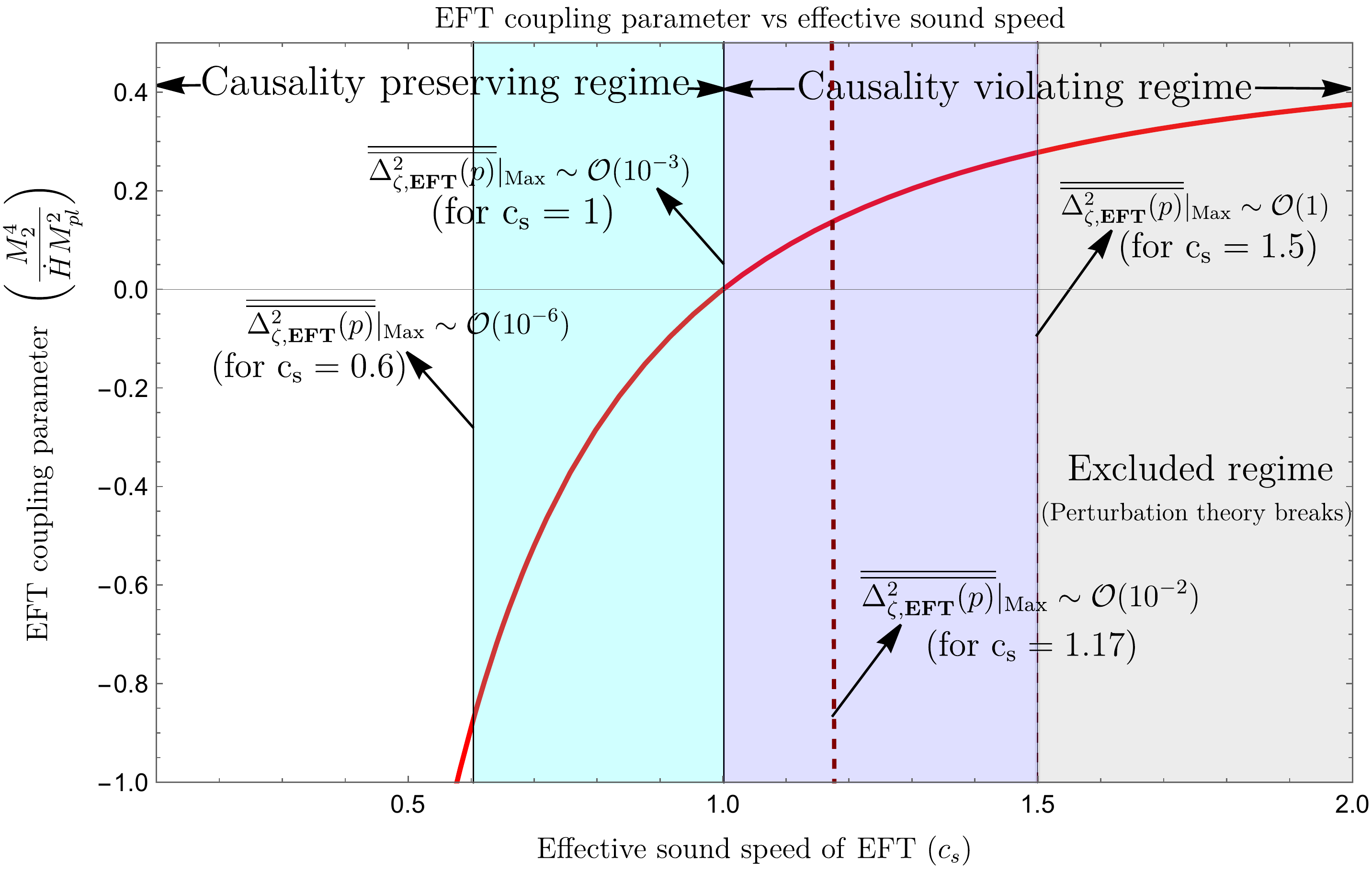}
        \label{KX1}
    }
    	\caption[Optional caption for list of figures]{Behaviour of the EFT coupling parameter from the present EFT set up with respect to the effective sound speed $c_s$ are plotted. Plots show that $c_s\gtrsim 1$ is the allowed range of effective sound speed, out of which $c_s=1.17$ gives the best outcome to have ${\cal O}(10^{-2})$ amplitude of the corresponding spectrum. At $c_s=1.5$ the amplitude reaches at ${\cal O}(1)$, at which the perturbation theory breaks.} 
    	\label{Spectrum5}
    \end{figure*}
    \begin{figure*}[htb!]
    	\centering
    \subfigure[ $M_{\rm PBH}$ vs $c_s$.]{
      	\includegraphics[width=8cm,height=8cm] {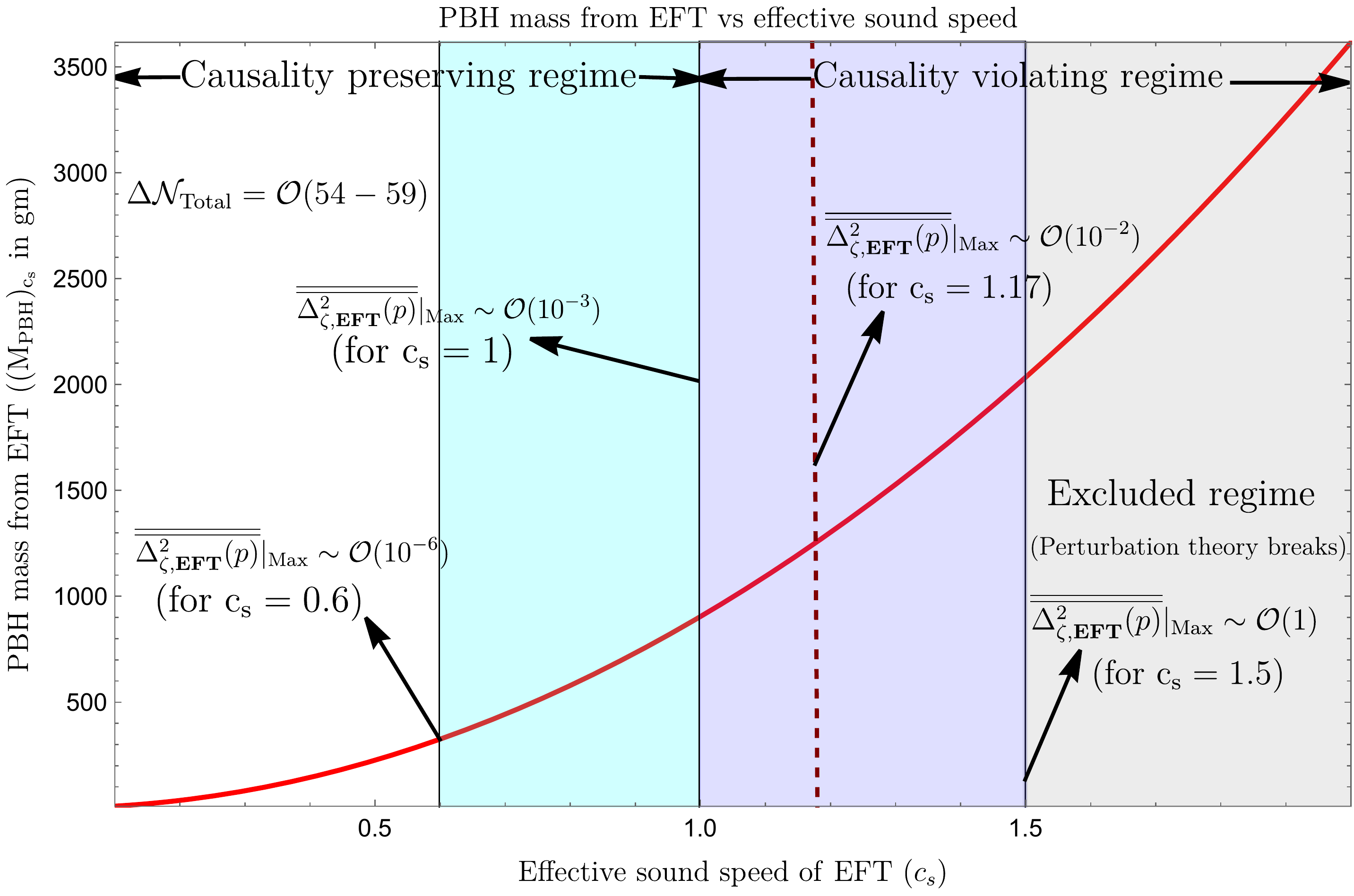}
        \label{Kv1}
    }
    \subfigure[$M_{\rm PBH}$ vs $c_s$.]{
       \includegraphics[width=8cm,height=8cm] {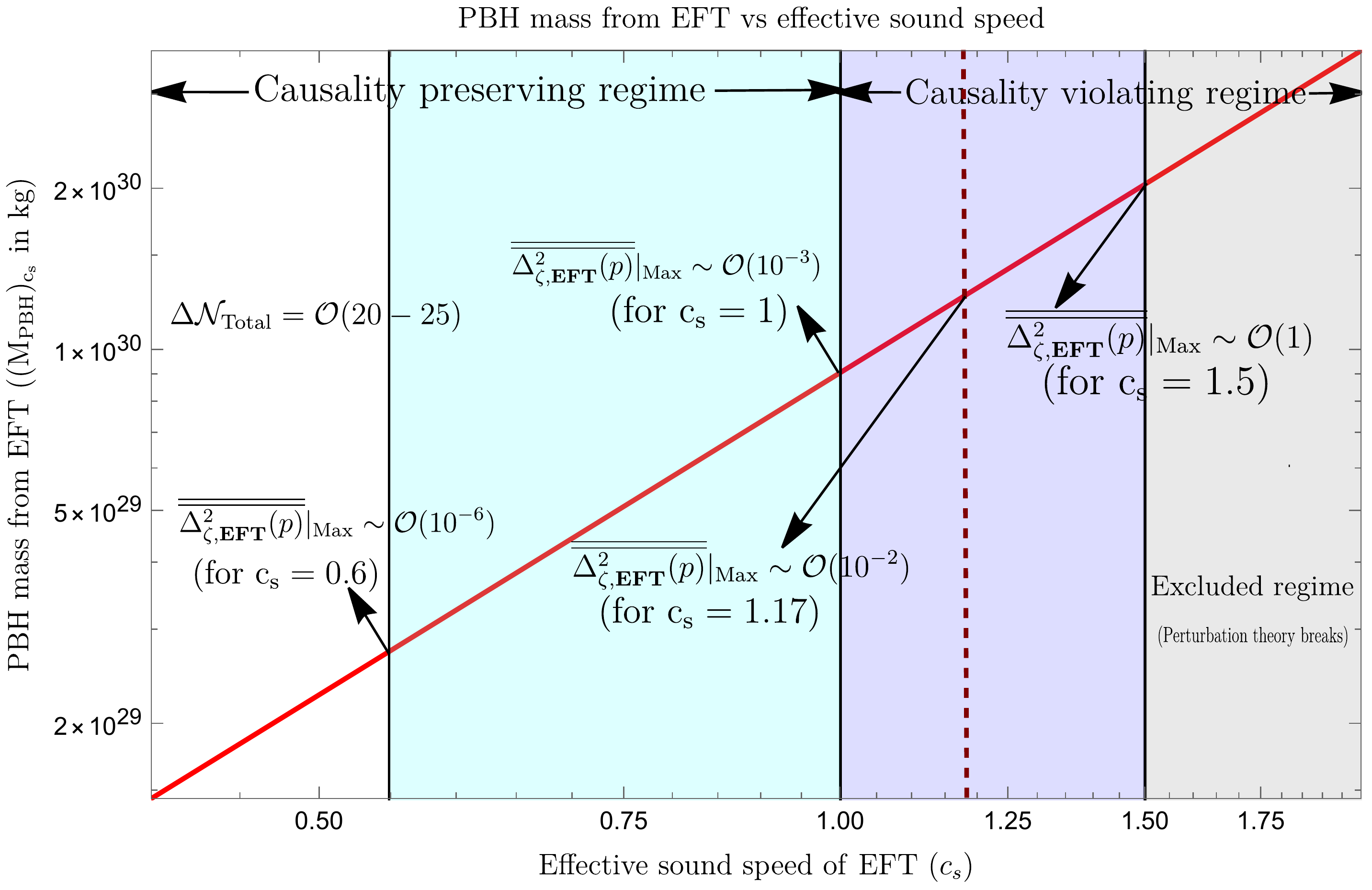}
        \label{Kv2}
       }
    	\caption[Optional caption for list of figures]{Behaviour of the PBH mass from the present EFT set up with respect to the effective sound speed $c_s$ are plotted for two cases, where \ref{Kv1} the total number of e-foldings are fixed at $\Delta{\cal N}_{\rm Total}\sim {\cal O}(54-59)$ and \ref{Kv2} $\Delta{\cal N}_{\rm Total}\sim {\cal O}(20-25)$, respectively. In these representative plots we fix the pivot scale at $p_*=0.02\;{\rm Mpc}^{-1}$, SR to USR sharp transition scale at for the first plot, $k_s=10^{21}\;{\rm Mpc}^{-1}$ and for the second plot, $k_s=10^{6}\;{\rm Mpc}^{-1}$. Plots show that $c_s\gtrsim 1$ is the allowed range of effective sound speed, out of which $c_s=1.17$ gives the best outcome to have ${\cal O}(10^{-2})$ amplitude of the corresponding spectrum.  Both of the plots are same for the Framework I and Framework II, studied in this paper and from the perspective of PBH formation there is no difference observed from our computation. Here  \ref{Kv1} supports small mass PBH formation having sufficient number of e-foldings for inflation. On the other hand,  \ref{Kv2} supports large mass PBH formation having insufficient number of e-foldings for inflation. } 
    	\label{Spectrum6}
    \end{figure*}

We found from our analysis performed in this paper for the Framework I and Framework II that the prolonged USR period is not allowed for PBH formation in both the cases. Also we found that at the level of PBH formation one cannot able to distinguish the outcomes obtained from both of these mentioned frameworks. Within this short span one can further give the estimation of PBH mass in terms of the sound speed $c_s$ for the underlying Goldstone EFT set up, using the following expression:
\bea \label{sd1}\boxed{\left(\frac{M_{\rm PBH}}{M_{\odot}}\right)_{c_s}=\left(\frac{M_{\rm PBH}}{M_{\odot}}\right)_{c_s=1}\times c^{2}_s},\eea 
where for the canonical models of inflation ($c_s=1$), we have:
\bea \left(\frac{M_{\rm PBH}}{M_{\odot}}\right)_{c_s=1}=4.52\times {10}^{-31}\bigg(\frac{\gamma}{0.2}\bigg)\bigg(\frac{g_*}{106.75}\bigg)^{-1/6}.\eea 
\begin{table}
\footnotesize
\begin{tabular}{|c|c|c|c|c|c|c|c|}
 \hline\hline
 \multicolumn{8}{|c|}{A detailed list of $P(X,\phi)$ models and its connection with Goldstone EFT} \\
 \hline\hline
 Model name & $P(X,\phi)$ function & $\displaystyle \frac{M^4_2}{\dot{H}M^2_{pl}}$ & $\displaystyle \frac{M^3_1}{HM^2_{pl}}$ & $\displaystyle \frac{M^4_3}{H^2M^2_{pl}}$   & $c_s$ & $\displaystyle{\cal M}=\frac{M_{\rm PBH}}{10^2{\rm gm}}$ & Comment\\
 \hline\hline
 & & & & & & &  \\
DBI   & $\displaystyle -\frac{\Lambda^4}{f(\phi)}\sqrt{1-\frac{f(\phi)}{\Lambda^4}X}+\frac{\Lambda^4}{f(\phi)}-V(\phi)$  & $(-0.89)-0$  & $(-0.1)-0.1$ &   $(-0.5)-0.05$ &   $0.6-1$ &  $0.36-9.04$ & Causal (\ding{56})\\
& & & & & & &  \\
\hline\hline
 & & & & & & &  \\
Tachyon   & $\displaystyle -V(\phi)\sqrt{1-2\alpha^{'}X}$  & $(-0.89)-0$  & $(-0.2)-0.1$ &   $(-0.4)-0$ &   $0.6-1$ &  $0.36-9.04$ & Causal (\ding{56})\\
& & & & & & &  \\
\hline\hline
 & & & & & & &  \\
GTachyon   & $\displaystyle -V(\phi)\left(1-2\alpha^{'}X\right)^{q}$\;($q\leq \frac{1}{2}$, $\frac{1}{2}<q<1$)     & $(-0.89)-0$  & $(-0.2)-0$ &   $(-0.4)-0$ &   $0.6-1$ &  $0.36-9.04$ & Causal (\ding{56})
\\
   & \quad\quad\quad\quad\quad\;($q> \frac{1}{2}$)     & $0-0.28$  & $0-0.07$ &   $0-0.09$ &   $1-1.5$ &  $9.04-20.3$ & Acausal (\ding{52})\\
& & & & & & &  \\
\hline\hline
  & & & & & & &  \\
$K$   & $\displaystyle \gamma_n X^n-V(\phi)$\;($1.89>n>1$)  & $(-0.89)-0$  & $(-0.4)-0$ &   $(-0.3)-0$ &   $0.6-1$ &  $0.36-9.04$ & Causal (\ding{56}) \\
  & \quad\quad\quad\quad\quad\;\;($1>n>0.72$)  & $0-0.28$  & $0-0.1$ &   $0-0.2$ &   $1-1.15$ &  $9.04-20.3$ & Acausal (\ding{52})\\
& & & & & & &  \\
\hline\hline
 Canonical& & & & & &  & \\
 Single    & $\displaystyle X-V(\phi)$  & $0$  & $0$ &   $0$ &   $1$ &  $9.04$ & Causal (\ding{52})\\
field & & & & & &  & \\
 \hline\hline
\end{tabular}
\caption{Constraints on $P(X,\phi)$ theories in the light of PBH formation. Here we have quoted the constraints for any general effective potential which is allowed by the corresponding physical frameworks and sufficient e-folds of inflation have to be achieved. We have explicitly mentioned the allowed range of the EFT parameters $\displaystyle M^4_2/\dot{H}M^2_{pl}$, $\displaystyle M^3_1/HM^2_{pl}$ and $\displaystyle M^4_3/H^2M^2_{pl}$ as well as the allowed PBH mass range from each of the $P(X,\phi)$ models of inflation. Here \ding{56} and \ding{52} correspond to the not-allowed and allowed models of $P(X,\phi)$ models of inflation in the light of PBH formation having a small mass. In this estimations we use $k_s=10^{21}{\rm Mpc}^{-1}$ and $k_e=10^{22}{\rm Mpc}^{-1}$, which actually give rise to the small mass PBHs.  Consequently, we have maintained the constraint on the total number of the e-foldings, $\Delta {\cal N}_{\rm Total}\sim {\cal O}(54-59)$ (Considering both Framework I and II), which is necessarily required to achieve sufficient inflation in the present context. The possibility of large mass PBHs is not explored in this table as it is not able to generate sufficient e-foldings for inflation.}
\label{tab:1}
\end{table}
Here $M_{\odot}\sim 2\times 10^{30}{\rm kg}$ is the solar mass, which is used for the final estimation of the PBH mass. 
For $c_s=1$ we have used the following formula for PBH mass:
\be \label{ddf}\left(\frac{M_{\rm PBH}}{M_{\odot}}\right)_{c_s=1}=1.13\times 10^{15}\bigg(\frac{\gamma}{0.2}\bigg)\bigg(\frac{g_*}{106.75}\bigg)^{-1/6}\bigg(\frac{k_s}{p_*}\bigg)^{-2}.\ee
where we fix $k_s=10^{21}\;{\rm Mpc}^{-1}$ for the sharp transition scale and pivot scale at $p_*=0.02\;{\rm Mpc}^{-1}$. Here $\gamma\sim 0.2$ and relativistic d.o.f. $g_*\sim 106.75$ for Standard Model and $g_*\sim 226$ for SUSY d.o.f. Choosing $k_s=10^{21}\;{\rm Mpc}^{-1}$, $k_e=10^{22}\;{\rm Mpc}^{-1}$ along with $k_e/k_s=10$ necessarily required to validate the perturbative approximation in the one-loop contribution and to generate sufficient number of e-foldings, which is $\Delta{\cal N}_{\rm Total}=59$ for the Framework I and $\Delta{\cal N}_{\rm Total}=54$ for the Framework II for inflation. We found from our analysis performed in this section that the estimated PBH mass from the Goldstone EFT set up is extreme small and lying within the window, $10^{2}{\rm gm}\lesssim M_{\rm PBH}\lesssim 10^{3} \rm gm$ for the effective sound speed $0.6<c_s<1.17$ to have total number of e-foldings $\Delta{\cal N}_{\rm Total}=59$ (Framework I) and $\Delta{\cal N}_{\rm Total}=54$ (Framework II).

We now consider here another situation where we fix the wave numbers at $k_s=10^{6}\;{\rm Mpc}^{-1}$, $k_e=10^{7}\;{\rm Mpc}^{-1}$ where $k_e/k_s=10$ necessarily constraint is maintained to validate the perturbative approximation in the one-loop contribution. However, in this case for the Framework II we have to end inflation strictly at $k_{\rm end}=10^{9}\;{\rm Mpc}^{-1}$ for the Framework I and $k_e=10^{7}\;{\rm Mpc}^{-1}$ for the Framework II to obtained a correct renormalized and DRG resummed power spectrum. In this case the total number of e-foldings are given by, $\Delta{\cal N}_{\rm Total}=25$ for the Framework I and $\Delta{\cal N}_{\rm Total}=20$ for the Framework II, which are not sufficient enough to achieve inflation. For this specific case, when $k_s=10^{6}\;{\rm Mpc}^{-1}$, using equation (\ref{ddf}) the PBH mass is computed for the canonical model ($c_s=1$) is:
\be \left(\frac{M_{\rm PBH}}{M_{\odot}}\right)_{c_s=1}=0.46\bigg(\frac{\gamma}{0.2}\bigg)\bigg(\frac{g_*}{106.75}\bigg)^{-1/6}.\ee
This result implies that if we don't want sufficient inflation in terms of the total number of e-foldings then the estimated PBH mass from the Goldstone EFT setup is large and using equation (\ref{sd1}) we found that it is lying within the window, $10^{29}{\rm kg}\lesssim M_{\rm PBH}\lesssim 10^{30} \rm kg$ for the effective sound speed $0.6<c_s<1.17$. But to generate large mass PBHs sacrificing the longevity of the inflationary paradigm is not a good choice. For this reason, this possibility is strictly discarded from the theoretical as well as observational ground.

In this analysis, the corresponding effective sound speed varies within the range, $1<c_s<1.17$ for the following EFT parameter coupling range:
\bea && 0\lesssim M^4_2/\dot{H}M^2_p\lesssim 0.13,\\ 
 && 0\lesssim \bar{M}^3_2/HM^2_p\lesssim 0.08,\\ 
 && 0\lesssim M^4_3/H^2M^2_p\lesssim 0.05.\eea
In figure (\ref{KX1}), figure (\ref{Kv1}), and figure (\ref{Kv2}), we have plotted the behaviour of the EFT coupling parameter $M^4_2/\dot{H}M^2_p$ and PBH mass from the present EFT set up $(M_{\rm PBH})_{c_s}$ with respect to the effective sound speed $c_s$. Particularly in figure (\ref{Kv1}) and figure (\ref{Kv2}) we fix the pivot scale at $p_*=0.02\;{\rm Mpc}^{-1}$, sharp transition scale at $k_s=10^{21}\;{\rm Mpc}^{-1}$ and $k_s=10^{6}\;{\rm Mpc}^{-1}$, respectively. All of these plots show that $c_s\gtrsim 1$ is the allowed range of effective sound speed, out of which $c_s=1.17$ gives the best outcome to have ${\cal O}(10^{-2})$ amplitude of the corresponding spectrum. At $c_s=1.5$ the amplitude reaches ${\cal O}(1)$, at which the perturbative approximations completely break down in the present analysis. Additionally, it is important to note that, both of the outcomes are exactly same for the Framework I and Framework II, and from the perspective of PBH formation there is no difference observed from our computation.  Comparing figure (\ref{Kv1}) and figure (\ref{Kv2}) we found that large mass PBHs within the range ${\cal O}(10^{29}-10^{30}){\rm kg}$ can be generated with insufficient number of total e-foldings $\Delta{\cal N}_{\rm Total}\sim {\cal O}(20-25)$ for inflation. On the other hand, to accommodate a sufficient total number of e-foldings $\Delta{\cal N}_{\rm Total}\sim {\cal O}(54-59)$ for inflation one can only able to generate small mass PBHs lying within a very tiny window, ${\cal O}(10^{2}-10^{3}){\rm gm}$. In both cases, the favoured range of effective sound speed will line within the window $1\leq c_s\leq 1.17$ out of which only for $c_s=1.17$ one can able to achieve enhanced amplitude of the one-loop corrected power spectrum ${\cal O}(10^{-2})$.

The other two EFT parameters, $\bar{M}^3_2/HM^2_p$ and $M^4_3/H^2M^2_p$ are the byproducts of the one-loop contributions from the single SR for the Framework II and SRI, SRII phases for the Framework I studied in this paper. We found that the results obtained in this paper are not extremely sensitive to both of these parameters as the SR contribution is suppressed compared to USR contribution in the one-loop corrected expression for the primordial power spectrum for scalar modes. Just by observing the effect in the amplitude of the one-loop corrected spectrum one cannot able to fix both of these parameters from the first principle. However we know that the one-loop contribution is computed from the third order perturbative action for the scalar curvature perturbation and using this action one can able to directly compute the three point function and the associated effect in the primordial non-Gaussianity \cite{Maldacena:2002vr,Seery:2005wm,Senatore:2009gt,Chen:2006nt,Chen:2010xka,Chen:2009zp,Chen:2009we,Chen:2008wn,Chen:2006xjb,Choudhury:2012whm,Agarwal:2012mq,Holman:2007na,Creminelli:2005hu,Behbahani:2011it,Smith:2009jr,Cheung:2007sv,Creminelli:2006rz,Creminelli:2006gc,Kalaja:2020mkq,Meerburg:2019qqi,Lee:2016vti,Maldacena:2011nz,Werth:2023pfl}. To  generate feasible large non-Gaussian effects from the quoted $P(X,\phi)$ theory models of inflation we have fixed the two EFT parameters, $\bar{M}^3_2/HM^2_p$ and $M^4_3/H^2M^2_p$ in the certain specified range. But as we know from the canonical single field models of inflationary paradigm large non-Gaussian effects cannot be generated, we have fixed  these two EFT parameters, $\bar{M}^3_2/HM^2_p=0$ and $M^4_3/H^2M^2_p=0$ for the $P(X,\phi)=X-V$ model.

To justify our analysis performed in this paper we did a detailed computation in the Appendix \ref{app:P} which helps us to make a clear connection between Goldstone EFT set up and the 
$P(X,\phi)$ theory models of inflation. We have shown that the EFT parameter $M^4_2/\dot{H}M^2_p$ can be directly expressed in terms of the derivatives of the $P(X,\phi)$ with respect to the kinetic contribution $X$ at the background level. This helps us to constrain the EFT parameter $M^4_2/\dot{H}M^2_p$ for various types of $P(X,\phi)$ models of single field inflation including the canonical single field slow roll framework. In the Appendix \ref{app:P} we have discussed the constraints on the various model parameters as well as the PBH mass obtained from the various types of $P(X,\phi)$ models of single field inflation. These results are explicitly quoted in the Table (\ref{tab:1}) for completeness. In this table we have quoted the constraints for any general effective potential $V(\phi)$ which is allowed by the corresponding underlying physical frameworks and sufficient e-folds of inflation has to be achieved for each models. We have explicitly mentioned the allowed range of the EFT parameters $\displaystyle M^4_2/\dot{H}M^2_{pl}$, $\displaystyle M^3_1/HM^2_{pl}$ and $\displaystyle M^4_3/H^2M^2_{pl}$ as well as the allowed PBH mass range from each of the $P(X,\phi)$ models of inflation lying within the range of the effective sound speed, $0.6<c_s<1.5$. From this table we also found that the range, $1<c_s<1.17$ is more favoured where the underlying EFT prefers breakdown of causality due to having maximum amount of enhancement during PBH formation in the one-loop corrected primordial power spectrum for scalar modes. For the better understanding purpose we use the symbols \ding{56} and \ding{52} to indicate the not-allowed and allowed models of $P(X,\phi)$ models of inflation in the light of PBH formation having small mass. From this analysis we have found that within the framework Goldstone EFT which can be easily connected in terms of all possible known $P(X,\phi)$ models of single field inflation which includes both the canonical as well as the non-canonical frameworks large mass PBH cannot be strictly generated if we include the one-loop contribution in the primordial power spectrum for scalar modes in the both Framework I and Framework II. This in turn establish a No-go theorem for all possible types of single field inflation which can be mapped in terms of $P(X,\phi)$ models of single field inflationary paradigm.
    \begin{figure*}[htb!]
    	\centering
{
      	\includegraphics[width=14cm,height=10cm] {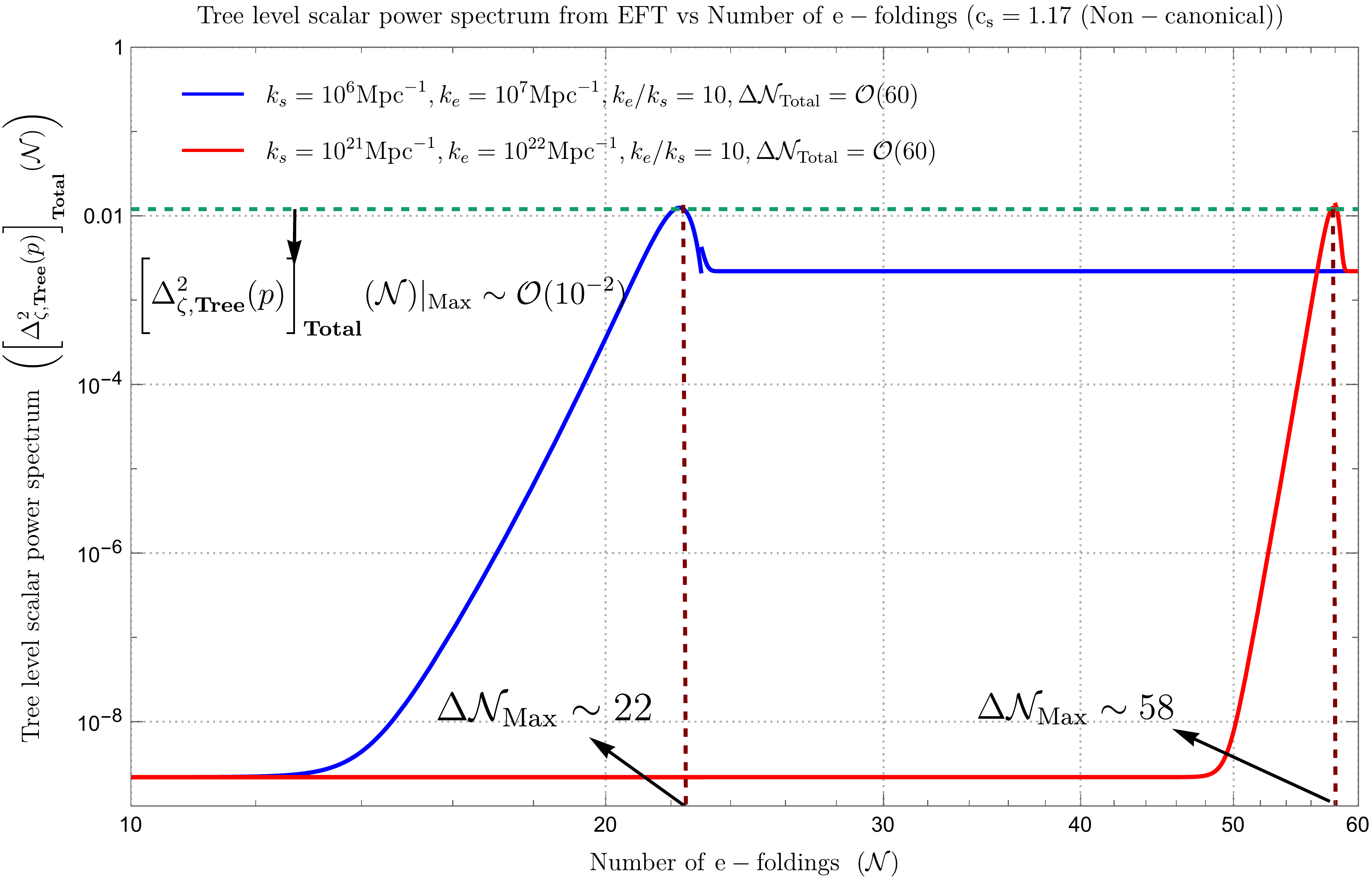}
        \label{D1}
    }
    	\caption[Optional caption for list of figures]{Behaviour of dimensionless tree level total power spectrum for scalar modes from the present EFT setup is plotted with respect to the number of e-foldings. In these representative plots we fix the pivot scale at $p_*=0.02\;{\rm Mpc}^{-1}$, sharp transition scales at $k_s=10^{6}\;{\rm Mpc}^{-1}$ (for blue) and $k_s=10^{21}\;{\rm Mpc}^{-1}$ (for red) at $c_s=1.17$ where the peak of the spectrum for both the cases appeared at ${\cal O}(10^{-2})$.} 
    	\label{SpectrumTree}
    \end{figure*}
    \begin{figure*}[htb!]
    	\centering
{
      	\includegraphics[width=14cm,height=10cm] {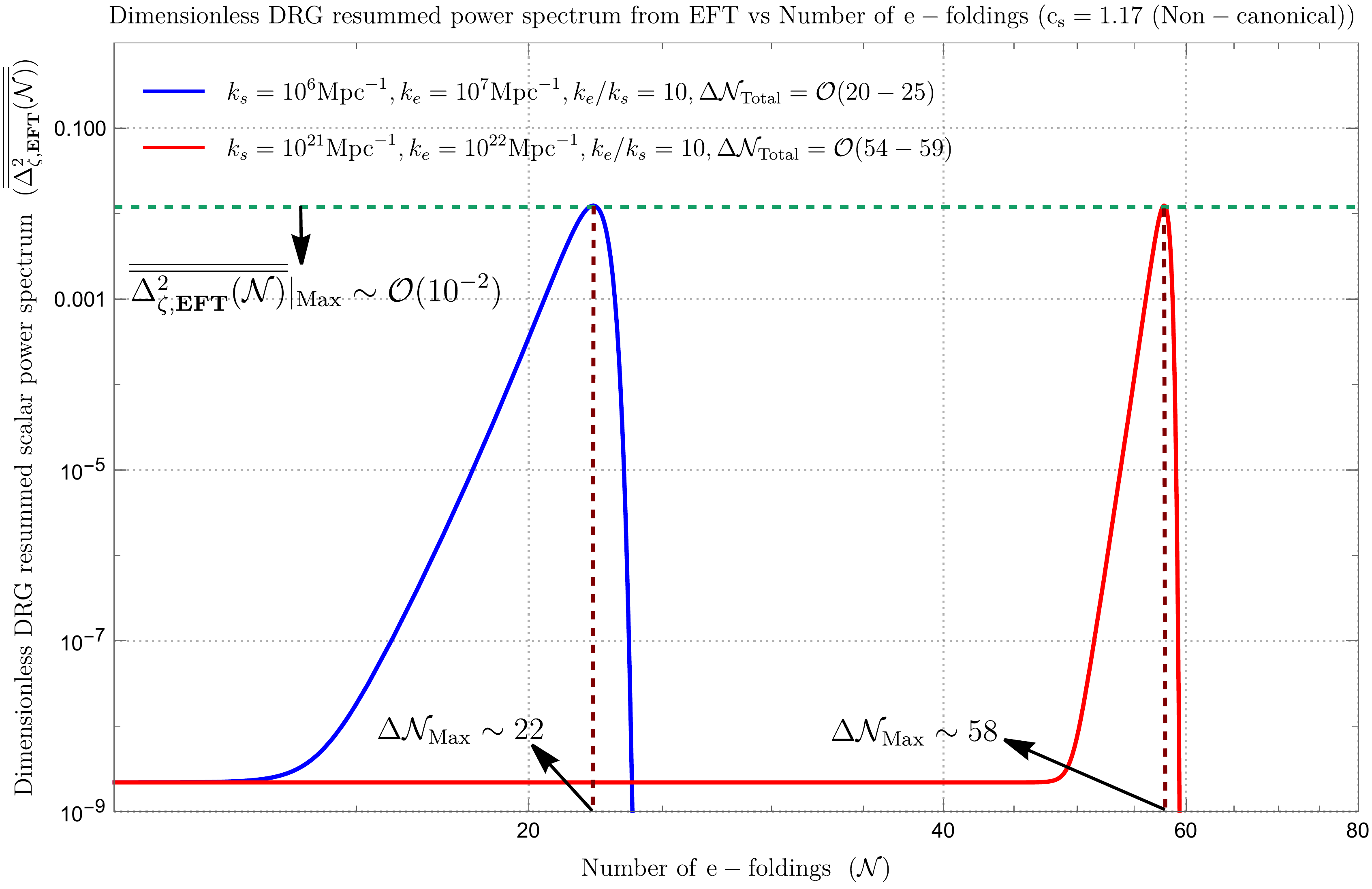}
        \label{D1}
    }
    	\caption[Optional caption for list of figures]{Behaviour of dimensionless DRG resummed power spectrum for scalar modes from the present EFT setup is plotted with respect to the number of e-foldings. In these representative plots we fix the pivot scale at $p_*=0.02\;{\rm Mpc}^{-1}$, sharp transition scales at $k_s=10^{6}\;{\rm Mpc}^{-1}$ (for blue) and $k_s=10^{21}\;{\rm Mpc}^{-1}$ (for red) at $c_s=1.17$ where the peak of the spectrum for both the cases appeared at ${\cal O}(10^{-2})$.} 
    	\label{SpectrumDRG}
    \end{figure*}
Now let us finally mention clearly what are the things that we have learned from the analysis performed in this paper point-wise:
\begin{itemize}

    \item It is important to mention that the generation of PBH mass is solely dependent on the sharp transition scale $k_s$ for both Framework I and Framework II where SR to USR sharp transition occurs. The estimation of the PBH mass is not dependent on how the USR phase or the inflationary paradigm ends.

    \item It is important to note down that to generate large mass PBHs from the present set-up position of the SR to USR sharp transition i.e. the explicit value of the sharp transition scale $k_s$ is extremely important for both Framework I and Framework II.

    \item We found from our analysis that if we fix the wave numbers $k_s=10^{6}{\rm Mpc}^{-1}$ and the pivot scale at $p_*=0.02{\rm Mpc}^{-1}$, then it is possible to generate large mass PBH having mass, $M_{\rm PBH}\sim {\cal O}(10^{29}-10^{30}){\rm kg}$ within the allowed window of effective sound speed $0.6<c_s<1.17$. In this specific case number of e-folds allowed for the SR phase (first SR phase for the Framework I and single SR phase for the Framework II) is $\Delta{\cal N}_{\rm SR}= \ln(k_s/p_*)\sim 18$. Also we found a strict restriction that $k_e/k_s=10$, which is necessarily needed to incorporate the perturbative approximation during the computation of the one-loop contribution from USR phase in the primordial power spectrum for scalar modes. This further implies that the number of e-folds allowed for the USR phase (both for Framework I and Framework II) is $\Delta{\cal N}_{\rm USR}= \ln(k_e/k_s)\sim 2$. For the Framework II, we then have total $\Delta{\cal N}_{\rm Total}=\Delta{\cal N}_{SR}+\Delta{\cal N}_{\rm USR}=18+2=20$ e-folds which is insufficient to achieve inflation from the present set up. For the Framework I, we have found that to have a correct renormalized and DRG resummed power spectrum the second SR phase (SRII), as well as the inflation, has to strictly follow the constraint, $k_{\rm end}/k_{\rm e}=10^{2},$ using which the number of e-folds during the SRII phase can be computed as, $\Delta{\cal N}_{\rm SRII}= \ln(k_{\rm end}/k_{\rm e})\sim 5$. This further implies that in the case of Framework II, the total number of allowed e-folds will be $\Delta{\cal N}_{\rm Total}=\Delta{\cal N}_{\rm SRI}+\Delta{\cal N}_{\rm USR}+\Delta{\cal N}_{\rm SRII}=18+2+5=25$, which is also insufficient to achieve inflation from the present set up. So this implies that for both frameworks large mass PBHs can be generated by spoiling the necessary requirement for inflation in terms of a number of e-foldings. This is obviously not preferable from the theoretical perspective and can be discarded immediately. To justify this outcome clearly we have plotted the behaviour of the dimensionless tree-level total and DRG resummed power spectrum for the scalar modes with respect to the number of e-foldings in figures (\ref{SpectrumTree}) and (\ref{SpectrumDRG}) respectively. \textcolor{black}{In this plot (\ref{SpectrumTree}), we have explicitly shown that the peak of the spectrum appears at ${\cal O}(10^{-2})$ for the first (blue) situation at $\Delta {\cal N}_{\rm Total}\sim {\cal O}(60)$ with $\Delta {\cal N}_{\rm Max}\sim 22$ considering both the possibilities studied in Framework I and Framework II. Here the first (blue) one corresponds to the outcome of the specific case where we introduce the sharp transition scale at $k_s=10^{6}{\rm Mpc}^{-1}$. This possibility can able to generate large mass PBHs within the range, ${\cal O}(10^{29}-10^{30}){\rm kg}$ as clearly shown in figure (\ref{Kv2}). In this particular case, inflation is able to sustain a sufficient number of total e-foldings for the tree-level case, this possibility is giving rise to desirable results at least for the tree-level case. However, things are going to drastically change when we consider DRG resumed normalized power spectrum in the present context of discussion}. In the plot (\ref{SpectrumDRG}), we have explicitly shown that the peak of the spectrum appears at ${\cal O}(10^{-2})$ for the first (blue) situation at $\Delta {\cal N}_{\rm Total}\sim {\cal O}(20-25)$ with $\Delta {\cal N}_{\rm Max}\sim 22$ considering both the possibilities studied in Framework I and Framework II. Here the first (blue) one corresponds to the outcome of the specific case where we introduce the sharp transition scale at $k_s=10^{6}{\rm Mpc}^{-1}$, where we have maintained all the previously mentioned restrictions on the wave number corresponds to end of inflation as well the end of USR phase. This possibility can able to generate large mass PBHs within the range, ${\cal O}(10^{29}-10^{30}){\rm kg}$ as clearly shown in figure (\ref{Kv2}). However, as we have already mentioned since inflation cannot able to sustain a sufficient number of total e-foldings for the DRG ressumed case, this possibility is not desirable in the present context of the discussion.

    \item On the other hand, our analysis shows that if we fix the wave numbers $k_s=10^{21}{\rm Mpc}^{-1}$ and the pivot scale at $p_*=0.02{\rm Mpc}^{-1}$, then only it is possible to generate small mass PBH having mass, $M_{\rm PBH}\sim {\cal O}(10^{2}-10^{3}){\rm gm}$ within the allowed window of effective sound speed $0.6<c_s<1.17$. In this specific case number of e-folds allowed for the SR phase (first SR phase for the Framework I and single SR phase for the Framework II) is $\Delta{\cal N}_{\rm SR}= \ln(k_s/p_*)\sim 52$. In this case also we found the same strict restriction that $k_e/k_s=10$, which is necessarily needed to incorporate the perturbative approximation during the computation of the one-loop contribution from USR phase in the primordial power spectrum for scalar modes. This further implies that the number of e-folds allowed for the USR phase (both for Framework I and Framework II) is $\Delta{\cal N}_{\rm USR}= \ln(k_e/k_s)\sim 2$. For the Framework II, we then have total $\Delta{\cal N}_{\rm Total}=\Delta{\cal N}_{SR}+\Delta{\cal N}_{\rm USR}=52+2=54$ e-folds which is sufficient enough to achieve inflation from the present set up. For the Framework I, we have found that to have a correct DRG resummed power spectrum the second SR phase (SRII) as well as the inflation has follow the constraint, $k_{\rm end}/k_{\rm e}=10^{2},$ using which the number of e-folds during the SRII phase can be computed as, $\Delta{\cal N}_{\rm SRII}= \ln(k_{\rm end}/k_{\rm e})\sim 5$. This further implies that in the case of Framework II, the total number of allowed e-folds will be $\Delta{\cal N}_{\rm Total}=\Delta{\cal N}_{\rm SRI}+\Delta{\cal N}_{\rm USR}+\Delta{\cal N}_{\rm SRII}=52+2+5=59$, which is also sufficient enough to achieve inflation from the present set up. So this implies that for both frameworks small mass PBHs can be generated by maintaining the necessary requirement for inflation in terms of the number of e-foldings. This is obviously much preferable from the theoretical perspective and can be considered for future investigations. To justify this outcome clearly, we have plotted the behaviour of the dimensionless tree=level total and DRG resummed power spectrum for the scalar modes with respect to the number of e-foldings in figures (\ref{SpectrumTree}) and (\ref{SpectrumDRG}). \textcolor{black}{In this plot (\ref{SpectrumTree}), we have explicitly shown that the peak of the spectrum appears at ${\cal O}(10^{-2})$ for the second (red) situation at $\Delta {\cal N}_{\rm Total}\sim {\cal O}(54-59)$ with $\Delta {\cal N}_{\rm Max}\sim 58$ considering both the possibilities studied in Framework I and Framework II. Here the second (red) one corresponds to the outcome of the specific case where we introduce the sharp transition scale at $k_s=10^{21}{\rm Mpc}^{-1}$. This possibility can able to generate small mass PBHs within the range, ${\cal O}(10^{2}-10^{3}){\rm gm}$ as clearly shown in figure (\ref{Kv1}). Since a sufficient number of e-foldings is achieved for inflation in this case, this possibility might be useful for future investigations and satisfy the constraints at the tree-level case analysis. We have found that the tree-level plot can able to accommodate a constant non-vanishing amplitude due to having an extremely short-lived SRII region for Framework I and a small phase after USR and end of inflation for Framework II. On the other hand, in the case of DRG resummed case, we have found a sharp falling feature for both Frameworks I and II due to having additional stringent constraints from renormalization and resummation.} In the plot (\ref{SpectrumDRG}), we have explicitly shown that the peak of the spectrum appears at ${\cal O}(10^{-2})$ for the second (red) situation at $\Delta {\cal N}_{\rm Total}\sim {\cal O}(54-59)$ with $\Delta {\cal N}_{\rm Max}\sim 58$ considering both the possibilities studied in Framework I and Framework II. Here the second (red) one corresponds to the outcome of the specific case where we introduce the sharp transition scale at $k_s=10^{21}{\rm Mpc}^{-1}$, where we have maintained all the previously mentioned restrictions on the wave number corresponds to end of inflation as well the end of USR phase. This possibility can able to generate small mass PBHs within the range, ${\cal O}(10^{2}-10^{3}){\rm gm}$ as clearly shown in figure (\ref{Kv1}). Since a sufficient number of e-foldings is achieved for inflation in this case, this possibility might be useful for future investigations. 

    \item \textcolor{black}{Performing careful comparison between the figures (\ref{SpectrumTree}) and (\ref{SpectrumDRG}), one can clearly observe that there is a drastic change has been observed from the outcomes obtained from the PBHs generated of the mass range, ${\cal O}(10^{29}-10^{30}){\rm kg}$ at the scale $k_s\sim 10^6{\rm Mpc}^{-1}$ for the tree-level and DRG resummed loop level result. For the tree-level case, we have found that the sufficient number of e-foldings $\Delta {\cal N}_{\rm Total}\sim{\cal O}(60)$ has been achieved. On the other hand for the DRG resummed loop level result, we have found that number of e-foldings $\Delta {\cal N}_{\rm Total}\sim{\cal O}(20-25)$ become insufficient for inflation due to having sharp fall in the spectrum directly appearing from the additional constraints from renormalization and resummation. On the other hand, one can further observe that no such drastic change has been observed from the outcomes obtained from the PBHs generated of the mass range, ${\cal O}(10^{2}-10^{3}){\rm gm}$ at the scale $k_s\sim 10^{21}{\rm Mpc}^{-1}$ for the tree-level and DRG resummed loop level result. Most importantly, in both of these cases, we have found that the sufficient number of e-foldings $\Delta {\cal N}_{\rm Total}\sim{\cal O}(60)$ has been achieved. This clearly implies that the tree-level result and the DRG resummed result considering the loop effects almost give similar consequences both for Frameworks I and II as far as the PBHs generated of the mass range, ${\cal O}(10^{2}-10^{3}){\rm gm}$ is concerned in the present context of the discussion. By seeing such compatibility of the result obtained from the tree-level computation and DRG resummed one-loop level computation one can safely trust the obtained result for the PBHs mass range, ${\cal O}(10^{2}-10^{3}){\rm gm}$.}

    \item We have also found that the PBH mass very mildly depends on the effective sound speed parameter within the allowed range, $1<c_s<1.17$. This is true for both the large as well as the small mass PBH generation from the present Goldstone EFT setup, where the details of the Framework I and Framework II have been explicitly considered.

    \item Last but not the least, we found from our analysis that the peak of the power spectrum appears at ${\cal O}(10^{-2})$, which is only possible from our set up if we allow acasality/superluminality at $c_s\sim 1.17$ within the present EFT set up. This particular number is extremely important for PBH formation, which is achieved in the present set up.
\end{itemize}

\section{Comparison between Framework I and Framework II in the light of PBH formation}\label{s9}

In this section our prime objective is to compare the findings obtained from Framework I and Framework II, which will be extremely helpful to understand the individual novelty of each of the frameworks in the present context of discussion. The detailed comparison is appended below point-wise:
\begin{itemize}
    \item To generate large mass PBHs within the range, ${\cal O}(10^{29}-10^{30}){\rm kg}$ within the preferred effective sound speed window, $1<c_s<1.17$, the total number of allowed e-foldings from the Framework I is, $\Delta{\cal N}_{\rm Total}=\Delta{\cal N}_{\rm SRI}+\Delta{\cal N}_{\rm USR}+\Delta{\cal N}_{\rm SRII}=18+2+5=25$. On the other hand, for the same purpose, the total number of allowed e-foldings from Framework II is, $\Delta{\cal N}_{\rm Total}=\Delta{\cal N}_{\rm SR}+\Delta{\cal N}_{\rm USR}=18+2=20$. In both frameworks to generate large mass PBHs a sufficient number of e-foldings is not achieved for inflation.

    \item To generate small mass PBHs within the range, ${\cal O}(10^{2}-10^{3}){\rm gm}$ within the preferred effective sound speed window, $1<c_s<1.17$, total number of allowed e-foldings from the Framework I is, $\Delta{\cal N}_{\rm Total}=\Delta{\cal N}_{\rm SRI}+\Delta{\cal N}_{\rm USR}+\Delta{\cal N}_{\rm SRII}=52+2+5=59$. On the other hand, for the same purpose, the total number of allowed e-foldings from Framework II is, $\Delta{\cal N}_{\rm Total}=\Delta{\cal N}_{\rm SR}+\Delta{\cal N}_{\rm USR}=52+2=54$. In both frameworks to generate small mass PBHs a sufficient number of e-foldings is achieved for inflation.

    \item To generate large mass PBHs in the mentioned range from Framework I, the sharp transition scale from SRI to USR phase is fixed at $k_s=10^{6}{\rm Mpc}^{-1}$, end of USR phase is at $k_e=10^{7}{\rm Mpc}^{-1}$, which allows maintaining the strict criteria, $k_e/k_s=10$ to validate perturbative approximation during the one-loop computation from the USR phase. For the Framework I to get a correct DRG resummed power spectrum SRII phase and inflation has to end at $k_{\rm end}=10^{9}{\rm Mpc}^{-1}$, which helps us to maintain another requirement $k_{\rm end}/k_e=10^{2}$. On the other hand, for the same purpose as Framework II, the sharp transition scale from SR to USR phase is fixed at $k_s=10^{6}{\rm Mpc}^{-1}$, end of USR phase, as well as inflation, ends at $k_e=10^{7}{\rm Mpc}^{-1}$, where $k_e/k_s=10$ condition is strictly maintained. This possibility is discarded due to not having sufficient e-foldings for inflation.

    \item To generate small mass PBHs in the mentioned range from Framework I, the sharp transition scale from SRI to USR phase is fixed at $k_s=10^{21}{\rm Mpc}^{-1}$, end of USR phase is at $k_e=10^{22}{\rm Mpc}^{-1}$, which allows maintaining the strict criteria, $k_e/k_s=10$ to validate perturbative approximation during the one-loop computation from the USR phase. For the Framework I to get a correct DRG resummed power spectrum SRII phase and inflation has to end at $k_{\rm end}=10^{24}{\rm Mpc}^{-1}$, which helps us to maintain another requirement $k_{\rm end}/k_e=10^{2}$. On the other hand, for the same purpose as Framework II, the sharp transition scale from SR to USR phase is fixed at $k_s=10^{21}{\rm Mpc}^{-1}$, end of USR phase, as well as inflation, ends at $k_e=10^{22}{\rm Mpc}^{-1}$, where $k_e/k_s=10$ condition is strictly maintained. Here a sufficient number of e-folds has been achieved for inflation and further possibilities can be studied from this possibility.

    \item Analytically Framework I and Framework II give some distinguishable features, but numerically both frameworks produce almost the same outcome in terms of PBH mass formation. Due to having additional strict restrictions from the DRG resummation method one has to end the inflation soon because the spectrum falls sharply after reaching the peak value ${\cal O}(10^{-2})$. For this reason, in the case of Framework I, the contribution from the SRII phase is insignificant; the latter implies an additional $5$  e-foldings only.
\end{itemize}

\section{Comparison among various outcomes: Sharp vs Smooth Transition}
\label{comp}

Let us now discuss the outcomes of additional research that have been conducted to look at the same subject from a different angle. To be thorough, let us compare our findings to those of other authors. The authors recently pointed out in ref. \cite{Riotto:2023gpm,Firouzjahi:2023ahg,Firouzjahi:2023aum}  that by using a smooth transition from SRI to USR phase and USR to SRII phase, it is possible to neglect the effect from large amplitude fluctuation as seen in the one-loop corrected power spectrum for scalar modes. In these research, it is also said that it is feasible to produce enormous mass PBHs at the size, $k_{\rm PBH}=k_s\sim 10^{5}{\rm Mpc}^{-1}$, which can contain the necessary number of e-foldings to confirm inflation. More research on the same topic may be found in the references. \cite{Franciolini:2023lgy,Cheng:2023ikq,Tasinato:2023ukp,Choudhury:2023hvf,Choudhury:2023kdb,Choudhury:2023jlt,Motohashi:2023syh}. A sudden or smooth transition, in particular, has a significant impact on the predicted PBH mass and the accompanying PBH abundance. Without taking into consideration the renormalization \cite{Riotto:2023gpm,Firouzjahi:2023ahg,Firouzjahi:2023aum,Franciolini:2023lgy,Cheng:2023ikq,Tasinato:2023ukp,Choudhury:2023hvf,Choudhury:2023kdb,Motohashi:2023syh}, the one-loop corrected power spectrum is essentially unaffected for sharp transitions, but in the presence of renormalization and resummation final outcome becomes very sensitive and not at all permitting big mass PBHs creation from the underlying physical setup \cite{Choudhury:2023vuj,Choudhury:2023jlt}. The authors have explicitly demonstrated that the final result of the one-loop corrected spectrum is severely suppressed for the smooth transition in refs. \cite{Riotto:2023gpm,Firouzjahi:2023ahg,Firouzjahi:2023aum}.
However, all the previous works performed with smooth transition \cite{Riotto:2023gpm,Firouzjahi:2023ahg,Firouzjahi:2023aum} and some of the studies with sharp transition \cite{Kristiano:2022maq,Riotto:2023hoz,Kristiano:2023scm,Franciolini:2023lgy,Cheng:2023ikq,Motohashi:2023syh} have not studied the critical issue of renormalization and resummation in order to reach a final conclusion about the negligible one-loop correction and the large mass PBHs production. The strongest point of our study was accomplished in an EFT configuration with a sharp transition from SRI to USR phase and USR to SRII phase for Framework I and from SR to USR phase for Framework II, where we did the detailed study of renormalization and resummation to arrive at the final conclusion of our paper. We established all of the offered reasons in favour of our analysis in the preceding section of the study, using comprehensive calculation. 

\section{\textcolor{black}{Renormalization scheme dependence and validity of the derived result}}
\label{rs}
\textcolor{black}{Continuing the discussion elaborated in the previous sections of this paper it is important to point out that after performing the analysis stated before we have found out that the explicit mathematical form of the counter term is dependent on the renormalization scheme under consideration. We have explicitly shown that in the late-time and adiabatic/wavefunction renormalization schemes, the final determined form of the counter terms is different. However, we also found  that at least for the two mentioned  renormalization schemes the final computed results for the one-loop momentum integrals show exact equivalence, wherein in both cases one is able to completely remove quadratic UV divergence, and the result is expressed in terms of logarithmic IR divergent contribution. Such IR dependence is further smoothened via the implementation of the power spectrum renormalization scheme, which we need to implement after applying late-time or adiabatic/wave function renormalization i.e. after the complete removal of the UV divergent quadratic harmful contribution. The final conclusion we are drawing in this paper is completely based on the specific schemes of the renormalization performed in this related discussion. Apart from the mentioned schemes, (1) Late-time (LT) scheme \cite{Choudhury:2023jlt}, (2) Adiabatic-Wave function (AWF) scheme \cite{Choudhury:2023vuj}, (3) Power Spectrum (PS) scheme \cite{Choudhury:2023vuj,Choudhury:2023jlt}, 
there exist various other successful schemes of renormalization  in the literature on quantum field theory of curved space-time , which are, (4) Minimal subtraction (MS) scheme \cite{tHooft:1973mfk,Weinberg:1973xwm,Collins:1984xc} and the related modified minimal subtraction ($\overline{\rm MS}$) scheme \cite{tHooft:1973mfk,Weinberg:1973xwm,Collins:1984xc}, (5) On-shell scheme \cite{Peskin:1995ev}, (6) Bogoliubov-Parasiuk-Hepp-Zimmermann (BPHZ) scheme \cite{Dyson:1949ha,Kraus:1997bi,Piguet:1986ug,Zimmermann:1968mu,Zimmermann:1969jj}, (7) Bogoliubov-Parasiuk-Hepp-Zimmermann-Lowenstein (BPHZL) scheme \cite{Lowenstein:1975rg,Lowenstein:1975ps}, (8) Dimensional Renormalization (DR) scheme \cite{Binetruy:1980xn,Coquereaux:1979eq,Belusca-Maito:2020ala}, (9) Algebraic Renormalization (AR) scheme \cite{Adler:1969er,Batalin:1981jr,Becchi:1973gu,tHooft:1972tcz,Piguet:1995er} and there are many more in the list. We have not performed our analysis for the possibilities mentioned in (4) to (9) and have not cross-checked the applicability, correctness, and validity of the conclusion drawn in this paper toward proving a strong no-go theorem on the PBH mass. It would be interesting to carry out such an analysis in the near future. Hence at this moment, we can say that at least after considering the mentioned schemes as stated in (1) to (3) one can completely remove the quadratic UV divergence, smoothen the contribution of logarithmic IR divergence, and finally provide a no-go theorem on PBH mass by putting a constraint on the duration of the USR phase to maintain the perturbative approximations within the framework studied in this paper. For this reason, we are also not claiming any further strong statements regarding the applicability and correctness of the proposed no-go theorem on the PBH mass, which can only be made after performing the analysis for all possible classes of the mentioned renormalization schemes. However, last but not least one additional strong point we need explicitly mention before concluding in the next section. In our work we have provided the resummed version of the one-loop corrected power spectrum by making use of the DRG resummation method. This is the improved version of the well-known RG resummation method where performing the sum over all possible secular contributions i.e. using the repetitive structure in the higher order loop diagrams one can able to construct a finite and controllable amplitude of the scalar power spectrum which is perfectly consistent with cosmological $\beta$-functions and the related slow-roll hierarchy at the CMB pivot scale of the computation. As an immediate outcome, the related PBH production phenomena from the DRG resummed spectrum is completely consistent with the Renormalization Group (RG) flow and the proposed no-go theorem respects this fact in terms of the formation of PBH mass in the present context. To make further strong comments regarding the PBH production related to the present one-loop correction it would be better to incorporate the previously mentioned schemes of renormalization followed by DRG resummation, which we are planning to work on in the near future in great detail.}  

\section{Conclusion}
\label{s10}
    \begin{figure*}[htb!]
    	\centering
{
      	\includegraphics[width=16cm,height=15cm] {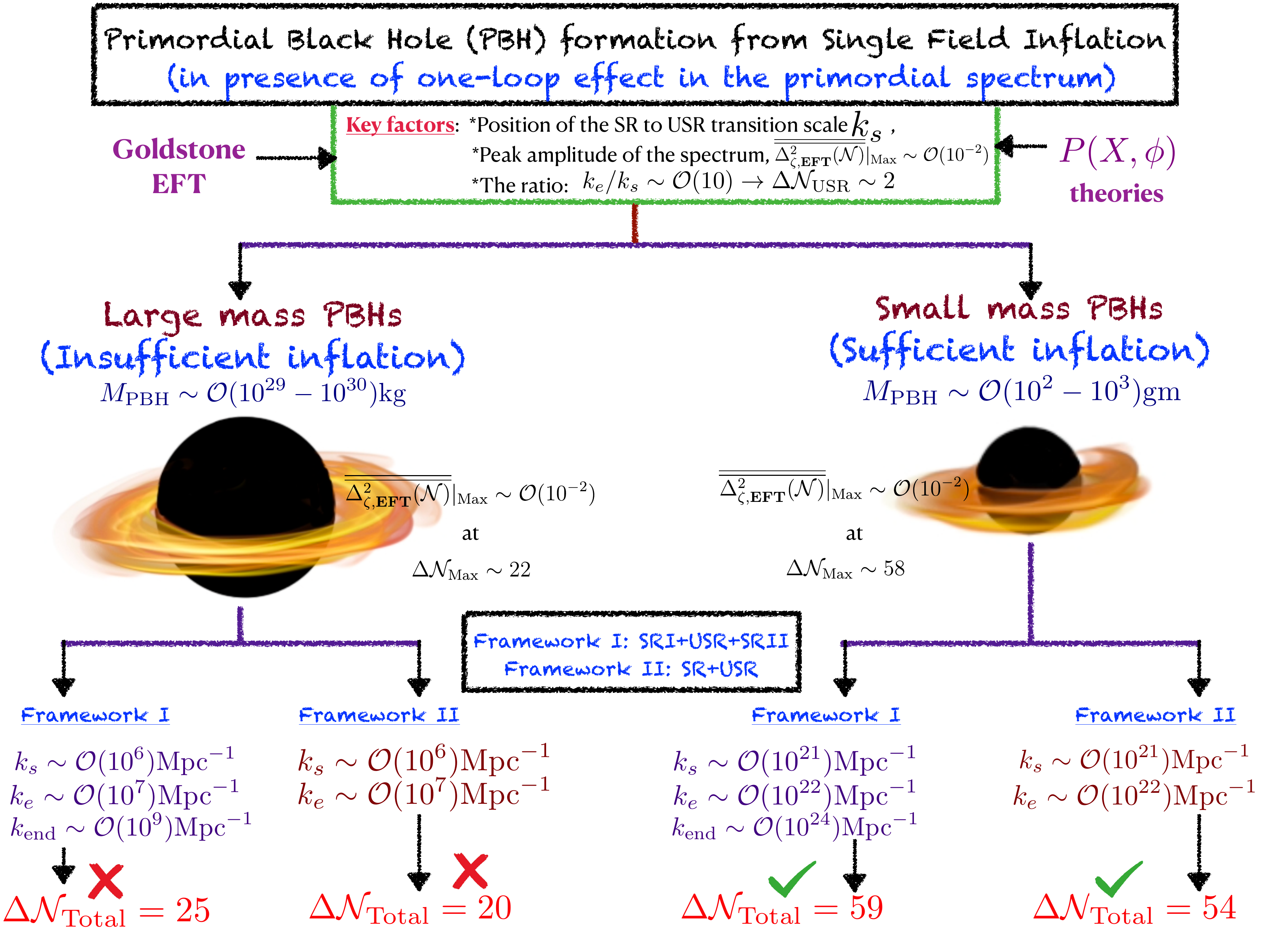}
    }
    	\caption[Optional caption for list of figures]{Schematic diagram of the total process of Primordial Black Hole (PBH) formation in the presence of one-loop correction in the primordial power spectrum for scalar modes within the framework of single field inflationary paradigm. } 
    	\label{Spectrum9}
    \end{figure*}
   
In this paper, we analyzed
in great detail, the prospects of PBH formation in the framework of a single field
inflation. To keep the discussion general and applicable to all classes of $P(X,\phi)$ theories, we used the EFT setting which allows us to draw generic model-independent conclusions. The results of our investigations have been discussed in great detail in the previous sections. We summarize here the main findings of our analysis. Let us recall that there are three important factors that play a decisive role in our analysis: (1) the size of the amplitude peak of the power spectrum necessary for PBH formation; (2) the location of the SR to USR sharp transition scale specified by $k_s$ that decides the PBH mass and (3) the ratio, $k_e/k_s\sim \mathcal{O}(10)$ (where $k_e$ relates to the end of USR for Framework I and end of USR and inflation both for the Framework II) which corresponds to approximately $2$ e-folds imposed by the validity of perturbation theory. 
If we confine to tree-level calculation, we can have PBH with large masses suitable to astrophysical (supermassive black holes)  and cosmological (dark matter and baryon asymmetry) requirements in the framework under consideration. However, the inclusion of one-loop corrections severely restricts the prospects of the formation of PBHs with generic masses. For the allowed range of sound speed, namely, $1<c_s<1.17$ (corresponding to gravitational coupling, $0\lesssim M^4_2/\dot{H}M^2_p\lesssim 0.13$), we examined the range of $k_e$ and $k_s$ constrained by the applicability of perturbative treatment. Additionally, for Framework I, there is another constraint due to having an additional SR phase after the USR phase, which is that it can only provide $k_{\rm end}/k_e\sim \mathcal{O}(10^2)$, means only $5$ e-folds are allowed for this SRII phase. We have demonstrated that in this case, the requirement of large PBH masses diminishes the prospects of inflation itself
limiting the number of e-folds much below the required ones. If we adhere to generic number of e-folds($\Delta \mathcal{N}\sim 60$), the corresponding PBH masses are small, lying within a very tiny window, $10^{2}{\rm gm}\lesssim M_{\rm PBH}\lesssim 10^{3} \rm gm$  of hardly any astrophysical/cosmological use.  We have checked in detail that adding a slow roll phase after $USR$  before the end of inflation (Framework I) does not significantly change our conclusions. It should be emphasized that the accuracy on which our conclusions are based goes beyond one loop; indeed, we used DRG re-summed power spectrum.
The details of all the crucial findings and constraints are summarized in the representative schematic figure (\ref{Spectrum9}),  provided for the sake of completeness.  
Furthermore, we can see from our analysis that both canonical and a-causal frameworks ($c_s\geq 1$) are allowed, with the a-causal one being preferred due to the maximum enhancement of the PBH power spectrum to the peak value of $\mathcal{O}(10^{-2})$. These findings strictly rule out the possibility of generating large mass PBH from all possible models of single field inflation (canonical and non-canonical), and with the help of the general model-independent EFT framework. \textcolor{black}{Here our analysis is perfectly valid at least for sharp transitions where the final constraints from the quantum loop effects are computed with the help of either Late-Time (LT) or Adiabatic-Wave function (AF) scheme followed by Power Spectrum (PS) renormalization schemes, where in the end a resummed spectrum is computed with the help of DRG resummation technique. For a better understanding  and to make a further strong comment on the applicability and the correctness in the broader sense one needs to compute the quantum loop corrections in the presence of other available renormalization schemes followed by the DRG resummation method. We are planning to pursue such a involved computation in our future projects.} It would be interesting to examine field theoretic frameworks beyond $P(X,\phi)$, such as ghost-free theories with higher derivative terms, EFT in the presence of heavy fields \cite{Arkani-Hamed:2015bza,Noumi:2012vr} and EFT of multi-fields \cite{Senatore:2010wk,Khosravi:2012qg,Shiu:2011qw}, to check whether the stringent constraints imposed by loop corrections are relaxed; we defer the same to our future investigations.




	\section*{Acknowledgements}
SC would like to thank the work-friendly environment of The Thanu Padmanabhan Centre For Cosmology and Science Popularization (CCSP), Shree Guru Gobind Singh Tricentenary (SGT) University, Gurugram, Delhi-NCR for providing tremendous support in
research and offering the Assistant Professor (Senior Grade) position.  SC would like to especially thank Soumitra SenGupta for inviting at IACS, Kolkata during the work. SC
thanks Shamik Banerjee for inviting as a plenary speaker to give a talk on "Quantum loop effects on Primordial Black
Hole Formation from Effective Field Theory of single field inflation" in the conference Current Topics in String Theory
at NISER, Bhubaneswar, India. SC also thanks Misao Sasaki, Vicharit Yingcharoenrat from IPMU, Tokyo University,
Japan and Rong-Gen Cai, Zong-Kuan Guo, Li Li, and Shi PSC, from the Institute of Theoretical Physics, Chinese
Academy of Sciences, Beijing, China for inviting to give online seminars in their Cosmology journal clubs on the
same topic. SC also thanks all
the members of our newly formed virtual international non-profit consortium Quantum Aspects of the SpaceTime \& Matter (QASTM) for elaborative discussions.  SC would like to thank The National Academy of Sciences (NASI), Prayagraj, India for being elected as a member of the academy. SP is supported by the INSA Senior scientist position at NISER, Bhubaneswar through the Grant number INSA/SP/SS/2023. The work of MS is supported by Science and Engineering Research Board (SERB), DST, Government of India under the Grant Agreement number CRG/2022/004120 (Core Research Grant). MS is also partially supported by the Ministry of Education and Science of the Republic of Kazakhstan, Grant
No. 0118RK00935 and CAS President's International Fellowship Initiative (PIFI). \textcolor{black}{The authors would like to sincerely thank the anonymous referee for her/his useful comments, which helped to improve the quality as well as the presentation of the paper.} Last but not
least, we would like to acknowledge our debt to the people belonging to the various parts
of the world for their generous and steady support for research in natural sciences.

\newpage
\section{Appendix}

\subsection{Goldstone modes from EFT and decoupling limit}
\label{s2}

The Goldstone mode ($\pi(t, {\bf x})$) transforms as follows under the time diffeomorphism symmetry:
\bea
\pi(t, {\bf x})\rightarrow\tilde{\pi}(t, {\bf x})=\pi(t, {\bf x})-\xi^{0}(t,{\bf x}).
\eea
   where the local parameter is represented by $\xi^{0}(t,{\bf x})$. These Goldstone modes serve as an analogue for the scalar modes' function in cosmic perturbation in this study. Next, the fixing condition for the appropriate unitary gauge is provided by:
   \be \pi(t,{\bf x})=0~~~~\Rightarrow~~~~\tilde{\pi}(t,{\bf x})=-\xi^{0}(t,{\bf x})~.\ee
   Now it is important to discuss the transformations caused by the broken time diffeomorphism symmetry for the space-time metric, the Ricci tensor, the Ricci scalar, the perturbation on the extrinsic curvature, the time-dependent coefficients, and the slowly varying Hubble parameter:
   \begin{enumerate}
	\item \underline{{\bf Space-time metric:}} \\ The contravariant and the covariant metrics transform as follows when the time diffeomorphism symmetry is broken:
\bea
		&&{g}^{00}\longrightarrow
	(1+\dot{\pi}(t,{\bf x}))^2 {g}^{00}+2(1+\dot{\pi}(t,{\bf x})){g}^{0 i}\partial_{i}\pi(t,{\bf x})+{g}^{ij}\partial_{i}\pi(t,{\bf x})\partial_{j}\pi(t,{\bf x}),\quad\quad\quad\\ 
		&&{g}^{0i}\longrightarrow
		(1+\dot{\pi}(t,{\bf x})){g}^{0i}+{g}^{ij}\partial_j \pi(t,{\bf x}),\\
		&&{g}^{ij}\longrightarrow{g}^{ij}.
	\\
			&&g_{00}\longrightarrow  (1+\dot{\pi}(t,{\bf x}))^2 g_{00},\\
		&&g_{0i}\longrightarrow (1+\dot{\pi}(t,{\bf x})){g}_{0i}+{g}_{00}\dot{\pi}(t,{\bf x})\partial_{i}\pi(t,{\bf x}),\\ && g_{ij}\longrightarrow{g}_{ij}+{g}_{0j}\partial_{i}\pi(t,{\bf x})+{g}_{i0}\partial_{j}\pi(t,{\bf x}).
	\eea
	\item \underline{{\bf Ricci scalar and Ricci tensor:}}  \\ The spatial component of the Ricci tensor and the Ricci scalar on a 3-hypersurface transform as follows under the broken time diffeomorphism symmetry:
\bea
			 &&{}^{(3)}R\longrightarrow\displaystyle {}^{(3)}R+\frac{4}{a^2}H(\partial^2\pi(t,{\bf x})),\\
			 &&{}^{(3)}R_{ij}\longrightarrow{}^{(3)}R_{ij}+H(\partial_{i}\partial_{j}\pi(t,{\bf x})+\delta_{ij}\partial^2\pi(t,{\bf x})).\quad
			\quad\eea
	\item \underline{{\bf Extrinsic curvature:}} \\  The trace and the spatial, time, and mixed components of the extrinsic curvature transform as follows under the broken time diffeomorphism symmetry:
\bea
	 &&\delta K\longrightarrow \displaystyle \delta K-3\pi\dot{H}-\frac{1}{a^2}(\partial^2\pi(t,{\bf x})),\\
		&&\delta K_{ij}\longrightarrow\delta K_{ij}-\pi(t,{\bf x})\dot{H}h_{ij}-\partial_{i}\partial_{j}\pi(t,{\bf x}),\\
		&&\delta K^{0}_{0}\longrightarrow\delta K^{0}_{0},\\
		&&\delta K^{0}_{i}\longrightarrow\delta K^{0}_{i},\\
		&&\delta K^{i}_{0}\longrightarrow\delta K^{i}_{0}+2Hg^{ij}\partial_j\pi(t,{\bf x}).
	\eea
	\item \underline{{\bf Time-dependent Wilson coefficients:}}  \\  The time-dependent EFT coefficients, following canonical normalisation $\pi_c(t,{\bf x})=Q^2(t)\pi(t,{\bf x})$, transform as follows under the broken time diffeomorphism symmetry:
\bea		
		&&Q(t)\longrightarrow Q(t+\pi(t,{\bf x}))\displaystyle \sum^{\infty}_{n=0}\underbrace{\frac{\pi^{n}_c(t,{\bf x})}{n!Q^{2n}(t)}}_{{\bf \ll 1}}\frac{d^{n}Q(t)}{dt^n}\approx Q(t)~.\quad\quad\eea
The time-dependent EFT action coefficients are represented here by $Q(t)$.

\item \underline{{\bf Hubble parameter:}} \\  When the temporal diffeomorphism symmetry is broken, the Hubble parameter changes to:
\bea
	H(t)\longrightarrow H(t+\pi(t,{\bf x}))&=&\displaystyle \sum^{\infty}_{n=0}\frac{\pi^{n}}{n!}\frac{d^{n}H(t)}{dt^n}=\displaystyle\left[1-\underbrace{\pi(t,{\bf x}) H(t) \epsilon-\frac{\pi^2(t,{\bf x})H(t)}{2}\left(\dot{\epsilon}-2\epsilon^2\right)+\cdots}_{{\bf sub-leading~contributions}}\right]H(t)~.\quad\quad\quad\eea
\end{enumerate}
We must now comprehend the decoupling limit in greater depth in order to build the EFT action. The gravity and Goldstone modes' mixing contributions in this limit are easily disregarded. Let's start with the EFT operator $-\dot{H}M_{pl}^2g^{00}$, which is necessary for further computation, to demonstrate the veracity of this assertion.

This operator undergoes the following transformation under the broken time diffeomorphism symmetry:
\begin{widetext}
\bea &&-\dot{H}M_{pl}^2g^{00}\longrightarrow -\dot{H}M_{pl}^2\bigg[ (1+\dot{\pi}(t,{\bf x}))^2g^{00}
+\left(2(1+\dot{\pi}(t,{\bf x}))\partial_i \pi(t,{\bf x}) g^{0i}+g^{ij}\partial_i\pi(t,{\bf x}) \partial_j \pi(t,{\bf x})\right)\bigg].\eea
\end{widetext}
The temporal part of the metric can be expressed as follows now that it has been perturbed:
\be g^{00}=\bar{g}^{00}+\delta g^{00},\ee
where $\bar{g}^{00}=-1$ denotes the temporal component of the background quasi-de Sitter metric and $\delta g^{00}$ denotes the perturbation. The remaining contributions are a kinetic contribution $M_{pl}^2\dot{H}\dot{\pi}^2\bar{g^{00}}$ and a mixing contribution $M_{pl}^2\dot{H}\dot{\pi}\delta g^{00}$.
A canonical normalised metric perturbation is also used,  \be \delta g^{00}_c=M_{pl}\delta g^{00}, \ee and mixing contribution is given by,  
\be M_{pl}^2\dot{H}\dot{\pi}\delta g^{00}= \sqrt{\dot{H}}\dot{\pi}_c\delta g^{00}_c.\ee  In the decoupling limit, one may conveniently ignore this mixing term above the energy scale, \be E_{mix}=\sqrt{\dot{H}}.\ee  Another option is to combine the following two crucial contributions:
\begin{enumerate}
    \item Term contains the operator $M_{pl}^2\dot{H}\dot{\pi}^2\delta{g^{00}}$,

    \item  Term contains the operator $\pi M_{pl}^2\ddot{H}\dot{\pi}\bar{g}^{00}$,
\end{enumerate}
which may be represented as follows after canonical normalization,  
\bea &&M_{pl}^2\dot{H}\dot{\pi}^2\delta{g^{00}}=\dot{\pi}_c^2\delta{g^{00}_c}/M_{pl}, \\
&&\pi M_{pl}^2\ddot{H}\dot{\pi}\bar{g}^{00}=\ddot{H}\pi_c\dot{\pi}_c\bar{g}^{00}/\dot{H}.\eea
The contribution from the $M_{pl}^2\dot{H}\dot{\pi}\delta{g^{00}}$ term can be disregarded for $E>E_{mix}$. The following simplification applies to the decoupling limit:
\bea -\dot{H}M_{pl}^2g^{00}\rightarrow -\dot{H}M_{pl}^2g^{00}\left[\dot{\pi}^2-\frac{1}{a^2}(\partial_i\pi)^2\right].\eea
The streamlined expression indicated above can be used for additional calculations because it is mentioned directly in the EFT action. 
  
\subsection{Connecting generalized $P(X,\phi)$ theory with Goldstone modes of EFT in the framework of Cosmological Perturbation Theory }\label{app:P}	

In this subsection our objective is to make a clear connection between the generalized $P(X,\phi)$ theory with the Goldstone modes of the EFT set-up which appears as an outcome of time diffeomorphism breaking in the unitary gauge of cosmological perturbation. In the decoupling limit the mixing contributions of the metric fluctuations can be neglected very easily and one can write the following for the inflaton perturbation $\delta\phi(t,{\bf x})$ in an unperturbed space-time:
\bea X=\bar{X}+\delta X,\eea
where the background and the perturbation on $X$ can be written as:
\bea \bar{X}=-\frac{1}{2}\dot{\bar{\phi}}^2_0(t)\quad\quad{\rm and}\quad\quad \delta X=\dot{\bar{\phi}}_0(t)\partial_{t}\left(\delta \phi(t,{\bf x})\right)-\frac{1}{2}(\partial_{\mu}\left(\delta \phi(t,{\bf x})\right))^2\eea
Further using the connecting relationship between inflaton perturbation $\delta\phi(t,{\bf x})$ and the Goldstone mode $\pi(t,{\bf x})$, which is given by:
\bea \delta\phi(t,{\bf x})=\dot{\bar{\phi}}_0(t)\pi(t,{\bf x}),\eea and ignoring the derivative contributions of $\dot{\bar{\phi}}_0(t)$ (slow-roll suppressed) we get the following simplified result of the perturbation on $X$, which is given by:
\bea \delta X\approx 2\bar{X}\bigg[\dot{\pi}-\frac{1}{2}\left(\partial_{\mu}\pi\right)^2\bigg]=2\bar{X}\bigg[\dot{\pi}+\frac{1}{2}\dot{\pi}^2-\frac{1}{2a^2}\left(\partial_{i}\pi\right)^2\bigg].\eea
Further we consider the Taylor series expansion of the function $P(X,\phi)$, which gives:
\bea P(X,\phi)=P(\bar{X},\bar{\phi}_0)+P_{,\bar{X}}\delta X+\frac{1}{2!}P_{,\bar{X}\bar{X}}\left(\delta X\right)^2+\frac{1}{3!}P_{,\bar{X}\bar{X}\bar{X}}\left(\delta X\right)^3+\cdots,\eea
which can be expressed in terms of the Goldstone fluctuation by the following expression:
\bea P(X,\phi)=-\bar{X}P_{,\bar{X}}\left(\partial_{\mu}\pi\right)^2+2\bar{X}^2P_{,\bar{X}\bar{X}}\bigg(\dot{\pi}^2-\dot{\pi}\left(\partial_{\mu}\pi\right)^2+\cdots\bigg)+\frac{4}{3}\bar{X}^3 P_{,\bar{X}\bar{X}\bar{X}}\bigg(\dot{\pi}^3+\cdots\bigg)+\cdots,\eea
Now using the background equation of motion we get:
\bea \label{back} \bar{X}P_{,\bar{X}}=-M^2_{pl}\dot{H},\eea
which will going to fix the first expansion coefficient in the above mentioned Taylor series. However, the rest of the coefficient in the above mentioned series expansion cannot be fixed. Using a particular cosmological paradigm one can put constraints on these coefficients.

One can further recast the above $P(X,\phi)$ function in terms of the following simplified form of Goldstone action:
\bea \label{pxpi}
P(X,\phi)=-M^2_{pl}\dot{H}\left(\partial_{\mu}\pi\right)^2+2M^4_2\bigg(\dot{\pi}^2-\dot{\pi}\left(\partial_{\mu}\pi\right)^2+\cdots\bigg)+\frac{4}{3}M^3_4\bigg(\dot{\pi}^3+\cdots\bigg)+\cdots,\eea
where all the coefficients except the first term (since it is already fixed from background equation of motion) can be expressed in terms of the derivatives of the $P(X,\phi)$ function as:
\bea M^4_n=\bar{X}^n \left(\frac{d^nP}{d\bar{X}^n}\right)\quad\forall n=2,3,4,\cdots.\eea
For this reason we can immediately write:
\bea \label{coeff} M^4_2=\bar{X}^2 \left(\frac{d^2P}{d\bar{X}^2}\right)=\bar{X}^2P_{,\bar{X}\bar{X}},\quad M^4_3=\bar{X}^3 \left(\frac{d^3P}{d\bar{X}^3}\right)=\bar{X}^3P_{,\bar{X}\bar{X}\bar{X}}.\eea
Let us now comment on the numerical values of these EFT coefficients which helps us to connect with some known physical frameworks:
\begin{enumerate}
    \item First of all, if we fix the coefficients $M_2=0$ and $M_3=0$ then we can explain the canonical single field slow roll inflationary paradigm where the single scalar field kinetic term $X$ and the effective potential $V(\phi)$. We need to additionally use the decoupling limit approximation, which fix, $\dot{H}\rightarrow 0$ and $M_{pl}\rightarrow \infty$ such that the following constraint is satisfied:
    \bea M^2_{pl}\dot{H}={\rm Constant}.\eea
    In this case the the perturbations are treated to be completely Gaussian and no small non-Gaussian as well as any higher order contributions will not appear in the one-loop corrected power spectrum and in the computation of higher point cosmological correlation functions.

    \item From the structure of the obtained equation (\ref{pxpi}) it is clearly observed that in this computation the $M_2$ coefficient is treated to be correction to the leading order coefficient $-M^2_{pl}\dot{H}$, because both are associated with the Goldstone operator $\dot{\pi}^2$ in the present context. However with the spatial part of the gradient operator of the Goldstone mode, $(\partial_i\pi)^2/a^2$ only the background dependent fixed coefficient, $M^2_{pl}\dot{H}$ is associated. Coefficient $M_2$ is not attached with such contributions. This is the direct outcome of the time diffeomorphism symmetry breaking and as a byproduct an effective non-trivial sound speed $c_s$ is generated, in terms of which the second order contribution to the Goldstone action can be further written in the following simplified form:
    \bea {\cal L}^{(2)}_{\pi}=-\frac{M^2_{pl}\dot{H}}{c^2_s}\bigg(\dot{\pi}^2-c^2_s\frac{(\partial_i\pi)^2}{a^2}\bigg),\eea
    where the effective sound speed parameter $c_s$ is defined as:
    \bea \label{cs1} c_s\equiv \frac{1}{\displaystyle \sqrt{1-\frac{2M^4_2}{\dot{H}M^2_{pl}}}}.\eea
\end{enumerate}
Now replacing the value of $M^2_{pl}\dot{H}$ with the background equation of motion as stated in equation (\ref{back}) and using the definition of the higher order Taylor expansion coefficients as mentioned in equation (\ref{coeff}), the sound speed can be further expressed in terms of the $P(X,\phi)$ function and in terms of its derivatives as:
\bea \label{cs2} c_s=\frac{1}{\displaystyle \sqrt{1+\frac{2\bar{X}^2P_{,\bar{X}\bar{X}}}{\bar{X}P_{,\bar{X}}}}}=\sqrt{\frac{P_{,\bar{X}}}{P_{,\bar{X}}+2\bar{X}P_{,\bar{X}\bar{X}}}} \eea
Now comparing equation (\ref{cs1}) and equation (\ref{cs2}) one can express the EFT coefficient $M_2$ in terms of the derivatives of the $P(X,\phi)$ function, which is given by the following expression:
\be \boxed{\frac{M^4_2}{\dot{H}M^2_{pl}}=\frac{1}{2}\Bigg(1-\frac{P_{,\bar{X}}}{P_{,\bar{X}}+2\bar{X}P_{,\bar{X}\bar{X}}}\Bigg)=\Bigg(\frac{\bar{X}P_{,\bar{X}\bar{X}}}{P_{,\bar{X}}+2\bar{X}P_{,\bar{X}\bar{X}}}\Bigg)}.\ee
The above equation is extremely helpful for connecting a given $P(X,\phi)$ theory with the Goldstone EFT action in terms of the coefficients and it will also going tell us that how the constraints on the sound speed will restrict the model parameters.

\subsection{PBH mass constraints on various $P(X,\phi)$ models}\label{app:PBH}

To visualize the clear connection between $P(X,\phi)$ models with the Goldstone EFT model within the framework of single field inflation and to know about the constraints on PBH mass we are now going to investigate this issue in detail. 

\subsubsection{Dirac-Born-Infeld (DBI) model} 

    Let us start with the Dirac-Born-Infeld (DBI) model where the $P(X,\phi)$ function is given by the following expression:
    \bea P(X,\phi)=-\frac{\Lambda^4}{f(\phi)}\sqrt{1-\frac{f(\phi)}{\Lambda^4}X}+\frac{\Lambda^4}{f(\phi)}-V(\phi)\quad\quad {\rm where}\quad\quad X=-\frac{1}{2}\left(\partial_{\mu}\phi\right)^2.\eea
    Here $f(\phi)$, $\Lambda^4$ and $V(\phi)$ depends on the details of the underlying String Theory framework on $D3$ brane. 
    From this model the effective sound speed $c_s$ and the EFT coefficient $M_2$ can be evaluated as:
    \bea c_s=\frac{1}{\displaystyle \sqrt{1-2\frac{f(\bar{\phi}_0)}{\Lambda^4}\bar{X}}}\quad\quad {\rm and}\quad \quad \frac{M^4_2}{\dot{H}M^2_{pl}}=\frac{f(\bar{\phi}_0)}{\Lambda^4}\bar{X}.\eea 
    It is expected that for this model the effective sound speed has to be less than unity and from our analysis in this paper we found that $c_s>0.6$. Which implies the following constraint is valid for the DBI model of inflation:
 \bea -0.89\frac{\Lambda^4}{f(\bar{\phi}_0)}<\bar{X}<0.\eea

  Now in the context of DBI model the estimation of PBH mass is given by the following expression:
\bea \left(M_{\rm PBH}\right)_{\rm DBI}=9.04\times 10^2{\rm gm}\times \frac{1}{\displaystyle \bigg(1-2\frac{f(\bar{\phi}_0)}{\Lambda^4}\bar{X}\bigg)}\times \bigg(\frac{\gamma}{0.2}\bigg)\bigg(\frac{g_*}{106.75}\bigg)^{-1/6},\eea 
where we fix $k_s=10^{21}\;{\rm Mpc}^{-1}$ and pivot scale at $p_*=0.02\;{\rm Mpc}^{-1}$. Now fixing the efficiency factor $\gamma=0.2$ and degrees of freedom $g_*=106.75$ we get the following bound on the PBH mass from the DBI inflationary model:
\bea 0.36\times 10^2{\rm gm}<\left(M_{\rm PBH}\right)_{\rm DBI}<9.04\times 10^2{\rm gm}.\eea

\subsubsection{Tachyon model} 

    Next we discuss about the Tachyon model where the $P(X,\phi)$ function is given by the following expression:
    \bea P(X,\phi)=-V(\phi)\sqrt{1-2\alpha^{'}X}\quad\quad {\rm where}\quad\quad X=-\frac{1}{2}\left(\partial_{\mu}\phi\right)^2.\eea
    Here $\alpha^{'}$ is the Regge slope parameter which is inverse of the string tension $T$ and $V(\phi)$ depends on the details of the String Theory model under consideration.
    From this model the effective sound speed $c_s$ and the EFT coefficient $M_2$ can be evaluated as:
    \bea c_s=\frac{1}{\displaystyle \sqrt{1-2\alpha^{'}\bar{X}}}\quad\quad {\rm and}\quad \quad \frac{M^4_2}{\dot{H}M^2_{pl}}=\alpha^{'}\bar{X}.\eea 
    It is expected that for this model the effective sound speed has to be less than unity and from our analysis in this paper we found that $c_s>0.6$. Which implies the following constraint is valid for the Tachyon model of inflation:
 \bea -\frac{0.89}{\alpha^{'}}<\bar{X}<0.\eea

  Now in the context of Tachyon model the estimation of PBH mass is given by the following expression:
\bea \left(M_{\rm PBH}\right)_{\rm Tachyon}=9.04\times 10^2{\rm gm}\times \frac{1}{\displaystyle \bigg(1-2\alpha^{'}\bar{X}\bigg)}\times \bigg(\frac{\gamma}{0.2}\bigg)\bigg(\frac{g_*}{106.75}\bigg)^{-1/6},\eea 
where we fix $k_s=10^{21}\;{\rm Mpc}^{-1}$ and pivot scale at $p_*=0.02\;{\rm Mpc}^{-1}$. Now fixing the efficiency factor $\gamma=0.2$ and degrees of freedom $g_*=106.75$ we get the following bound on the PBH mass from the Tachyon inflationary model:
\bea 0.36\times 10^2{\rm gm}<\left(M_{\rm PBH}\right)_{\rm Tachyon}<9.04\times 10^2{\rm gm}.\eea

\subsubsection{GTachyon model} 

     Next we discuss about the GTachyon model where the $P(X,\phi)$ function is given by the following expression:
    \bea P(X,\phi)=-V(\phi)\left(1-2\alpha^{'}X\right)^{q}\quad\quad {\rm where}\quad\quad X=-\frac{1}{2}\left(\partial_{\mu}\phi\right)^2.\eea
    This is treated to be the generalized version of the Tachyon model which can be derived from String Theory set up. 
    For this model inflation can happen in branch of values of the index $q$, which are (1) $q<1/2$, (2) $1/2<q<1$ and (3) $q>1$ in the present context. The result for $q=1/2$ will exactly matches with the predictions of Tachyon model as discussed above.
    
    From this model the effective sound speed $c_s$ and the EFT coefficient $M_2$ can be evaluated as:
    \bea c_s=\sqrt{\frac{ \displaystyle 1+2\alpha^{'}(1-2q)\bar{X}}{\displaystyle 1-2\alpha^{'}\bar{X} }}\quad\quad {\rm and}\quad \quad \frac{M^4_2}{\dot{H}M^2_{pl}}=\bigg(\frac{2\alpha^{'}(1-q)\bar{X}}{1+2\alpha^{'}(1-2q)\bar{X}}\bigg).\eea 
    It is expected that for this model the effective sound speed has to be less than unity for (1) $q<1/2$, (2) $1/2<q<1$ and from our analysis in this paper we found that $c_s>0.6$. Which implies the following constraint is valid for the GTachyon model of inflation for the $q$ parameter, described in the region, $q<1/2$, and $1/2<q<1$, given by:
 \bea -\frac{0.89}{2\alpha^{'}\left[\left(1-q\right)+0.89\left(1-2q\right)\right]}<\bar{X}<0 \quad\quad\quad\quad {\rm for}\quad q<\frac{1}{2}\quad {\rm and}\quad \frac{1}{2}<q<1.\eea
 However for $q>1$ it can violate causality and exceeds the sound speed beyond unity. Which implies the following constraint is valid for the GTachyon model of inflation for the $q$ parameter, described in the region, $q>1$, given by:
 \bea 0<\bar{X}<\frac{0.28}{2\alpha^{'}\left[\left(1-q\right)-0.28\left(1-2q\right)\right]} \quad\quad\quad\quad {\rm for}\quad q>1.\eea

  Now in the context of GTachyon model the estimation of PBH mass is given by the following expression in terms of any allowed values of $q$ parameter for GTachyon:
\bea \left(M_{\rm PBH}\right)_{\rm GTachyon}=9.04\times 10^2{\rm gm}\times \bigg(\frac{ \displaystyle 1+2\alpha^{'}(1-2q)\bar{X}}{\displaystyle 1-2\alpha^{'}\bar{X} }\bigg)\times \bigg(\frac{\gamma}{0.2}\bigg)\bigg(\frac{g_*}{106.75}\bigg)^{-1/6},\eea 
where we fix $k_s=10^{21}\;{\rm Mpc}^{-1}$ and pivot scale at $p_*=0.02\;{\rm Mpc}^{-1}$. Now fixing the efficiency factor $\gamma=0.2$ and degrees of freedom $g_*=106.75$ we get the following bound on the PBH mass from the GTachyon inflationary model:
\bea && 0.36\times 10^2{\rm gm}<\left(M_{\rm PBH}\right)_{\rm GTachyon}<9.04\times 10^2{\rm gm}\quad\quad\quad\quad {\rm for}\quad q<\frac{1}{2}\quad {\rm and}\quad \frac{1}{2}<q<1,\\
&& 9.04\times 10^2{\rm gm}<\left(M_{\rm PBH}\right)_{\rm GTachyon}<2.03\times 10^3{\rm gm}\quad\quad\quad\quad {\rm for}\quad q>1.\eea

\subsubsection{$K$ inflation model} 
 
 Next we discuss about the $K$ inflation model where the $P(X,\phi)$ function is given by the following expression:
    \bea P(X,\phi)=K(X)-V(\phi)\quad\quad {\rm where}\quad\quad X=-\frac{1}{2}\left(\partial_{\mu}\phi\right)^2.\eea
    Here $V(\phi)$ is a given effective potential which will attribute to single field inflation. However, the generalized kinetic term $K(X)$ is a general function of $X$ and any arbitrary form of this function is allowed which is consistent with observational constraints. Here we consider a specific example of this function, which is given by the following expression:
    \bea K(X)=\gamma_n X^n,\eea
    where $n$ is a integer number and $\gamma_n$ is an overall dimensionfull constant for the $n$-th contribution in the kinetic interaction. In the simplest model we have not consider the sum over the index $n$ and appearing as a monomial type of kinetic interaction in the present context of discussion.

     From this model the effective sound speed $c_s$ and the EFT coefficient $M_2$ can be evaluated as:
    \bea c_s=\frac{1}{\sqrt{2n-1}}\quad\quad {\rm and}\quad \quad \frac{M^4_2}{\dot{H}M^2_{pl}}=\bigg(\frac{n-1}{2n-1}\bigg).\eea 
    To have real finite value of the effective sound speed here one have an additional constraint on this index $n$, which is the following:
    \bea n>\frac{1}{2}.\eea
    It is expected that for this model the effective sound speed has to be less than unity for certain value of $n$ and from our analysis in this paper we found that $c_s>0.6$. Which implies the following constraint is valid for the $K$-inflation model, given by:
 \bea 1.89>n>1.\eea
 However causality is violated and exceeds the sound speed beyond unity when:
 \bea 1>n>0.72.\eea

  Now in the context of $K$-inflation model the estimation of PBH mass is given by the following expression in terms of any allowed values of $n$ parameter:
\bea \left(M_{\rm PBH}\right)_{\rm K}=9.04\times 10^2{\rm gm}\times \bigg(\frac{ 1}{\displaystyle 2n-1}\bigg)\times \bigg(\frac{\gamma}{0.2}\bigg)\bigg(\frac{g_*}{106.75}\bigg)^{-1/6},\eea 
where we fix $k_s=10^{21}\;{\rm Mpc}^{-1}$ and pivot scale at $p_*=0.02\;{\rm Mpc}^{-1}$. Now fixing the efficiency factor $\gamma=0.2$ and degrees of freedom $g_*=106.75$ we get the following bound on the PBH mass from the GTachyon inflationary model:
\bea && 9.04\times 10^2{\rm gm}<\left(M_{\rm PBH}\right)_{\rm K}<2.51\times 10^3{\rm gm}\quad\quad\quad\quad {\rm for}\quad 1>n>0.68,\\
&& 3.96\times 10^2{\rm gm}<\left(M_{\rm PBH}\right)_{\rm K}<9.04\times 10^2{\rm gm}\quad\quad\quad\quad {\rm for}\quad 1.64>n>1.\eea

    \subsubsection{Canonical single field model}

    Finally, we discuss about the canonical single field inflation model where the $P(X,\phi)$ function is given by the following expression:
    \bea P(X,\phi)=X-V(\phi)\quad\quad {\rm where}\quad\quad X=-\frac{1}{2}\left(\partial_{\mu}\phi\right)^2.\eea
    Here $V(\phi)$ is a given effective potential that will attribute to single field inflation. 

     From this model the effective sound speed $c_s$ and the EFT coefficient $M_2$ can be evaluated as:
    \bea c_s=1\quad\quad {\rm and}\quad \quad \frac{M^4_2}{\dot{H}M^2_{pl}}=0.\eea 
   
  Now in the context of the canonical single-field inflation model, the estimation of PBH mass is given by the following expression:
\bea \left(M_{\rm PBH}\right)_{\rm K}=9.04\times 10^2{\rm gm}\times \bigg(\frac{\gamma}{0.2}\bigg)\bigg(\frac{g_*}{106.75}\bigg)^{-1/6},\eea 
where we fix $k_s=10^{21}\;{\rm Mpc}^{-1}$ and pivot scale at $p_*=0.02\;{\rm Mpc}^{-1}$. Now fixing the efficiency factor $\gamma=0.2$ and degrees of freedom $g_*=106.75$ we get the following value of the PBH mass from the canonical single field inflationary model:
\bea \left(M_{\rm PBH}\right)_{\rm K}=9.04\times 10^2{\rm gm}.\eea

\newpage
\subsection{Computation of time-dependent integration one-loop integrals in the USR regime}\label{app:A}	

In this present calculation, we consider two sharp transitions at the conformal time scales, $\tau=\tau_s$ (SRII to USR) and $\tau=\tau_e$ (USR to SRII), where we define the following equation:
\bea \label{rrf} \partial_{\tau}\left(\frac{\eta(\tau)}{c^2_s(\tau)}\right)\approx\frac{\Delta \eta(\tau)}{c^2_s(\tau)}\bigg(\underbrace{\delta(\tau-\tau_e)}_{\bf USR\rightarrow SRII}-\underbrace{\delta(\tau-\tau_s)}_{\bf SRI\rightarrow USR}\bigg).\eea
Further using equation (\ref{rrf}) we now write down expressions for the two-point correlation functions for the primordial scalar perturbation in the USR period in a more compact fashion, which is given by the following expression:
    \bea &&\int^{0}_{-\infty}d\tau\;\frac{1}{c^2_s(\tau)}\partial_{\tau}\left(\frac{\eta(\tau)}{c^2_s(\tau)}\right)\; {\cal J}_{1}(\tau)
 =\bigg(\frac{\Delta \eta(\tau_e)}{c^2_s}\; {\cal J}_{1}(\tau=\tau_e)-\frac{\Delta \eta(\tau_s)}{c^2_s}\; {\cal J}_{1}(\tau=\tau_s)\bigg)-\underbrace{\int^{0}_{-\infty}d\tau\;\left(\frac{\eta(\tau)}{c^2_s(\tau)}\right)\; {\cal J}^{'}_{1}(\tau)}_{\approx 0}\nonumber\\&&\quad\quad\quad\quad\quad\quad\quad\quad\quad\quad\quad\quad\quad\quad\approx\bigg(\frac{\Delta \eta(\tau_e)}{c^2_s}\; {\cal J}_{1}(\tau_e)-\frac{\Delta \eta(\tau_s)}{c^2_s}\; {\cal J}_{1}(\tau_s)\bigg),\\
 &&\int^{0}_{-\infty}d\tau_1\;\int^{0}_{-\infty}d\tau_2\;\frac{1}{c^2_s(\tau_1)}\frac{1}{c^2_s(\tau_2)}\partial_{\tau_1}\left(\frac{\eta(\tau_1)}{c^2_s(\tau_1)}\right)\;\partial_{\tau_2}\left(\frac{\eta(\tau_2)}{c^2_s(\tau_2)}\right)\; {\cal J}_{2}(\tau_1,\tau_2)\nonumber\\
&&=\int^{0}_{-\infty}d\tau_2\frac{1}{c^2_s(\tau_2)}\partial_{\tau_2}\left(\frac{\eta(\tau_2)}{c^2_s(\tau_2)}\right)\;\bigg(\frac{\Delta \eta(\tau_e)}{c^4_s}\; {\cal J}_{2}(\tau_1=\tau_e,\tau_2)-\frac{\Delta \eta(\tau_s)}{c^4_s}\; {\cal J}_{2}(\tau_1=\tau_s,\tau_2)\bigg)\nonumber\\
&&\quad\quad-\underbrace{\int^{0}_{-\infty}d\tau_1\;\int^{0}_{-\infty}d\tau_2\;\frac{1}{c^2_s(\tau_1)}\frac{1}{c^2_s(\tau_2)}\left(\frac{\eta(\tau_1)}{c^2_s(\tau_1)}\right)\;\partial_{\tau_2}\left(\frac{\eta(\tau_2)}{c^2_s(\tau_2)}\right)\; \partial_{\tau_1}{\cal J}_{2}(\tau_1,\tau_2)}_{\approx 0}\\&&\approx \bigg(\frac{\Delta \eta(\tau_e)}{c^8_s}\; {\cal J}_{2}(\tau_1=\tau_e,\tau_2=\tau_e)-\frac{\Delta \eta(\tau_s)}{c^8_s}\; {\cal J}_{2}(\tau_1=\tau_s,\tau_2=\tau_s)\bigg)\nonumber\\&&-\underbrace{\int^{0}_{-\infty}d\tau_2\left(\frac{\eta(\tau_2)}{c^4_s(\tau_2)}\right)\;\bigg(\frac{\Delta \eta(\tau_e)}{c^4_s}\; \partial_{\tau_2}{\cal J}_{2}(\tau_1=\tau_e,\tau_2)-\frac{\Delta \eta(\tau_s)}{c^4_s}\; \partial_{\tau_2}{\cal J}_{2}(\tau_1=\tau_s,\tau_2)\bigg)}_{\approx 0}\approx \bigg(\frac{\Delta \eta(\tau_e)}{c^8_s}\; {\cal J}_{2}(\tau_e)-\frac{\Delta \eta(\tau_s)}{c^8_s}\; {\cal J}_{2}(\tau_s)\bigg),\nonumber
 \eea
where it is important to note that the integral kernels ${\cal J}_2$ and ${\cal J}_1$ are given by the following expressions:
\bea {\cal J}_2(\tau_1,\tau_2):=\footnotesize\left\{
	\begin{array}{ll}
		\displaystyle -\frac{M^4_{ pl}}{4}a^2(\tau_1)\epsilon(\tau_1)a^2(\tau_2)\epsilon(\tau_2)\nonumber\\
 \displaystyle\times\int \frac{d^{3}{\bf k}_1}{(2\pi)^3} \int \frac{d^{3}{\bf k}_2}{(2\pi)^3} \int \frac{d^{3}{\bf k}_3}{(2\pi)^3} \int \frac{d^{3}{\bf k}_4}{(2\pi)^3} \int \frac{d^{3}{\bf k}_5}{(2\pi)^3} \int \frac{d^{3}{\bf k}_6}{(2\pi)^3}\nonumber\\
  \displaystyle\times \delta^3\bigg({\bf k}_1+{\bf k}_2+{\bf k}_3\bigg) \delta^3\bigg({\bf k}_4+{\bf k}_5+{\bf k}_6\bigg)\nonumber\\
 \displaystyle\times \langle \hat{\zeta}_{\bf p}\hat{\zeta}_{-{\bf p}}\hat{\zeta}^{'}_{{\bf k}_1}(\tau_1)\hat{\zeta}_{{\bf k}_2}(\tau_1)\hat{\zeta}_{{\bf k}_3}(\tau_1)\hat{\zeta}^{'}_{{\bf k}_4}(\tau_2)\hat{\zeta}_{{\bf k}_5}(\tau_2)\hat{\zeta}_{{\bf k}_6}(\tau_2)\rangle
  & \mbox{from equation} 
  (\ref{g2})  \\ 
			 \displaystyle-\frac{M^4_{ pl}}{4}a^2(\tau_1)\epsilon(\tau_1)a^2(\tau_2)\epsilon(\tau_2)\nonumber\\
 \displaystyle\times\int \frac{d^{3}{\bf k}_1}{(2\pi)^3} \int \frac{d^{3}{\bf k}_2}{(2\pi)^3} \int \frac{d^{3}{\bf k}_3}{(2\pi)^3} \int \frac{d^{3}{\bf k}_4}{(2\pi)^3} \int \frac{d^{3}{\bf k}_5}{(2\pi)^3} \int \frac{d^{3}{\bf k}_6}{(2\pi)^3}\nonumber\\
  \displaystyle\times \delta^3\bigg({\bf k}_1+{\bf k}_2+{\bf k}_3\bigg) \delta^3\bigg({\bf k}_4+{\bf k}_5+{\bf k}_6\bigg)\nonumber\\
 \displaystyle\times \langle \hat{\zeta}_{\bf p}\hat{\zeta}_{-{\bf p}}\hat{\zeta}^{'}_{{\bf k}_1}(\tau_1)\hat{\zeta}_{{\bf k}_2}(\tau_1)\hat{\zeta}_{{\bf k}_3}(\tau_1)\hat{\zeta}^{'}_{{\bf k}_4}(\tau_2)\hat{\zeta}_{{\bf k}_5}(\tau_2)\hat{\zeta}_{{\bf k}_6}(\tau_2)\rangle^{\dagger}
  & \mbox{from equation} 
  (\ref{g3})
  \\ 
			 \displaystyle-\frac{M^4_{ pl}}{4}a^2(\tau_1)\epsilon(\tau_1)a^2(\tau_2)\epsilon(\tau_2)\nonumber\\
 \displaystyle\times\int \frac{d^{3}{\bf k}_1}{(2\pi)^3} \int \frac{d^{3}{\bf k}_2}{(2\pi)^3} \int \frac{d^{3}{\bf k}_3}{(2\pi)^3} \int \frac{d^{3}{\bf k}_4}{(2\pi)^3} \int \frac{d^{3}{\bf k}_5}{(2\pi)^3} \int \frac{d^{3}{\bf k}_6}{(2\pi)^3}\nonumber\\
  \displaystyle\times \delta^3\bigg({\bf k}_1+{\bf k}_2+{\bf k}_3\bigg) \delta^3\bigg({\bf k}_4+{\bf k}_5+{\bf k}_6\bigg)\nonumber\\
 \displaystyle\times \langle \hat{\zeta}^{'}_{{\bf k}_1}(\tau_1)\hat{\zeta}_{{\bf k}_2}(\tau_1)\hat{\zeta}_{{\bf k}_3}(\tau_1)\hat{\zeta}_{\bf p}\hat{\zeta}_{-{\bf p}}\hat{\zeta}^{'}_{{\bf k}_4}(\tau_2)\hat{\zeta}_{{\bf k}_5}(\tau_2)\hat{\zeta}_{{\bf k}_6}(\tau_2)\rangle
  & \mbox{from equation} 
  (\ref{g4})
	\end{array}
\right.\eea

\bea {\cal J}_1(\tau):=\footnotesize\left\{
	\begin{array}{ll}
		\displaystyle -\frac{iM^2_{ pl}}{2}a^2(\tau)\epsilon(\tau)\times\int \frac{d^{3}{\bf k}_1}{(2\pi)^3} \int \frac{d^{3}{\bf k}_2}{(2\pi)^3} \int \frac{d^{3}{\bf k}_3}{(2\pi)^3} \nonumber\\
  \displaystyle \times\delta^3\bigg({\bf k}_1+{\bf k}_2+{\bf k}_3\bigg) \times \langle \hat{\zeta}_{\bf p}\hat{\zeta}_{-{\bf p}}\hat{\zeta}^{'}_{{\bf k}_1}(\tau)\hat{\zeta}_{{\bf k}_2}(\tau)\hat{\zeta}_{{\bf k}_3}(\tau)\rangle& \mbox{from equation} 
  (\ref{g0})  \\ 
			\displaystyle -\frac{iM^2_{ pl}}{2}a^2(\tau)\epsilon(\tau)\times\int \frac{d^{3}{\bf k}_1}{(2\pi)^3} \int \frac{d^{3}{\bf k}_2}{(2\pi)^3} \int \frac{d^{3}{\bf k}_3}{(2\pi)^3} \nonumber\\
  \displaystyle \times\delta^3\bigg({\bf k}_1+{\bf k}_2+{\bf k}_3\bigg) \times \langle \hat{\zeta}_{\bf p}\hat{\zeta}_{-{\bf p}}\hat{\zeta}^{'}_{{\bf k}_1}(\tau)\hat{\zeta}_{{\bf k}_2}(\tau)\hat{\zeta}_{{\bf k}_3}(\tau)\rangle^{\dagger} \quad\quad\quad& \mbox{from equation} 
  (\ref{g1})
	\end{array}
\right.
  \eea

Also in the present computation we have utilized the fact that the first slow-roll parameter $\epsilon$ is also constant in the SR and USR regime. For this reason one can immediately have the constraint,
$\epsilon^{'}(\tau)\approx 0$ which implies ${\cal H}^{''}(\tau)\approx 2{\cal H}^{'}(\tau)$. As a result, we have the following crucial immediate facts: 
$\partial_{\tau}{\cal J}_1(\tau)\approx 0$,
$\partial_{\tau_1}{\cal J}_2(\tau_1,\tau_2)\approx 0$,
$\partial_{\tau_2}{\cal J}_2(\tau_1=\tau_e,\tau_2)\approx 0$,
$\partial_{\tau_2}{\cal J}_2(\tau_1=\tau_s,\tau_2)\approx 0$.
Each of the consequences are extremely helpful to simplify the one-loop contribution to the two-point scalar correlation in the USR region.

\subsection{Computation of cut-off regulated one-loop momentum integrals}\label{app:B}

In this appendix our prime objective explicitly evaluate the full contributions of the momentum integrals using which in the late time scale, $\tau\rightarrow 0$ we quantify the one-loop contribution to the primordial scalar power spectrum from the SRI, SRII and USR region. 

\subsubsection{One-loop momentum integrals in USR period}
\paragraph{A. First contribution:}
Let us first write the contribution in terms of a momentum dependent integral appearing in the USR region:
\bea  \label{gk1} &&{\bf I}(\tau):=\int^{k_e}_{k_s}\frac{dk}{k}\;\left|{\cal G}_{\bf k}(\tau)\right|^{2},\eea
where we define a new function ${\cal G}_{\bf k}(\tau)$, which is defined as:
		\bea \label{hhgx} {\cal G}_{\bf k}(\tau)&=&\bigg[\alpha^{(2)}_{\bf k}\left(1+ikc_s\tau\right)\; e^{-ikc_s\tau}-\beta^{(2)}_{\bf k}\left(1-ikc_s\tau\right)\; e^{ikc_s\tau}\bigg].\eea
  Here the Bogoliubov coefficients $\alpha^{(2)}_{\bf k}$ and $\beta^{(2)}_{\bf k}$ in the USR region is given by the following expressions:
  \bea \alpha^{(2)}_{\bf k}&=&1-\frac{3}{2ik^{3}c^{3}_s\tau^{3}_s}\left(1+k^{2}c^{2}_s\tau^{2}_s\right),
  \quad\quad\quad
\beta^{(2)}_{\bf k}=-\frac{3}{2ik^{3}c^{3}_s\tau^{3}_s}\left(1+ikc_s\tau_s\right)^{2}\; e^{-2ikc_s\tau_s}.\eea
After substituting the specific form of the function ${\cal G}_{\bf k}(\tau)$ in equation (\ref{gk1}), we found the following simplified relation:
\bea \label{gk2} &&{\bf I}(\tau)= {\bf I}_1(\tau)+{\bf I}_2(\tau)+{\bf I}_3(\tau)+{\bf I}_4(\tau),\eea
where the four individual contributions ${\bf I}_i(\tau)\forall i=1,2,3,4$, can be defined as well as explicitly computed as:
\bea {\bf I}_1(\tau)&=&\int^{k_e}_{k_s}\frac{dk}{k}\;\bigg(1+\frac{9}{4}\frac{\left(1+k^2c^2_s\tau^2_s\right)^2}{k^6c^6_s\tau^6_s}\bigg) \left(1+k^2c^2_s\tau^2\right),\\
{\bf I}_2(\tau)&=&\int^{k_e}_{k_s}\frac{dk}{k}\;\bigg(\frac{9}{4}\frac{\left(1+k^2c^2_s\tau^2_s\right)^2}{k^6c^6_s\tau^6_s}\bigg) \left(1+k^2c^2_s\tau^2\right),\\
{\bf I}_3(\tau)&=&-\int^{k_e}_{k_s}\frac{dk}{k}\;\bigg(1+\frac{3}{2i}\frac{\left(1+k^2c^2_s\tau^2_s\right)}{k^3c^3_s\tau^3_s}\bigg)\bigg(-\frac{3}{2i} \frac{\left(1+ikc_s\tau_s\right)^2}{k^3c^3_s\tau^3_s}\bigg)\left(1-ikc_s\tau\right)^2\; e^{2ikc_s(\tau-\tau_s)},\\
{\bf I}_4(\tau)&=&-\int^{k_e}_{k_s}\frac{dk}{k}\;\bigg(1-\frac{3}{2i}\frac{\left(1+k^2c^2_s\tau^2_s\right)}{k^3c^3_s\tau^3_s}\bigg)\bigg(\frac{3}{2i} \frac{\left(1-ikc_s\tau_s\right)^2}{k^3c^3_s\tau^3_s}\bigg)\left(1+ikc_s\tau\right)^2\; e^{-2ikc_s(\tau-\tau_s)}.\eea
After performing a bit of algebraic steps we get the following results of the above mentioned integrals:
\bea \label{c1}{\bf I}_1(\tau)&=&\bigg[\frac{1}{2}\left(k^2_e-k^2_s\right)c^2_s\tau^2+\left(1+\frac{9}{4}\left(\frac{\tau}{\tau_s}\right)^2\right) \ln\left(\frac{k_e}{k_s}\right)-\frac{9}{8}\frac{1}{c^4_s\tau^4_s}\bigg(1+\left(\frac{\tau}{\tau_s}\right)^2\bigg)\bigg(\frac{1}{k^4_e}-\frac{1}{k^4_s}\bigg)\nonumber\\
&&\quad\quad\quad\quad\quad\quad\quad\quad\quad\quad\quad\quad -\frac{9}{8}\frac{1}{c^2_s\tau^2_s}\bigg(1+2\left(\frac{\tau}{\tau_s}\right)^2\bigg)\bigg(\frac{1}{k^2_e}-\frac{1}{k^2_s}\bigg)-\frac{3}{8}\frac{1}{c^6_s\tau^6_s}\bigg(\frac{1}{k^6_e}-\frac{1}{k^6_s}\bigg)\bigg],\\
\label{c2}{\bf I}_2(\tau)&=&\bigg[\frac{9}{4}\left(\frac{\tau}{\tau_s}\right)^2 \ln\left(\frac{k_e}{k_s}\right)-\frac{9}{8}\frac{1}{c^4_s\tau^4_s}\bigg(1+\frac{1}{2}\left(\frac{\tau}{\tau_s}\right)^2\bigg)\bigg(\frac{1}{k^4_e}-\frac{1}{k^4_s}\bigg)\nonumber\\
&&\quad\quad\quad\quad\quad\quad\quad\quad\quad\quad\quad\quad -\frac{9}{8}\frac{1}{c^2_s\tau^2_s}\bigg(1+2\left(\frac{\tau}{\tau_s}\right)^2\bigg)\bigg(\frac{1}{k^2_e}-\frac{1}{k^2_s}\bigg)-\frac{3}{8}\frac{1}{c^6_s\tau^6_s}\bigg(\frac{1}{k^6_e}-\frac{1}{k^6_s}\bigg)\bigg],\\
\label{c3}{\bf I}_3(\tau)&=&-\frac{1}{16 c^6_s \tau _s^6}\bigg[\bigg(36 c^6_s \tau^2 \tau _s^4+8 c^6_s \tau _s^6-8 c^6_s \tau^6 \bigg)\bigg(\text{Ei}\left(2 i c_s k_e \left(\tau-\tau _s\right)\right)-\text{Ei}\left(2 i c_s k_s \left(\tau-\tau _s\right)\right)\bigg)\nonumber\\
&&+ \bigg(\frac{12 c^6_s \tau^2 \tau _s^5}{\tau-\tau _s}\bigg(e^{2 i c_s k_e \left(\tau-\tau _s\right)}-e^{2 i c_s k_s \left(\tau-\tau _s\right)}\bigg)\nonumber\\
&&-4 i c^5_s \left(\tau^4 \tau _s+\tau^3 \tau _s^2+\tau^2 \tau _s^3+7 \tau \tau _s^4+\tau^5+\tau _s^5\right)\bigg(\frac{e^{2 i c_s k_e \left(\tau-\tau _s\right)}}{k_e}-\frac{e^{2 i c_s k_s \left(\tau-\tau _s\right)}}{k_s}\bigg)\nonumber\\
&&-2 c^4_s \left(2 \tau^3 \tau _s+3 \tau^2 \tau _s^2+28 \tau \tau _s^3+\tau^4-7 \tau _s^4\right)\bigg(\frac{e^{2 i c_s k_e \left(\tau-\tau _s\right)}}{k^2_e}-\frac{e^{2 i c_s k_s \left(\tau-\tau _s\right)}}{k^2_s}\bigg)\nonumber\\
&&+2 i c^3_s \left(3 \tau^2 \tau _s+6 \tau \tau _s^2+\tau^3-14 \tau _s^3\right)\bigg(\frac{e^{2 i c_s k_e \left(\tau-\tau _s\right)}}{k^3_e}-\frac{e^{2 i c_s k_s \left(\tau-\tau _s\right)}}{k^3_s}\bigg)\nonumber\\
&&+3 c^2_s \left(-8 \tau \tau _s+\tau^2-2 \tau _s^2\right)\bigg(\frac{e^{2 i c_s k_e \left(\tau-\tau _s\right)}}{k^4_e}-\frac{e^{2 i c_s k_s \left(\tau-\tau _s\right)}}{k^4_s}\bigg)\nonumber\\
&&+12 i c_s \left(\tau-\tau _s\right)\bigg(\frac{e^{2 i c_s k_e \left(\tau-\tau _s\right)}}{k^5_e}-\frac{e^{2 i c_s k_s \left(\tau-\tau _s\right)}}{k^5_s}\bigg)-6\bigg(\frac{e^{2 i c_s k_e \left(\tau-\tau _s\right)}}{k^6_e}-\frac{e^{2 i c_s k_s \left(\tau-\tau _s\right)}}{k^6_s}\bigg)\bigg)\bigg],\\
\label{c4}{\bf I}_4(\tau)&=&-\frac{1}{16 c^6_s \tau _s^6}\bigg[\bigg(36 c^6_s \tau^2 \tau _s^4+8 c^6_s \tau _s^6-8 c^6_s \tau^6 \bigg)\bigg(\text{Ei}\left(-2 i c_s k_e \left(\tau-\tau _s\right)\right)-\text{Ei}\left(-2 i c_s k_s \left(\tau-\tau _s\right)\right)\bigg)\nonumber\\
&&+ \bigg(\frac{12 c^6_s \tau^2 \tau _s^5}{\tau-\tau _s}\bigg(e^{-2 i c_s k_e \left(\tau-\tau _s\right)}-e^{-2 i c_s k_s \left(\tau-\tau _s\right)}\bigg)\nonumber\\
&&+4 i c_s^5 \left(\tau ^4 \tau _s+\tau ^3 \tau _s^2+\tau ^2 \tau _s^3+7 \tau  \tau _s^4+\tau _s^5+\tau ^5\right)\bigg(\frac{e^{-2 i c_s k_e \left(\tau-\tau _s\right)}}{k_e}-\frac{e^{-2 i c_s k_s \left(\tau-\tau _s\right)}}{k_s}\bigg)\nonumber\\
&&-2 c_s^4 \left(2 \tau ^3 \tau _s+3 \tau ^2 \tau _s^2+28 \tau  \tau _s^3-7 \tau _s^4+\tau ^4\right)\bigg(\frac{e^{-2 i c_s k_e \left(\tau-\tau _s\right)}}{k^2_e}-\frac{e^{-2 i c_s k_s \left(\tau-\tau _s\right)}}{k^2_s}\bigg)\nonumber\\
&&-2 i c_s^3 \left(3 \tau ^2 \tau _s+6 \tau  \tau _s^2-14 \tau _s^3+\tau ^3\right)\bigg(\frac{e^{-2 i c_s k_e \left(\tau-\tau _s\right)}}{k^3_e}-\frac{e^{-2 i c_s k_s \left(\tau-\tau _s\right)}}{k^3_s}\bigg)\nonumber\\
&&+3 c_s^2 \left(-8 \tau  \tau _s-2 \tau _s^2+\tau ^2\right)\bigg(\frac{e^{-2 i c_s k_e \left(\tau-\tau _s\right)}}{k^4_e}-\frac{e^{-2 i c_s k_s \left(\tau-\tau _s\right)}}{k^4_s}\bigg)\nonumber\\
&&-12 i c_s \left(\tau -\tau _s\right)\bigg(\frac{e^{-2 i c_s k_e \left(\tau-\tau _s\right)}}{k^5_e}-\frac{e^{-2 i c_s k_s \left(\tau-\tau _s\right)}}{k^5_s}\bigg)-6\bigg(\frac{e^{-2 i c_s k_e \left(\tau-\tau _s\right)}}{k^6_e}-\frac{e^{-2 i c_s k_s \left(\tau-\tau _s\right)}}{k^6_s}\bigg)\bigg)\bigg].\eea
Now to understand the behaviour of the obtained results in the USR regime let us add the all the contributions obtained in equation (\ref{c1}), equation (\ref{c2}), equation (\ref{c3}) and equation (\ref{c4}) together, which will give rise to the following simplified results:
\bea \label{c11}{\bf I}(\tau)&=&{\bf I}_1(\tau)+{\bf I}_2(\tau)+{\bf I}_3(\tau)+{\bf I}_4(\tau)\nonumber\\
&=&\Bigg\{\bigg[\frac{1}{2}\left(k^2_e-k^2_s\right)c^2_s\tau^2+\left(1+\frac{9}{2}\left(\frac{\tau}{\tau_s}\right)^2\right) \ln\left(\frac{k_e}{k_s}\right)-\frac{9}{4}\frac{1}{c^4_s\tau^4_s}\bigg(1+\frac{3}{4}\left(\frac{\tau}{\tau_s}\right)^2\bigg)\bigg(\frac{1}{k^4_e}-\frac{1}{k^4_s}\bigg)\nonumber\\
&&\quad\quad\quad\quad\quad\quad\quad\quad\quad\quad\quad\quad -\frac{9}{4}\frac{1}{c^2_s\tau^2_s}\bigg(1+2\left(\frac{\tau}{\tau_s}\right)^2\bigg)\bigg(\frac{1}{k^2_e}-\frac{1}{k^2_s}\bigg)-\frac{3}{4}\frac{1}{c^6_s\tau^6_s}\bigg(\frac{1}{k^6_e}-\frac{1}{k^6_s}\bigg)\bigg]\nonumber\eea
\bea
&&-\frac{1}{16 c^6_s \tau _s^6}\bigg[\bigg(36 c^6_s \tau^2 \tau _s^4+8 c^6_s \tau _s^6-8 c^6_s \tau^6 \bigg)\nonumber\\
&&\quad\quad\quad\quad\times\bigg(\text{Ei}\left(2 i c_s k_e \left(\tau-\tau _s\right)\right)+\text{Ei}\left(-2 i c_s k_e \left(\tau-\tau _s\right)\right)-\text{Ei}\left(2 i c_s k_s \left(\tau-\tau _s\right)\right)-\text{Ei}\left(-2 i c_s k_s \left(\tau-\tau _s\right)\right)\bigg)\nonumber\\
&&+ \bigg(\frac{24 c^6_s \tau^2 \tau _s^5}{\tau-\tau _s}\bigg(\cos \left(c_s k_e \left(\tau-\tau _s\right)\right)-\cos \left(c_s k_s \left(\tau-\tau _s\right)\right)\bigg)\nonumber\\
&&+8 c_s^5 \left(\tau ^4 \tau _s+\tau ^3 \tau _s^2+\tau ^2 \tau _s^3+7 \tau  \tau _s^4+\tau _s^5+\tau ^5\right)\bigg(\frac{\sin \left(c_s k_e \left(\tau-\tau _s\right)\right)}{k_e}-\frac{\sin \left(c_s k_s \left(\tau-\tau _s\right)\right)}{k_s}\bigg)\nonumber\\
&&-4 c_s^4 \left(2 \tau ^3 \tau _s+3 \tau ^2 \tau _s^2+28 \tau  \tau _s^3-7 \tau _s^4+\tau ^4\right)\bigg(\frac{\cos \left(c_s k_e \left(\tau-\tau _s\right)\right)}{k^2_e}-\frac{\cos \left(c_s k_s \left(\tau-\tau _s\right)\right)}{k^2_s}\bigg)\nonumber\\
&&-4 c_s^3 \left(3 \tau ^2 \tau _s+6 \tau  \tau _s^2-14 \tau _s^3+\tau ^3\right)\bigg(\frac{\sin \left(c_s k_e \left(\tau-\tau _s\right)\right)}{k^3_e}-\frac{\sin \left(c_s k_s \left(\tau-\tau _s\right)\right)}{k^3_s}\bigg)\nonumber\\
&&+6 c_s^2 \left(-8 \tau  \tau _s-2 \tau _s^2+\tau ^2\right)\bigg(\frac{\cos \left(c_s k_e \left(\tau-\tau _s\right)\right)}{k^4_e}-\frac{\cos \left(c_s k_s \left(\tau-\tau _s\right)\right)}{k^4_s}\bigg)\nonumber\\
&&-24 c_s \left(\tau -\tau _s\right)\bigg(\frac{\sin \left(c_s k_e \left(\tau-\tau _s\right)\right)}{k^5_e}-\frac{\sin \left(c_s k_s \left(\tau-\tau _s\right)\right)}{k^5_s}\bigg)\nonumber\\
&&-6\bigg(\frac{\cos \left(c_s k_e \left(\tau-\tau _s\right)\right)}{k^6_e}-\frac{\cos \left(c_s k_s \left(\tau-\tau _s\right)\right)}{k^6_s}\bigg)\bigg)\bigg]\Bigg\}.\eea
Now in the super-horizon late time limiting scale we have found the following result which will contribute to the one-loop integral:
\bea \boxed{{\bf I}(\tau_e)={\bf I}(\tau_s)\approx {\cal O}(1)+\ln\left(\frac{k_e}{k_s}\right)\approx \ln\left(\frac{k_e}{k_s}\right)=\ln\left(\frac{k_{\rm UV}}{k_{\rm IR}}\right)}.\eea
To obtain this simplified result we have neglected the contributions from the oscillatory terms as well as inverse power law terms of $(k_s/k_e)(\ll 1)$, as all of them will not be able to significant change the overall behaviour of the scalar power spectrum in USR period. Additionally, a contribution from ${\cal O}(1)$ is neglected for the same purpose. Another crucial important fact for neglecting ${\cal O}(1)$ contribution from the above mentioned result is more deeper from the physical ground. To match the contributions appearing in the sub-horizon, horizon crossing and super-horizon we found that ${\cal O}(1)$ effect is inconsistent. Because such effect is completely absent in the sub-horizon scale where the quantum effects are dominant. From the horizon crossing to the super-horizon scale such effect appear. However, to get a consistent reliable contribution throughout all the scales one can immediately neglect such contribution from our calculation. Only the relevant effect comes from the logarithmically divergent contribution which will finally appear in the one-loop corrected result of the primordial power spectrum for scalar modes of perturbation in the USR period. It's also crucial to note that in order to extract the finite contributions from each of the aforementioned integrals, we have constrained the momentum integration within a window, $k_s<k<k_e$, by introducing two well known physical cut-offs: the IR cut-off $k_{\rm IR}=k_s$ and the UV cut-off $k_{\rm UV}=k_e$. 
\paragraph{B. Second contribution:}
Let us write the sub-leading contribution in terms of a momentum dependent integral appearing in the USR region:
\bea  \label{gslk1} &&{\bf F}(\tau):=\int^{k_e}_{k_s}\frac{dk}{k}\;|{\cal G}_{\bf k}(\tau)|^2\bigg(\frac{d\ln|{\cal G}_{\bf k}(\tau)|^2 }{d\ln k}\bigg)=\int^{k_e}_{k_s}d\ln k\;\bigg(\frac{d|{\cal G}_{\bf k}(\tau)|^2 }{d\ln k}\bigg)=\Bigg[|{\cal G}_{\bf k}(\tau)|^2\Bigg]^{k_e}_{k_s},\eea
where we have already defined the function ${\cal G}_{\bf k}$ in equation (\ref{hhgx}). After substituting the specific form of the function ${\cal G}_{\bf k}(\tau)$ in equation (\ref{gslk1}), we found the following simplified relation:
\bea \label{gksim2} &&{\bf F}(\tau)= {\bf F}_1(\tau)+{\bf F}_2(\tau)+{\bf F}_3(\tau)+{\bf F}_4(\tau),\eea
where the four individual contributions ${\bf F}_i(\tau)\forall i=1,2,3,4$, can be defined as well as explicitly computed as:
\bea {\bf F}_1(\tau)&=&\Bigg[\bigg(1+\frac{9}{4}\frac{\left(1+k^2_ec^2_s\tau^2_s\right)^2}{k^6_ec^6_s\tau^6_s}\bigg) \left(1+k^2_ec^2_s\tau^2\right)-\bigg(1+\frac{9}{4}\frac{\left(1+k^2_sc^2_s\tau^2_s\right)^2}{k^6_sc^6_s\tau^6_s}\bigg) \left(1+k^2_sc^2_s\tau^2\right)\Bigg],\\
{\bf F}_2(\tau)&=&\Bigg[\bigg(\frac{9}{4}\frac{\left(1+k^2_ec^2_s\tau^2_s\right)^2}{k^6_ec^6_s\tau^6_s}\bigg) \left(1+k^2_ec^2_s\tau^2\right)-\bigg(\frac{9}{4}\frac{\left(1+k^2_sc^2_s\tau^2_s\right)^2}{k^6_sc^6_s\tau^6_s}\bigg) \left(1+k^2_sc^2_s\tau^2\right)\Bigg],\\
{\bf F}_3(\tau)&=&-\Bigg[\bigg(1+\frac{3}{2i}\frac{\left(1+k^2_ec^2_s\tau^2_s\right)}{k^3_ec^3_s\tau^3_s}\bigg)\bigg(-\frac{3}{2i} \frac{\left(1+ik_ec_s\tau_s\right)^2}{k^3_ec^3_s\tau^3_s}\bigg)\left(1-ik_ec_s\tau\right)^2\; e^{2ik_ec_s(\tau-\tau_s)}\nonumber\\
&&\quad\quad\quad\quad\quad\quad-\bigg(1+\frac{3}{2i}\frac{\left(1+k^2_sc^2_s\tau^2_s\right)}{k^3_sc^3_s\tau^3_s}\bigg)\bigg(-\frac{3}{2i} \frac{\left(1+ik_sc_s\tau_s\right)^2}{k^3_sc^3_s\tau^3_s}\bigg)\left(1-ik_sc_s\tau\right)^2\; e^{2ik_sc_s(\tau-\tau_s)}\Bigg],\\
{\bf F}_4(\tau)&=&-\Bigg[\bigg(1-\frac{3}{2i}\frac{\left(1+k^2_ec^2_s\tau^2_s\right)}{k^3_ec^3_s\tau^3_s}\bigg)\bigg(\frac{3}{2i} \frac{\left(1-ik_ec_s\tau_s\right)^2}{k^3_ec^3_s\tau^3_s}\bigg)\left(1+ik_ec_s\tau\right)^2\; e^{-2ik_ec_s(\tau-\tau_s)}\nonumber\\
&&\quad\quad\quad\quad\quad\quad-\bigg(1-\frac{3}{2i}\frac{\left(1+k^2_sc^2_s\tau^2_s\right)}{k^3_sc^3_s\tau^3_s}\bigg)\bigg(\frac{3}{2i} \frac{\left(1-ik_sc_s\tau_s\right)^2}{k^3_sc^3_s\tau^3_s}\bigg)\left(1+ik_sc_s\tau\right)^2\; e^{-2ik_sc_s(\tau-\tau_s)}\Bigg].\eea
Now in the super-horizon late time limiting scale we have found the following results for the above mentioned functions:
\bea {\bf F}_1&=&\frac{9}{4}\Bigg[\frac{\left(1+k^2_ec^2_s\tau^2_s\right)^2}{k^6_ec^6_s\tau^6_s}-\frac{\left(1+k^2_sc^2_s\tau^2_s\right)^2}{k^6_sc^6_s\tau^6_s}\Bigg],\\
{\bf F}_2&=&\frac{9}{4}\Bigg[\frac{\left(1+k^2_ec^2_s\tau^2_s\right)^2}{k^6_ec^6_s\tau^6_s}-\frac{\left(1+k^2_sc^2_s\tau^2_s\right)^2}{k^6_sc^6_s\tau^6_s}\Bigg],\\
{\bf F}_3&=&-\Bigg[\bigg(1+\frac{3}{2i}\frac{\left(1+k^2_ec^2_s\tau^2_s\right)}{k^3_ec^3_s\tau^3_s}\bigg)\bigg(-\frac{3}{2i} \frac{\left(1+ik_ec_s\tau_s\right)^2}{k^3_ec^3_s\tau^3_s}\bigg)\; e^{-2ik_ec_s\tau_s}\nonumber\\
&&\quad\quad\quad\quad\quad\quad-\bigg(1+\frac{3}{2i}\frac{\left(1+k^2_sc^2_s\tau^2_s\right)}{k^3_sc^3_s\tau^3_s}\bigg)\bigg(-\frac{3}{2i} \frac{\left(1+ik_sc_s\tau_s\right)^2}{k^3_sc^3_s\tau^3_s}\bigg)\; e^{-2ik_sc_s\tau_s}\Bigg],\\
{\bf F}_4&=&-\Bigg[\bigg(1-\frac{3}{2i}\frac{\left(1+k^2_ec^2_s\tau^2_s\right)}{k^3_ec^3_s\tau^3_s}\bigg)\bigg(\frac{3}{2i} \frac{\left(1-ik_ec_s\tau_s\right)^2}{k^3_ec^3_s\tau^3_s}\bigg)\; e^{2ik_ec_s\tau_s}\nonumber\\
&&\quad\quad\quad\quad\quad\quad-\bigg(1-\frac{3}{2i}\frac{\left(1+k^2_sc^2_s\tau^2_s\right)}{k^3_sc^3_s\tau^3_s}\bigg)\bigg(\frac{3}{2i} \frac{\left(1-ik_sc_s\tau_s\right)^2}{k^3_sc^3_s\tau^3_s}\bigg)\; e^{2ik_sc_s\tau_s}\Bigg].\eea
After adding all of these above mentioned contributions in the super-horzon scale and performing a little bit of algebra we get the following simplified result:
\bea \boxed{{\bf F}(\tau_e)={\bf F}(\tau_s)=-{\cal O}(1)}.\eea
To obtain this simplified result we have neglected the contributions from the oscillatory terms as well as inverse power law terms of $(k_s/k_e)(\ll 1)$, as all of them will not be able to significant change the overall behaviour of the scalar power spectrum in USR period in the sub-leading order. Most importantly, this particular contribution is completely free any type of divergences and for this reason this is a suppressed contribution compared to the leading logarithmically divergent contribution when the effective sound speed $c_s<1$. On the other hand, for $c_s>1$ one can show that the second contribution dominates over the first one as computed in the previous subsection.

\paragraph{C. Quantifying the relative strength of the contributions:}
Now to understand the relative amplitude difference between these two contributions let us take the ratio of the momentum integrals, which is given by the following expression in the super-horizon scale: 
\bea \boxed{\frac{{\bf F}(\tau_e)}{{\bf I}(\tau_e)}=\frac{{\bf F}(\tau_s)}{{\bf I}(\tau_s)}=\frac{{\bf F}}{{\bf I}}\approx\frac{{\cal O}(1)}{\displaystyle\ln\left(\frac{k_s}{k_e}\right)}}.\eea
For the numerical purpose we choose, $k_e=10^{22}{\rm Mpc}^{-1}$ and $k_s=10^{21}{\rm Mpc}^{-1}$, for which we have $k_s/k_e=10^{-1}\ll 1$. Consequently, we have the following numerical contribution from the corresponding ratio for the given choices of the wave number, $k_e$ and $k_s$:
\bea \frac{{\bf F}(\tau_e)}{{\bf I}(\tau_e)}=\frac{{\bf F}(\tau_s)}{{\bf I}(\tau_s)}=\frac{{\bf F}}{{\bf I}}\approx-0.43<1.\eea
This estimation correctly justifies the fact that, the integral ${\bf F}$ is suppressed compared to the contribution coming from the integral ${\bf I}$. Also this estimation tells us that the contribution coming from ${\bf F}$ will add correction to ${\bf I}$. However, it seems like that the correction is though sub-leading but it is not very small. To give a correct estimation one needs to take care of the pre-factors which is appearing in the expression for the one-loop contribution coming from the USR period. To show this explicitly let us write down the contribution explicitly which we have quoted in the text portion, and it is given by the following expression:
\bea {\bf Z}_1:&=&\frac{1}{4}\bigg[\Delta^{2}_{\zeta,{\bf Tree}}(p)\bigg]_{\bf SR}\times \Bigg(\frac{\left(\Delta\eta(\tau_e)\right)^2}{c^8_s} \left(\frac{k_e}{k_s}\right)^{6}-\frac{\left(\Delta\eta(\tau_s)\right)^2}{c^8_s}\Bigg)\times{\bf I},\\
{\bf Z}_2:&=&\frac{1}{2}\bigg[\Delta^{2}_{\zeta,{\bf Tree}}(p)\bigg]_{\bf SR}\times\bigg(\frac{\left(\Delta\eta(\tau_e)\right)}{c^4_s}\left(\frac{k_e}{k_s}\right)^{6}-\frac{\left(\Delta\eta(\tau_s)\right)}{c^4_s}\bigg)\times{\bf F}.\eea
Now to take care of the pre-factors properly instead of taking the ratio of ${\bf F}$ and ${\bf I}$, let us take the ratio of ${\bf Z}_2$ and ${\bf Z}_1$, which is given by the following expression:
\bea \boxed{\frac{{\bf Z}_2}{{\bf Z}_1}=2\times \frac{\displaystyle \bigg(\frac{\left(\Delta\eta(\tau_e)\right)}{c^4_s}\left(\frac{k_e}{k_s}\right)^{6}-\frac{\left(\Delta\eta(\tau_s)\right)}{c^4_s}\bigg)}{\displaystyle \Bigg(\frac{\left(\Delta\eta(\tau_e)\right)^2}{c^8_s} \left(\frac{k_e}{k_s}\right)^{6}-\frac{\left(\Delta\eta(\tau_s)\right)^2}{c^8_s}\Bigg)}\times \frac{{\bf F}}{{\bf I}}}.\eea
Here for the numerical purpose we take $\Delta\eta(\tau_e)=1$ and $\Delta\eta(\tau_s)=-6$. We choose, $k_e=10^{22}{\rm Mpc}^{-1}$ and $k_s=10^{21}{\rm Mpc}^{-1}$, for which we have $k_s/k_e=10^{-1}\ll 1$. Additionally we consider the sound speed will lie within the widow, $0.6<c_s<1.5$. To study the effect of the sound speed let us consider three regions, where the sound speed lie within the window, $0.6< c_s< 1$, $c_s=1$ and $1< c_s< 1.5$. Then we found the following constraints on this ratio in the two consecutive regions:
\bea &&\underline{{\bf For}\;\; 0.6< c_s< 1 :}\quad\quad 0.12<\frac{{\bf Z}_2}{{\bf Z}_1}<0.86,\\
&&\underline{{\bf For}\;\; c_s= 1 :}\quad\quad\quad\quad\quad\quad\quad\;\;\;\frac{{\bf Z}_2}{{\bf Z}_1}\sim 0.86,\\
&&\underline{{\bf For}\;\; 1< c_s< 1.5 :}\quad\quad 0.86<\frac{{\bf Z}_2}{{\bf Z}_1}<4.35.\eea
\subsubsection{One-loop momentum integrals in SRI period}
Next, we will talk about the following integral which is appearing in the computation of the one-loop correction to the primordial power spectrum of scalar modes in the SRI region. Let us evaluate the following integral:
\bea\label{intSR} {\bf K}(\tau):=\int^{k_e}_{p_*}\frac{dk}{k}\;\left|{\cal M}_{\bf k}(\tau)\right|^{2},\eea
where we define a new function ${\cal M}_{\bf k}(\tau)$, which is defined as:
		\bea {\cal M}_{\bf k}(\tau)&=&\left(1+ikc_s\tau\right)\; e^{-ikc_s\tau}.\eea
After substituting the explicit form of the above function ${\cal M}_{\bf k}(\tau)$ in equation (\ref{intSR}), we get the following result:
\bea{\bf K}(\tau)=\int^{k_e}_{p_*}\frac{dk}{k}\;\left(1+k^2c^2_s\tau^2\right)=\bigg[\ln\left(\frac{k_e}{p_*}\right)+\frac{1}{2}\left(k^2_e-p^2_*\right)c^2_s\tau^2\bigg].\eea
Then in the super-horizon late time limiting scale we found the following simplified result:
\bea \boxed{{\bf K}(\tau_e)=\ln\left(\frac{k_e}{p_*}\right)=\ln\left(\frac{k_{\rm UV}}{p_*}\right)}.\eea
Here in SRI phase $p_*$ represents the pivot scale which is expected to be $p_*\ll k_s$ in the present framework under consideration. In the SRI region the final one-loop result will be only controlled by the above mentioned logarithmically divergent contribution. 

\subsubsection{One-loop momentum integrals in SRII period}
Let us first write the contribution in terms of a momentum dependent integral appearing in the SRII region:
\bea  \label{gkk1} &&{\bf O}(\tau):=\left(\frac{\tau_s}{\tau_e}\right)^6\int^{k_{\rm end}}_{k_e}\frac{dk}{k}\;\left|{\cal X}_{\bf k}(\tau)\right|^{2},\eea
where we define a new function ${\cal X}_{\bf k}(\tau)$, which is defined as:
		\bea {\cal X}_{\bf k}(\tau)&=&\bigg[\alpha^{(3)}_{\bf k}\left(1+ikc_s\tau\right)\; e^{-ikc_s\tau}-\beta^{(3)}_{\bf k}\left(1-ikc_s\tau\right)\; e^{ikc_s\tau}\bigg].\eea
  Here the Bogoliubov coefficients $\alpha^{(3)}_{\bf k}$ and $\beta^{(3)}_{\bf k}$ in the SRII region is given by the following expressions:
\bea \alpha^{(3)}_{\bf k}&=&-\frac{1}{4k^6c^6_s\tau^3_s\tau^3_e}\Bigg[9\left(kc_s\tau_s-i\right)^2\left(kc_s\tau_e+i\right)^2 e^{2ikc_s(\tau_e-\tau_s)}\nonumber\\
&&\quad\quad\quad\quad\quad\quad\quad\quad\quad\quad\quad\quad\quad\quad\quad\quad-
\left\{k^2c^2_s\tau^2_e\left(2kc_s\tau_e-3i\right)-3i\right\}\left\{k^2c^2_s\tau^2_s\left(2kc_s\tau_s+3i\right)+3i\right\}\Bigg],\\
\beta^{(3)}_{\bf k}&=&\frac{3}{4k^6c^6_s\tau^3_s\tau^3_e}\Bigg[\left(kc_s\tau_s-i\right)^2\left\{k^2c^2_s\tau^2_e\left(3-2ikc_s\tau_e\right)+3\right\}e^{-2ikc_s\tau_s}\nonumber\\
&&\quad\quad\quad\quad\quad\quad\quad\quad\quad\quad\quad\quad\quad\quad\quad\quad+i\left(kc_s\tau_e-i\right)^2\left\{3i+k^2c^2_s\tau^2_s\left(2kc_s\tau_s+3i\right)\right\}e^{-2ikc_s\tau_e}\Bigg].\eea
After substituting the specific form of the function ${\cal X}_{\bf k}(\tau)$ in equation (\ref{gkk1}), we found the following simplified relation:
\bea \label{gk21} &&{\bf O}(\tau)= \left(\frac{\tau_s}{\tau_e}\right)^6\Bigg[{\bf O}_1(\tau)+{\bf O}_2(\tau)+{\bf O}_3(\tau)+{\bf O}_4(\tau)\Bigg],\eea
where the four individual contributions ${\bf O}_i(\tau)\forall i=1,2,3,4$, can be defined as well as explicitly computed as:
\bea {\bf O}_1(\tau)&=&\int^{k_{\rm end}}_{k_e}\frac{dk}{k}\;\frac{1}{16k^{12}c^{12}_s\tau^6_s\tau^6_e}\Bigg|\Bigg[9\left(kc_s\tau_s-i\right)^2\left(kc_s\tau_e+i\right)^2 e^{2ikc_s(\tau_e-\tau_s)}\nonumber\\
&&\quad\quad\quad\quad\quad\quad\quad\quad-
\left\{k^2c^2_s\tau^2_e\left(2kc_s\tau_e-3i\right)-3i\right\}\left\{k^2c^2_s\tau^2_s\left(2kc_s\tau_s+3i\right)+3i\right\}\Bigg]\Bigg|^{2} \left(1+k^2c^2_s\tau^2\right),\\
{\bf O}_2(\tau)&=&\int^{k_{\rm end}}_{k_e}\frac{dk}{k}\;\frac{9}{16k^{12}c^{12}_s\tau^6_s\tau^6_e}\Bigg|\Bigg[\left(kc_s\tau_s-i\right)^2\left\{k^2c^2_s\tau^2_e\left(3-2ikc_s\tau_e\right)+3\right\}e^{-2ikc_s\tau_s}\nonumber\\
&&\quad\quad\quad\quad\quad\quad\quad\quad+i\left(kc_s\tau_e-i\right)^2\left\{3i+k^2c^2_s\tau^2_s\left(2kc_s\tau_s+3i\right)\right\}e^{-2ikc_s\tau_e}\Bigg]\Bigg|^{2} \left(1+k^2c^2_s\tau^2\right),\\
{\bf O}_3(\tau)&=&\int^{k_{\rm end}}_{k_e}\frac{dk}{k}\;\frac{3}{16k^{12}c^{12}_s\tau^6_s\tau^6_e}\Bigg[9\left(kc_s\tau_s-i\right)^2\left(kc_s\tau_e+i\right)^2 e^{2ikc_s(\tau_e-\tau_s)}\nonumber\\
&&\quad\quad\quad\quad\quad\quad\quad\quad-
\left\{k^2c^2_s\tau^2_s\left(2kc_s\tau_e-3i\right)-3i\right\}\left\{k^2c^2_s\tau^2_s\left(2kc_s\tau_s+3i\right)+3i\right\}\Bigg]\nonumber\eea
\bea
&&\quad\quad\quad\quad\quad\quad\quad\quad\times\Bigg[\left(kc_s\tau_s+i\right)^2\left\{k^2c^2_s\tau^2_e\left(3+2ikc_s\tau_e\right)+3\right\}e^{2ikc_s\tau_s}\nonumber\\
&&\quad\quad\quad\quad\quad\quad\quad\quad+i\left(kc_s\tau_e+i\right)^2\left\{-3i+k^2c^2_s\tau^2_s\left(2kc_s\tau_s-3i\right)\right\}e^{2ikc_s\tau_e}\Bigg]\left(1-ikc_s\tau\right)^2\; e^{2ikc_s(\tau-\tau_s)},\\
{\bf O}_4(\tau)&=&\int^{k_{\rm end}}_{k_e}\frac{dk}{k}\;\frac{3}{16k^{12}c^{12}_s\tau^6_s\tau^6_e}\Bigg[9\left(kc_s\tau_s+i\right)^2\left(kc_s\tau_e-i\right)^2 e^{-2ikc_s(\tau_e-\tau_s)}\nonumber\\
&&\quad\quad\quad\quad\quad\quad\quad\quad-
\left\{k^2c^2_s\tau^2_e\left(2kc_s\tau_e+3i\right)+3i\right\}\left\{k^2c^2_s\tau^2_s\left(2kc_s\tau_s-3i\right)-3i\right\}\Bigg]\nonumber\\
&&\quad\quad\quad\quad\quad\quad\quad\quad\times\Bigg[\left(kc_s\tau_s-i\right)^2\left\{k^2c^2_s\tau^2_e\left(3-2ikc_s\tau_e\right)+3\right\}e^{-2ikc_s\tau_s}\nonumber\\
&&\quad\quad\quad\quad\quad\quad\quad\quad+i\left(kc_s\tau_e-i\right)^2\left\{3i+k^2c^2_s\tau^2_s\left(2kc_s\tau_s+3i\right)\right\}e^{-2ikc_s\tau_e}\Bigg]\left(1+ikc_s\tau\right)^2\; e^{-2ikc_s(\tau-\tau_s)}.\quad\quad\quad\quad\eea
Explicit results can be easily computed from the above mentioned integral. However, due to having cumbersome expressions we avoid here quote the individual results in detail. Now in the super-horizon late time limiting scale we have found the following result which will contribute to the
one-loop integral:
\bea \boxed{{\bf O}(\tau_e)={\bf O}(\tau_{\rm end})\approx -\ln \left(\frac{k_e}{k_{\rm end}}\right)-\frac{27}{32}\Bigg\{1-\left(\frac{k_e}{k_{\rm end}}\right)^{12}\Bigg\}}.\eea
In the SRII phase $k_e\ll k_{\rm end}$, for this reason $k_e/k_{\rm end}\ll 1$ and such contributions are highly suppressed. For the numerical purpose we choose, $k_e=10^{22}{\rm Mpc}^{-1}$ and $k_{\rm end}=10^{24}{\rm Mpc}^{-1}$, for which we have $k_e/k_{\rm end}=10^{-2}\ll 1$. In the SRII region the final one-loop result will be controlled by a logarithmic contribution and a power law contributions. However, due to having the constraint $k_e/k_{\rm end}\ll 1$ the logarithmic term is more dominant than the power law term. Consequently, we have the following numerical contribution from the corresponding integral for the given choices of the wave number, $k_e$ and $k_{\rm end}$:
\bea {\bf O}(\tau_e)={\bf O}(\tau_{\rm end})\approx 3.76.\eea
It is important to note that, this particular contribution will not appear in the case of Framework II.


\newpage
\bibliographystyle{utphys}
\bibliography{references2}

\end{document}